\documentclass[12pt]{article}
\usepackage{amsmath}
\usepackage{graphicx}
\usepackage{enumerate}
\usepackage{natbib}
\usepackage{url} 

\usepackage{mathtools, comment}
\usepackage{graphicx, color, url}

\usepackage{amsmath}
\usepackage{dsfont}
\usepackage{bbm}
\usepackage{amssymb}
\usepackage{epstopdf}
\usepackage{boxedminipage}
\usepackage{amsfonts}
\usepackage{url}
\usepackage{amsmath}
\usepackage{amsbsy}
\usepackage{setspace}
\usepackage{bm}
\usepackage{wasysym}
\usepackage{textcomp}
\usepackage{tikz}
\usepackage{multirow}
\usepackage{tabularx,ragged2e,booktabs,caption}
\usepackage{subcaption}
\usepackage{graphicx}
\usetikzlibrary{trees}
\usepackage[colorlinks,citecolor=blue,urlcolor=blue,bookmarks=false,hypertexnames=true]{hyperref} 
\usepackage{amsthm}
\usepackage[a4paper]{geometry}
\usepackage{graphicx}
\usepackage[textwidth=8em,textsize=small]{todonotes}
\usepackage{natbib}
\usepackage{algorithm}
\usepackage{algpseudocode}

\newtheorem{theorem}{Theorem}
\newtheorem{assumption}{Assumption}
\newtheorem{corollary}{Corollary}
\newtheorem{proposition}{Proposition}
\newtheorem{lemma}{Lemma}
\newtheorem{example}{Example}
\newtheorem{definition}{Definition}
\newtheorem{remark}{Remark}

\newcommand{\beq}{\begin{equation}}
\newcommand{\eeq}{\end{equation}}
\newcommand{\beas}{\begin{eqnarray*}}
\newcommand{\eeas}{\end{eqnarray*}}
\newcommand{\bea}{\begin{eqnarray}}
\newcommand{\eea}{\end{eqnarray}}
\newcommand{\bei}{\begin{itemize}}
\newcommand{\eei}{\end{itemize}}
\newcommand{\ben}{\begin{enumerate}}
\newcommand{\een}{\end{enumerate}}
\newcommand{\bet}{\begin{theorem}}
\newcommand{\eet}{\end{theorem}}
\newcommand{\bel}{\begin{lemma}}
\newcommand{\eel}{\end{lemma}}
\newcommand{\bep}{\begin{proposition}}
\newcommand{\eep}{\end{proposition}}
\newcommand{\bed}{\begin{definition}}
\newcommand{\eed}{\end{definition}}
\newcommand{\bec}{\begin{corollary}}
\newcommand{\eec}{\end{corollary}}
\newcommand{\bex}{\begin{example}}
\newcommand{\eex}{\end{example}}

\newcommand{\EE}{\mathbb E}
\newcommand{\PP}{\mathbb P}
\newcommand{\II}{\mathbb I}

\def\limsup{\mathop{\overline{\rm lim}}}
\def\liminf{\mathop{\underline{\rm lim}}}

\def\II{{\mathbb I}}
\def\mD{{\mathcal D}}
\def\limsup{\mathop{\overline{\rm lim}}}
\def\liminf{\mathop{\underline{\rm lim}}}

\begin{document}

\def\spacingset#1{\renewcommand{\baselinestretch}%
{#1}\small\normalsize} \spacingset{1}


{
\title{A Burden Shared is a Burden Halved: A Fairness-Adjusted Approach to  Classification}
\author{Bradley Rava$^1$, Wenguang Sun$^{2}$, Gareth M. James$^3$ and Xin Tong$^4$} 

\date{}
\footnotetext[1]{University of Sydney Business School.}
\footnotetext[2]{Center for Data Science and School of Management, Zhejiang University.}
\footnotetext[3]{Goizueta Business School, Emory University.}
\footnotetext[4]{Faculty of Business and Economics, University of Hong Kong. \\ \medskip 

We are grateful to the associate editor and two referees whose meticulous and constructive feedback has substantially improved the clarity, presentation and theory of our manuscript. We would also like to thank Matteo Sesia, Zinan Zhao and Wangcheng Li for their valuable discussion and suggestions on methodology and theory. 
}

\maketitle
} 

\bigskip
\begin{abstract}

We investigate the fairness issue in classification, where automated decisions are made for individuals from different protected groups. In high-consequence scenarios, decision errors can disproportionately affect certain protected groups, leading to unfair outcomes. To address this issue, we propose a fairness-adjusted selective inference (FASI) framework and develop data-driven algorithms that achieve statistical parity by controlling the false selection rate (FSR) among protected groups. Our FASI algorithm operates by converting the outputs of black-box classifiers into R-values, which are both intuitive and computationally efficient. These R-values serve as the basis for selection rules that are provably valid for FSR control in finite samples for protected groups, effectively mitigating the unfairness in group-wise error rates. We demonstrate the numerical performance of our approach using both simulated and real data.

\end{abstract}

\noindent%
\vfill

\newpage
\spacingset{1} 

\section{Introduction} \label{section:intro}

In a broad range of applications, artificial intelligence (AI) systems are rapidly replacing human decision-making. Many of these scenarios are sensitive in nature, where the AI's decision, correct or not, can directly impact one's social or economic status. A few examples include a bank determining credit card limits, stores using facial recognition systems to detect shoplifters, and hospitals attempting to identify which of their patients has a specific disorder. Unfortunately, despite their supposedly unbiased approach to decision-making, there has been increasing evidence that AI algorithms often fail to treat equally people of different genders, races, religions, or other protected attributes. Whether this is due to the historical bias in one's training data, or otherwise, it is important, for both legal and policy reasons, that we make ethical use of data and ensure that decisions are made fairly for everyone regardless of their protected attributes. 

Despite the significant efforts in developing supervised learning algorithms to improve the prediction accuracy, making reliable and fair decisions in the classification setting remains a critical and challenging problem for two main reasons. Firstly, AI algorithms are often required to make classifications on all new observations without a careful assessment of associated uncertainty or ambiguity. This limitation highlights the need for a more flexible framework to handle intrinsically difficult classification tasks where a definitive decision carries high stakes. Such a framework should enable decision-makers to wait and gather additional information with greater confidence before making a final decision. Secondly, modern machine learning models, such as neural networks, are often highly complex, making it challenging, if not impossible, to explicitly quantify the uncertainty associated with their outputs or to provide guarantees on the fairness of the decisions. Therefore, developing methods that can ensure both risk control and fairness is crucial for AI systems to be reliable and trustworthy.
 
This article develops a ``fairness-adjusted selective inference'' (FASI) framework to address the critical issues of uncertainty assessment, error rate control and statistical parity in classification. We provide an \textit{indecision} option for observations which cannot be selected into any classes with confidence. These observations will then be separately evaluated. This practice often aligns with the policy objectives in many real world scenarios. For example, incorrectly classifying a low-risk individual as a recidivist or rejecting a well-deserving candidate for the loan request is much more expensive than turning the case over for a more careful review. A mis-classification is an error, the probability of which must be controlled to be small as its consequence can be severe. By contrast, the cost of an indecision is usually much less. For example, the ambiguity can be mitigated by  collecting additional contextual knowledge of the convicted individual or requesting more information from the loan applicant. Under the selective inference \citep{Ben10} framework, we only make definitive decisions on a \emph{selected subset} of all individuals; the less consequential indecision option is considered as a wasted opportunity rather than an error. A natural error rate notion under this framework is the \textit{False Selection Rate} (FSR), which is defined as the expected fraction of erroneous classifications among the selected subset of individuals. The goal is to develop decision rules that aim to control and equalize the FSR across protected groups, while minimizing the total wasted opportunities. 

A critical issue is that a classification rule that controls the overall FSR may have disparate impacts on different protected groups. We illustrate the point using the COMPAS data set (\citealp{Angetal16-COMPAS1, Dieetal16-COMPAS2}). The COMPAS algorithm has been widely used in the US to help inform courts about a defendant's recidivism likelihood, i.e., the likelihood of a convicted criminal recommitting a crime, so any prediction errors could have significant implications. 
The left panel of Figure~\ref{cc_intro.plot} shows the {\it False Selection Proportions} (FSP), i.e. the fraction of individuals who did not recommit a crime among those who were classified as recidivists. The classification rule was constructed via a Generalized Additive Model (GAM) \footnote{Although a GAM was utilized for illustration purposes, we emphasize that the same issue can arise regardless of the specific machine learning algorithm employed.} \citep{hastie2009elements, James2013} to achieve the target FSR of 25\%. We first split the COMPAS data into distinct training and test sets. The GAM was fitted using the training data set, and subsequently applied to the test set to predict whether a defendant was a recidivist.

We can see that the green bar, which provides the overall FSP for all races, is close to the target value. Moreover, the rule appears to be  ``fair'' for all individuals, regardless of their protected attributes, in the sense that the \emph{same} threshold has been applied to the confidence scores (i.e. estimated class probabilities) produced by the \emph{same} GAM fit. However, the blue and orange bars show that the FSPs for different racial groups differ significantly from 25\%, which is clearly not a desirable situation.   

\begin{figure}[t]
    \centering 
    \scalebox{1}{%
\begin{tikzpicture}[x=1pt,y=1pt]
\definecolor{fillColor}{RGB}{255,255,255}
\path[use as bounding box,fill=fillColor,fill opacity=0.00] (0,0) rectangle (361.35,130.09);
\begin{scope}
\path[clip] (  0.00, 24.34) rectangle (180.67,130.09);
\definecolor{drawColor}{RGB}{255,255,255}
\definecolor{fillColor}{RGB}{255,255,255}

\path[draw=drawColor,line width= 0.4pt,line join=round,line cap=round,fill=fillColor] (  0.00, 24.34) rectangle (180.67,130.09);
\end{scope}
\begin{scope}
\path[clip] ( 24.72, 44.29) rectangle (177.17,126.59);
\definecolor{fillColor}{RGB}{255,255,255}

\path[fill=fillColor] ( 24.72, 44.29) rectangle (177.17,126.59);
\definecolor{fillColor}{RGB}{213,94,0}

\path[fill=fillColor,fill opacity=0.75] (120.00,-296.10) rectangle (158.12,116.36);
\definecolor{fillColor}{RGB}{0,114,178}

\path[fill=fillColor,fill opacity=0.75] ( 81.89,-296.10) rectangle (120.00, 54.54);
\definecolor{fillColor}{RGB}{0,158,115}

\path[fill=fillColor,fill opacity=0.75] ( 43.78,-296.10) rectangle ( 81.89, 72.59);
\definecolor{drawColor}{RGB}{0,0,0}

\path[draw=drawColor,line width= 0.6pt,dash pattern=on 4pt off 4pt ,line join=round] ( 24.72, 77.96) -- (177.17, 77.96);
\end{scope}
\begin{scope}
\path[clip] (  0.00,  0.00) rectangle (361.35,130.09);
\definecolor{drawColor}{RGB}{0,0,0}

\path[draw=drawColor,line width= 0.4pt,line join=round] ( 24.72, 44.29) --
	( 24.72,126.59);
\end{scope}
\begin{scope}
\path[clip] (  0.00,  0.00) rectangle (361.35,130.09);
\definecolor{drawColor}{gray}{0.30}

\node[text=drawColor,anchor=base east,inner sep=0pt, outer sep=0pt, scale=  0.56] at ( 21.57, 46.11) {0.23};

\node[text=drawColor,anchor=base east,inner sep=0pt, outer sep=0pt, scale=  0.56] at ( 21.57, 61.07) {0.24};

\node[text=drawColor,anchor=base east,inner sep=0pt, outer sep=0pt, scale=  0.56] at ( 21.57, 76.03) {0.25};

\node[text=drawColor,anchor=base east,inner sep=0pt, outer sep=0pt, scale=  0.56] at ( 21.57, 90.99) {0.26};

\node[text=drawColor,anchor=base east,inner sep=0pt, outer sep=0pt, scale=  0.56] at ( 21.57,105.95) {0.27};

\node[text=drawColor,anchor=base east,inner sep=0pt, outer sep=0pt, scale=  0.56] at ( 21.57,120.92) {0.28};
\end{scope}
\begin{scope}
\path[clip] (  0.00,  0.00) rectangle (361.35,130.09);
\definecolor{drawColor}{gray}{0.20}

\path[draw=drawColor,line width= 0.4pt,line join=round] ( 22.97, 48.03) --
	( 24.72, 48.03);

\path[draw=drawColor,line width= 0.4pt,line join=round] ( 22.97, 63.00) --
	( 24.72, 63.00);

\path[draw=drawColor,line width= 0.4pt,line join=round] ( 22.97, 77.96) --
	( 24.72, 77.96);

\path[draw=drawColor,line width= 0.4pt,line join=round] ( 22.97, 92.92) --
	( 24.72, 92.92);

\path[draw=drawColor,line width= 0.4pt,line join=round] ( 22.97,107.88) --
	( 24.72,107.88);

\path[draw=drawColor,line width= 0.4pt,line join=round] ( 22.97,122.85) --
	( 24.72,122.85);
\end{scope}
\begin{scope}
\path[clip] (  0.00,  0.00) rectangle (361.35,130.09);
\definecolor{drawColor}{RGB}{0,0,0}

\path[draw=drawColor,line width= 0.4pt,line join=round] ( 24.72, 44.29) --
	(177.17, 44.29);
\end{scope}
\begin{scope}
\path[clip] (  0.00,  0.00) rectangle (361.35,130.09);
\definecolor{drawColor}{gray}{0.20}

\path[draw=drawColor,line width= 0.4pt,line join=round] (100.95, 42.54) --
	(100.95, 44.29);
\end{scope}
\begin{scope}
\path[clip] (  0.00,  0.00) rectangle (361.35,130.09);
\definecolor{drawColor}{gray}{0.30}

\node[text=drawColor,anchor=base,inner sep=0pt, outer sep=0pt, scale=  0.56] at (100.95, 37.29) {{\color{black}Unadjusted Method}};
\end{scope}
\begin{scope}
\path[clip] (  0.00,  0.00) rectangle (361.35,130.09);
\definecolor{drawColor}{RGB}{0,0,0}

\node[text=drawColor,rotate= 90.00,anchor=base,inner sep=0pt, outer sep=0pt, scale=  0.67] at (  8.32, 85.44) {False Selection Proportion};
\end{scope}
\begin{scope}
\path[clip] (180.67, 24.34) rectangle (361.35,130.09);
\definecolor{drawColor}{RGB}{255,255,255}
\definecolor{fillColor}{RGB}{255,255,255}

\path[draw=drawColor,line width= 0.4pt,line join=round,line cap=round,fill=fillColor] (180.67, 24.34) rectangle (361.35,130.09);
\end{scope}
\begin{scope}
\path[clip] (205.39, 44.29) rectangle (357.85,126.59);
\definecolor{fillColor}{RGB}{255,255,255}

\path[fill=fillColor] (205.39, 44.29) rectangle (357.85,126.59);
\definecolor{fillColor}{RGB}{213,94,0}

\path[fill=fillColor,fill opacity=0.75] (300.68,-296.10) rectangle (338.79, 79.98);
\definecolor{fillColor}{RGB}{0,114,178}

\path[fill=fillColor,fill opacity=0.75] (262.57,-296.10) rectangle (300.68, 73.94);
\definecolor{fillColor}{RGB}{0,158,115}

\path[fill=fillColor,fill opacity=0.75] (224.45,-296.10) rectangle (262.57, 80.84);
\definecolor{drawColor}{RGB}{0,0,0}

\path[draw=drawColor,line width= 0.6pt,dash pattern=on 4pt off 4pt ,line join=round] (205.39, 77.96) -- (357.85, 77.96);
\end{scope}
\begin{scope}
\path[clip] (  0.00,  0.00) rectangle (361.35,130.09);
\definecolor{drawColor}{RGB}{0,0,0}

\path[draw=drawColor,line width= 0.4pt,line join=round] (205.39, 44.29) --
	(205.39,126.59);
\end{scope}
\begin{scope}
\path[clip] (  0.00,  0.00) rectangle (361.35,130.09);
\definecolor{drawColor}{gray}{0.30}

\node[text=drawColor,anchor=base east,inner sep=0pt, outer sep=0pt, scale=  0.56] at (202.24, 46.11) {0.23};

\node[text=drawColor,anchor=base east,inner sep=0pt, outer sep=0pt, scale=  0.56] at (202.24, 61.07) {0.24};

\node[text=drawColor,anchor=base east,inner sep=0pt, outer sep=0pt, scale=  0.56] at (202.24, 76.03) {0.25};

\node[text=drawColor,anchor=base east,inner sep=0pt, outer sep=0pt, scale=  0.56] at (202.24, 90.99) {0.26};

\node[text=drawColor,anchor=base east,inner sep=0pt, outer sep=0pt, scale=  0.56] at (202.24,105.95) {0.27};

\node[text=drawColor,anchor=base east,inner sep=0pt, outer sep=0pt, scale=  0.56] at (202.24,120.92) {0.28};
\end{scope}
\begin{scope}
\path[clip] (  0.00,  0.00) rectangle (361.35,130.09);
\definecolor{drawColor}{gray}{0.20}

\path[draw=drawColor,line width= 0.4pt,line join=round] (203.64, 48.03) --
	(205.39, 48.03);

\path[draw=drawColor,line width= 0.4pt,line join=round] (203.64, 63.00) --
	(205.39, 63.00);

\path[draw=drawColor,line width= 0.4pt,line join=round] (203.64, 77.96) --
	(205.39, 77.96);

\path[draw=drawColor,line width= 0.4pt,line join=round] (203.64, 92.92) --
	(205.39, 92.92);

\path[draw=drawColor,line width= 0.4pt,line join=round] (203.64,107.88) --
	(205.39,107.88);

\path[draw=drawColor,line width= 0.4pt,line join=round] (203.64,122.85) --
	(205.39,122.85);
\end{scope}
\begin{scope}
\path[clip] (  0.00,  0.00) rectangle (361.35,130.09);
\definecolor{drawColor}{RGB}{0,0,0}

\path[draw=drawColor,line width= 0.4pt,line join=round] (205.39, 44.29) --
	(357.85, 44.29);
\end{scope}
\begin{scope}
\path[clip] (  0.00,  0.00) rectangle (361.35,130.09);
\definecolor{drawColor}{gray}{0.20}

\path[draw=drawColor,line width= 0.4pt,line join=round] (281.62, 42.54) --
	(281.62, 44.29);
\end{scope}
\begin{scope}
\path[clip] (  0.00,  0.00) rectangle (361.35,130.09);
\definecolor{drawColor}{gray}{0.30}

\node[text=drawColor,anchor=base,inner sep=0pt, outer sep=0pt, scale=  0.56] at (281.62, 37.29) {{\color{black}FASI}};
\end{scope}
\begin{scope}
\path[clip] (  0.00,  0.00) rectangle (361.35,130.09);
\definecolor{fillColor}{RGB}{255,255,255}

\path[fill=fillColor] (112.54,  0.00) rectangle (248.81, 24.34);
\end{scope}
\begin{scope}
\path[clip] (  0.00,  0.00) rectangle (361.35,130.09);
\definecolor{fillColor}{RGB}{0,158,115}

\path[fill=fillColor,fill opacity=0.75] (120.25,  4.21) rectangle (136.18, 20.13);
\end{scope}
\begin{scope}
\path[clip] (  0.00,  0.00) rectangle (361.35,130.09);
\definecolor{fillColor}{RGB}{0,114,178}

\path[fill=fillColor,fill opacity=0.75] (168.34,  4.21) rectangle (184.26, 20.13);
\end{scope}
\begin{scope}
\path[clip] (  0.00,  0.00) rectangle (361.35,130.09);
\definecolor{fillColor}{RGB}{213,94,0}

\path[fill=fillColor,fill opacity=0.75] (211.12,  4.21) rectangle (227.04, 20.13);
\end{scope}
\begin{scope}
\path[clip] (  0.00,  0.00) rectangle (361.35,130.09);
\definecolor{drawColor}{RGB}{0,0,0}

\node[text=drawColor,anchor=base west,inner sep=0pt, outer sep=0pt, scale=  0.56] at (140.39, 10.24) {All Races};
\end{scope}
\begin{scope}
\path[clip] (  0.00,  0.00) rectangle (361.35,130.09);
\definecolor{drawColor}{RGB}{0,0,0}

\node[text=drawColor,anchor=base west,inner sep=0pt, outer sep=0pt, scale=  0.56] at (188.47, 10.24) {Black   };
\end{scope}
\begin{scope}
\path[clip] (  0.00,  0.00) rectangle (361.35,130.09);
\definecolor{drawColor}{RGB}{0,0,0}

\node[text=drawColor,anchor=base west,inner sep=0pt, outer sep=0pt, scale=  0.56] at (231.25, 10.24) {Other};
\end{scope}
\end{tikzpicture}}
    	\caption{\label{cc_intro.plot}\small The selection of recidivists from a pool of criminal defendants (Broward County, Florida). The target FSR is 25\%. Left: the unadjusted approach. Right: the proposed FASI approach. }
\end{figure}

This article introduces a new notion of fairness that requires parity in FSR control across various protected groups. This aligns with the social and policy goals in a range of decision-making scenarios such as selecting recidivists or determining risky loan applicants, where the burden of erroneous classifications should be shared equally among different genders and races. However, the development of effective and fair FSR rules is challenging. First, controlling the error rate associated with a classifier, such as one built around the GAM procedure, critically depends on the accuracy of the scores. However, the assessment of the accuracy/uncertainty of these scores largely remains unknown. Second, we wish to provide practitioners with theoretical guarantees on the parity and validity for FSR control, regardless of the algorithm being used, including complex black-box classifiers. 

To address these issues, we develop a data-driven FASI algorithm specifically designed to control the FSRs of protected groups below a user-specified level $\alpha$. 
The right panel of Figure~\ref{cc_intro.plot} illustrates the FSPs of FASI on the recidivism data. All individual FSPs are controlled at 25\% approximately. FASI works by converting the confidence scores from a black-box algorithm to an R-value, which is intuitive, easy to compute, and comparable across different protected groups. We then show that selecting all observations with R-value no greater than $\alpha$ will result in an FSR of approximately $\alpha$. Hence, we can directly use this R-value to assign new observations a class label or, for observations with high R-values, assign them to the \textit{indecision} class.

This paper makes several contributions. Firstly, we introduce a novel notion of fairness within the selective inference framework, incorporating an indecision option. In high-consequence situations, it is sensible to exercise caution, by either withholding or separately evaluating such cases until additional evidence is gathered. This reduces the risk of making definitive decisions without sufficient support, thus promoting cautious and fair decisions in these complex scenarios. Secondly, a data-driven FASI Algorithm is developed based on the utilization of the R-value. This algorithm, which can be deployed with user-specified learning algorithms (e.g. random forest, neural networks), is intuitively appealing and easy to interpret. Thirdly, rigorous theoretical justifications are provided for the FASI algorithm. The theory on FSR control is established with mild assumptions on data exchangeability, accommodating scores generated by black-box algorithms. Finally, the empirical performance of FASI is investigated through extensive experimentation using simulated and real-world data sets, demonstrating the effectiveness and practical utility of the proposed approach.

The rest of the paper is structured as follows. In Section~\ref{section:prob_form} we  define the FSR and describe the problem formulation. Section~\ref{method:sec} introduces the R-value and FASI algorithm. The numerical results for simulated and real data are presented in Sections~\ref{section:simulation} and \ref{section:real_data}, respectively.   Section~\ref{si-general:sec} concludes the main article with a discussion of related works and possible extensions.  The Online Supplementary Material provides additional technical details about the methodology, proof of theorems, and supplementary numerical results.

\section{Problem Formulation}
\label{section:prob_form}

Suppose we observe a data set $\{ (X_i, A_i, Y_i): i \in \mathcal{D} \}$, where $\mathcal{D}=[n]\equiv\{1,...,n\}$ is an index set, $X_i\in \mathbb{R}^p$ is a $p$-dimensional vector of features, $A_i \in \mathcal A$ is an additional feature representing the protected or sensitive attribute, and $Y_i$ is a class label taking values in $\mathcal C=\{1, \ldots, C\}$. The goal is to predict the classes for $m$ new individuals indexed by $\mathcal{D}^{test}=\{n+1,...,n+m\}$, with observed features $\{(X_j, A_j): j\in \mathcal{D}^{test}\}$. Denote $\mathcal D^{test}_a=\{j\in\mathcal D^{test}: A_j=a\}$ for $a\in \mathcal A$. The predicted values for their class labels $\{Y_j: j\in \mathcal{D}^{test}\}$ are denoted by $\{\hat Y_j: j\in\mathcal{D}^{test}\}$.

\subsection{Background: predictive parity in classification}\label{section:fairness_lit_review_intro}

We focus on scenarios where an individual's membership to a particular protected group is known. Group-fairness approaches, which explicitly enforce fairness across groups, have been widely applied across various disciplines, ranging from medicine to the criminal justice system. To provide context for our fairness notion, we start with the widely used predictive parity or sufficiency principle in classification, as discussed in \cite{Cri03}, \cite{Bar17} and \cite{Cho17}. According to this principle, the probability of misclassifying an individual to class $c$ should be equal across all protected groups: 
    \begin{equation}\label{suff-princ}
    \PP(Y \neq \hat{Y} | \hat{Y}=c, A=a) \text{ are the same for all } a \in \mathcal{A}.
    \end{equation}
We highlight three primary issues related to machine learning methods developed under the sufficiency principle \citep{zeng2022fair, Pleetal17, Zafar_disparate_treatment}. First, the calibration by group method \citep{Bar17}, a popular approach for ensuring fair outcomes for subgroups, does not offer a theoretical guarantee on controlling the misclassification rate at a user-specified level. This lack of a guarantee can be particularly problematic in high-stakes decision-making situations. Second, current classification methods only focus on the accuracy of individual classifications, neglecting the complexities that arise when multiple individuals are classified simultaneously. This oversight regarding multiplicity can lead to severe inflation of misclassification errors.
Finally, concurrent state-of-the-art classifiers typically exhibit high complexity and analytical intractability, making it difficult to quantify the uncertainties around their predictions. Even when such theoretical analyses are feasible, they often involve strong assumptions about the underlying model and the accuracy of its outputs, which may not hold in practice. In response to these challenges, we propose a comprehensive approach comprising a selective classification framework (Section \ref{si-binary:sec}), a modified error rate criterion (Section \ref{fairness-si:sec}), and a novel class of model-free algorithms with strong theoretical guarantees (Sections \ref{fasi:sec}-\ref{why-fasi:sec}). Together, these  components provide a highly effective solution to the identified issues.

\subsection{A selective inference framework for binary classification}\label{si-binary:sec}

This article focuses on binary classification problems. The extension to the general multi-class setting is discussed briefly in Section~\ref{si-general:sec}.  

Consider an application scenario for predicting mortgage default, where \( Y = 2 \) indicates default and \( Y = 1 \) otherwise. A common practice is to produce \emph{confidence scores}, denoted \( \hat{S}^c(x,a) \) for \( c \in \mathcal{C} \equiv \{1, 2\} \), which are generated from a user-specified classifier and risk assessment software. We focus on scores corresponding to the estimated class probabilities of \( Y = c \) given the covariates \( (X, A) = (x, a) \). The scores satisfy \( \hat{S}^1(x, a) + \hat{S}^2(x, a) = 1 \). Suppose we need to classify \( m \) individuals with confidence scores \( \{\hat{S}_j^c \equiv \hat{S}^c(X_j, A_j) : c \in \mathcal{C}; \, j \in \mathcal{D}^{test} \} \) into ``high,'' ``medium,'' or ``low'' risk classes. It is natural to consider a class of rules in the form of
\begin{equation}\label{select-rule1}
\hat{Y}_j~=~\sum_{c \in \mathcal C\equiv\{1, 2\}} c \cdot \mathbb{I}(\hat S_j^c > t_c)~=~\mathbb{I}(\hat S_j^2 < 1 - t_1) + 2 \cdot \mathbb{I}(\hat S_j^2 > t_2), \quad \text{for } j \in \mathcal{D}^{test},
\end{equation}
where \( \mathbb{I}(\cdot) \) is the indicator function, and the thresholds $t_c$ satisfy \( t_c \in (0.5,1] \) for \( c \in \mathcal C\). 

\begin{remark}\rm{
The constraint \( t_c > 0.5 \) provides two benefits. First, it enhances interpretability by ensuring that a definitive class assignment occurs only when the confidence score exceeds 50\%. Second, it prevents overlapping selections: since \(S_j^1 + S_j^2 = 1\), it ensures unique class assignments, effectively avoid overlapping selections.

}
\end{remark}

The predicted label \( \hat{Y}_j \) takes three possible values in the action space $\Lambda = \{1, 2, 0\}$, indicating that an individual has low (\( \hat{Y}_j = 1 \)), high (\( \hat{Y}_j = 2 \)), and medium (\( \hat{Y}_j = 0 \)) risks of default, respectively. The value \( 0 \), referred to as an ``indecision’’ or ``reject option’’ in classification [cf. \citealp{HerWeg06, SunWei11, Lei14, Lee_21}], is used to express ``doubt,’’ indicating insufficient confidence to make a definitive decision. For example, an individual with \( \hat{Y}_j = 1 \) will be approved for a mortgage, an individual with \( \hat{Y}_j = 2 \) will be rejected, while an individual with \( \hat{Y}_j = 0 \) will receive a pending decision and be asked to provide additional information before resubmitting the application. 

\begin{remark}\rm{
We can interpret \eqref{select-rule1} as a \emph{selective inference} procedure that assigns individuals with extreme scores to high- or low-risk classes while returning an indecision for the remainder. Notably, in this framework the state space \(\mathcal{C} = \{1,2\}\) differs from the action space \(\Lambda = \{1, 2, 0\}\), contrasting with the standard classification setup which mandates $\mathcal C=\Lambda=\{1,2\}$. This flexible framework provides a useful interface for practitioners: assigning individuals to Class 1 offers economic benefits by preventing the misallocation of resources to low-risk candidates, thereby reducing study costs associated with unnecessary follow-up. Moreover, the selection for Class 2 is essential for identifying high-risk cases that require further intervention. The selective inference perspective can be employed to handle various types problems including outlier detection and multinomial classification; further discussion is provided in Sections \ref{RQP:sec} and \ref{subsec:R-multiclass} of the Appendix.}
\end{remark}

\subsection{False selection rate and the fairness issue}\label{fairness-si:sec}

In practice, it is desirable to avoid erroneous selections, which often have negative social or economic impacts. In the mortgage example, approving an individual who will truly default (i.e., $\hat Y=1$ but $Y=2$) would increase the financial burden of the lender, while rejecting an individual who will not  default (i.e., $\hat Y=2$ but $Y=1$) would lead to a loss of profit. In situations where $m$ is large, controlling the inflation of selection errors is a crucial task for policy makers. A practically useful notion is the false selection rate (FSR), which is defined as the expected fraction of erroneous decisions among all definitive decisions. We use the notation $\mbox{FSR}^{\mathcal C^\prime}$, where $\mathcal C^\prime\subset \mathcal C=\{1, 2\}$ is the set of class labels that we are interested in selecting. 

Consider the two-class classification problem with selection rule \eqref{select-rule1}. Denote $\mathcal S=\{j \in \mathcal{D}^{test}: \hat Y_{j}\neq 0\}$ the index set of the selected cases and $|\mathcal S|$ its cardinality. The FSR that combines the false selections from both classes is given by
\beq\label{FSR-12}
\mbox{FSR}^{\{1, 2\}}=\mathbb E\Bigg[ \frac{\sum_{j\in \mathcal D^{test}} \sum_{c\in\{1,2\}}\II(\hat Y_{j}=c, Y_j\neq c)}{|\mathcal S| \vee 1}  \Bigg],
\eeq
where $x\vee y=\max\{x, y\}$, and the expectation $\EE$ in Equation \eqref{FSR-12} [and later on in \eqref{FSR-c}-\eqref{FSR-a}] is taken over both the observed data $\{ (X_i, A_i, Y_i): i \in \mathcal{D} \}$ and test data $\{(X_j, A_j, Y_j): j\in \mathcal{D}^{test}\}$. 
Let $\mathcal S^c=\{j\in \mathcal{D}^{test}: \hat Y_{j}=c\}$ denote the index set of the cases assigned to class $c$. The FSRs evaluated for individual classes are defined as: 
\beq\label{FSR-c}
\mbox{FSR}^{\{c\}}=\mathbb E\left[ \frac{\sum_{j \in \mathcal{D}^{test}} \II(\hat Y_{j}=c, Y_{j}\neq c)}{|\mathcal S^c| \vee 1 }  \right], \quad c\in\mathcal C.  
\eeq

Incorporating the option of indecision facilitates the development of a decision rule that can control the FSRs at a user-specified level. However, attaining this objective is challenging within the conventional classification framework, which requires definitive decisions for every individual. As demonstrated in \cite{MeiRic06} and \cite{CaiSun17}, if the minimum condition on the classification boundary is not satisfied, it becomes impossible to simultaneously control both $\mbox{FSR}^1$ and $\mbox{FSR}^2$ at low levels.

The FSR is a general concept for selective inference  that encompasses  important special cases such as the misclassification rate, the false discovery rate (FDR, \citealp{BenHoc95}), among others. When both the state space and the action space are set to $\{1, 2\}$, thereby eliminating the possibility of indecision, then the FSR defined in \eqref{FSR-12} simplifies to the misclassification rate
$
\frac{1}{m}\mathbb E\left\{\sum_{j \in \mathcal{D}^{test}} (\hat Y_{j} \neq Y_{j})\right\}.
$
The connection between our FSR framework, one-class classification, and the FDR is discussed in detail in Section \ref{RQP:sec} of the Appendix.

In practical scenarios, minimizing the number of indecisions is highly desirable. To quantify this concept, we introduce the expected proportion of indecisions:
\begin{equation}\label{def:EPI}
\textstyle \text{EPI} = ({1}/{m})\mathbb{E}\big\{\sum_{j \in \mathcal{D}^{test}} \mathbb{I}(\hat Y_j = 0)\big\} = 1 - {\mathbb{E}(|\mathcal{S}|)}/{m}.
\end{equation}
Under the same FSR level, a smaller EPI corresponds to greater statistical power.

Next we turn to the important fairness issue in selective inference. A major concern is that the rate of erroneous decisions might be unequally shared between the protected groups, as illustrated in the COMPAS example. 
To address this issue, it is desirable to control the FSR for each protected attribute in $A$. Therefore, we aim to find a selective classification rule obeying the following constraint on group-wise FSRs:  
\begin{equation} \label{FSR-a} 
   \mbox{FSR}^{\{c\}}_a=\mathbb E\left[\frac{\sum_{j \in \mathcal{D}^{test}_a} \mathbb I(\hat Y_{j}=c, {Y}_{j}\neq c)}{\left\{ \sum_{j \in \mathcal{D}^{test}_a} \mathbb I(\hat Y_{j}=c)\right\}\vee 1}  \right]\leq \alpha_c, \quad  \mbox{for all $a\in \mathcal A$,} 
\end{equation}
where $\alpha_c$ is a user-specified tolerance level, $c\in\{1,2\}$. The fairness-adjusted error rate constraint \eqref{FSR-a} equally bounds the fraction of erroneous decisions among protected groups. We aim to develop a selective classification rule that solves the following constrained optimization problem: 
\beq\label{prob-form1}
\text{minimize the EPI subject to $\mbox{FSR}^{\{c\}}_a\leq\alpha_c$, \quad for $c\in\{1,2\}$ and
$a\in\mathcal A$.} 
\eeq
\begin{remark}\label{rem:asymptotic-theory}
\rm{
Although our problem formulation \eqref{prob-form1} only sets upper bounds for group-wise FSR levels, minimizing the EPI enforces the exhaustion of allowable FSR levels for each group, thereby asymptotically aligning all group-wise FSR levels with the designated nominal level. In Appendix \ref{sec:group-overall-FSR}, we also demonstrate that controlling group-wise FSRs [cf. Eq~\eqref{FSR-a}] asymptotically guarantees overall FSR [cf. Eq~\eqref{FSR-c}] control. These theoretical findings are consistently supported by our numerical studies. Achieving both group-wise FSR equalization and overall FSR control in finite samples, however, remains an open and challenging problem, which we leave for future research.  }
\end{remark}

\subsection{The construction of fair classifiers: issues and roadmap}

We investigate the important issue of what makes a ``fair'' classifier. In most classification tasks, the standard operation is to first construct a confidence score, and then secondly to turn this score into a decision by setting a threshold. Consider selection rule \eqref{select-rule1}. We present two approaches for constructing confidence scores. The notation \(S(x,a)\) is used instead of \(\hat S(x,a)\) to indicate the ideal setting in which an oracle, possessing knowledge of the true data-generating model, computes the scores analytically without estimation.

The two approaches, respectively referred to as the ``full covariate classifier'' (FCC) and  ``reduced covariate classifier'' (RCC), employs the following scores: 
\begin{eqnarray}\label{prob-score1}
S^{c,FCC}_j(x, a)& = & \PP\left(Y_{j}=c|X_{j}=x, A_{j}=a\right), \\
\label{prob-score2}
S^{c,RCC}_{j}(x) & = & \PP\left\{Y_{j}=c|X_{j}=x\right\},
\end{eqnarray}
\noindent for $c\in\{1,2\}$ and $j\in \mathcal{D}^{test}$. Consider the high-risk class \(c=2\). Then \(S^{2,\mathrm{FCC}}_j(x,a)\) denotes the (oracle) class probability of an individual belonging to class \(2\) based on all available covariates. In contrast, \(S^{2,\mathrm{RCC}}_j(x)\) is employed to estimate the same probability after removing the sensitive attribute from the covariate set. However, as will be demonstrated, both the FCC and RCC approaches may be inadequate for effectively addressing the fairness concern. 

\begin{figure}[tp]
    \centering 
    \scalebox{1}{%
    \input{Figures/intro_plot_sim}}
    	\caption{\label{intro.plot}\small For FCC and RCC, the degree of unfairness increases as $\pi_{2|M}$ and $\pi_{2|F}$ become more disparate. FASI ensures that the group-wise FSRs are effectively controlled and approximately equalized. } 
\end{figure}

Consider the mortgage example where we simulate a data set that contains a sensitive attribute ``gender''. The goal is to select individuals into the high risk class  with FSR control at $10\%$; the simulation setup is detailed in Section \ref{section:simulation}. We highlight here that the proportions of individuals with label ``2'' are different across the protected groups: for the male group, the proportion of individuals with label ``2'', denoted as $\pi_{2|M}$, is fixed at 50\%, whereas for the female group the  proportion $\pi_{2|F}$ varies from $15\%$ to $85\%$. 

We apply the FCC approach and plot the overall FSR and group-wise FSRs as functions of $\pi_{2|F}$ on the left panel of Figure \ref{intro.plot}. We can see that both the FCC and RCC control the overall FSR but not the group-wise FSRs. Hence these thresholding rules are harmful in the sense that the burden of erroneous decisions is not shared equally among the two gender groups. The RCC approach has two further drawbacks. Firstly, disregarding a sensitive attribute can result in significant power loss. Secondly, if feature $X$ is highly predictive of sensitive attribute $A$, then the RCC approach can still lead to unfair decisions due to the issue of \emph{surrogate encoding} \citep{Kusner_17, Long_21}. Concretely, in fairness research, surrogate encoding pertains to the circumstance where the sensitive attribute $A$ is absent from the list of predictors, but its information is encoded or concealed within other predictors, causing $A$ to still influence the outcome $Y$. We emphasize that the patterns in Figure \ref{intro.plot} are not specific to any particular classification algorithm but indicate a systematic bias. In our simulation, where perfect scores are available, the unfairness depicted in Figure \ref{intro.plot} still persists. In contrast, our proposed FASI algorithm, shown in the right panel of Figure \ref{intro.plot}, effectively controls the FSR and nearly equalizes the error rates across all protected groups.

\section{Methodology}\label{method:sec}

This section develops a fairness-adjusted selective inference (FASI) procedure for two-class classification with state space \(\mathcal{C} = \{1, 2\}\) and action space $\Lambda=\{0,1,2\}$. We focus on the error rate \(\mbox{FSR}^{\{c\}}_a\) defined in \eqref{FSR-a}, which is more relevant for addressing fairness issues in high-stakes decision-making scenarios. The methodologies for the more complex tasks of controlling \(\mbox{FSR}^{\{1,2\}}\) and performing multinomial classification are briefly discussed in Section \ref{si-general:sec} and Appendix \ref{subsec:R-multiclass}.

A major challenge in our methodological development is that many state-of-the-art machine learning algorithms are complex and offer no performance guarantees on their outputs. This limitation renders uncertainty quantification and error rate control challenging, if not intractable. To address this issue, we develop a model-free framework that is applicable to any black-box algorithm and relies solely on the exchangeability of the data points. 

\subsection{The R-value and FASI algorithm}\label{fasi:sec}

We first introduce a significance index, called the R-value, for ranking individuals and then discuss how the R-values can be employed for selective classification. 

The R-value is computed via the FASI algorithm, which consists of three steps: training, calibrating and thresholding. The observed data set $\{ (X_i, A_i, Y_i): i \in \mathcal{D} \}$ is randomly divided into a training set and a calibration set: $\mathcal D=\mathcal D^{train}\cup\mathcal D^{cal}$. Let $\mathcal D^{test}_a=\{i\in\mathcal D: A_i=a\}$ and $\mathcal D^{cal}_a=\{i\in\mathcal D^{cal}: A_i=a\}$ for $a\in\mathcal A$. Denote $n_a=|\mathcal D^{cal}_a|$ and $m_a=|\mathcal D^{test}_a|$. 

In the first step, we train score functions $\hat S^c(x, a)$, $c\in\{1,2\}$, using data $\{(X_i, A_i, Y_i): i\in\mathcal D^{train}\}$. The scores, representing estimated class probabilities, can be generated from any user-specified classifier satisfying $\hat {S}^1(x,a)+\hat {S}^2(x,a)=1$. We make no assumptions on the accuracy of these scores. 

In the second step, we use the scores $\{\hat S^c_i\coloneqq\hat{S}^c(X_i,A_i): i \in \mathcal D^{cal} \cup \mathcal D^{test}\}$ to calculate 
\beq\label{Q-kc}
Q_{k}^{c}=\sum_{a\in\mathcal A} \mathbb I(A_k=a)\cdot \frac{\big\{\sum_{i\in \mathcal D^{cal}_a} \mathbb I\big(\hat S_i^c\geq \hat S^c_{k}, Y_i\neq c\big)+1\big\}/(n_a+1)}{\big\{\sum_{i\in\mathcal D^{test}_a}\mathbb I\big(\hat S_i^c\geq  \hat S^c_{k}\big)\big\}/m_a} \wedge 1, 
\eeq
for $k\in \mathcal D^{cal}\cup\mathcal D^{test}$. The operation $\wedge$ indicates that $Q_k^c$ is set to $1$ if it exceeds $1$. As discussed in Section \ref{why-fasi:sec}, \(Q_k^c\) represents the estimated fraction of false selections among all selections using the cutoff \(\hat S^c_{k}\); a lower value of \(Q_k^c\) indicates greater confidence in classifying the \(k\)th individual to class \(c\). To enhance the algorithm's stability, one may modify \eqref{Q-kc} to include both the calibration and test data in the denominator: 
\beq\label{Q-kc-plus}
Q_{k}^{c}=\sum_{a\in\mathcal A} \mathbb I(A_k=a)\cdot \frac{\big\{\sum_{i\in \mathcal D^{cal}_a} \mathbb I\big(\hat S_i^c\geq \hat S^c_{k}, Y_i\neq c\big)+1\big\}/(n_a+1)}{\big\{\sum_{i\in \mathcal{D}^{cal}_a\cup\mathcal D^{test}_a}\mathbb I\big(\hat S_i^c\geq  \hat S^c_{k}\big)+1\big\}/(n_a+m_a+1)} \wedge 1.
\eeq
The subsequent steps of the algorithm are identical whether \eqref{Q-kc} or \eqref{Q-kc-plus} is used, so we denote both by \( Q_{k}^{c} \) and provide a unified discussion. 

\begin{remark}\rm{
The adjustment in Equation \eqref{Q-kc-plus} is useful when \( m_a \) is small; numerical evidence in Section \ref{rval_compare:sec} of the Appendix demonstrates the benefits of a larger sample size (i.e. $m_a+n_a$) in calibrating \( Q_k^c \). Although \eqref{Q-kc-plus} offers numerical advantages, it introduces additional theoretical complexity; accordingly, we develop separate theories for \eqref{Q-kc} and \eqref{Q-kc-plus} in Theorem \ref{FSR:thm} below.
}
\end{remark}

In practical scenarios, higher-scoring individuals may not consistently correspond to smaller $Q_k^c$ values. To eliminate this inconsistency, we propose the following monotonicity adjustment: 
\begin{equation}\label{Q-value-mod}
 \tilde Q_{j}^{c} = \sum_{a \in \mathcal{A}} \mathbb{I}(A_j = a) \cdot \min_{\{k \in \mathcal{D}^{cal}_a \cup \mathcal{D}^{test}_a: \hat{S}_{k}^{c} \leq \hat{S}_{j}^{c}\}} Q^{c}_{k}, \quad j \in \mathcal{D}^{test}. 
\end{equation}
The R-value, with more explanations provided in Remark \ref{rem:R-value} below, is defined as 
\begin{equation}\label{R-value-max}
R_j^{c} = \max\big\{ \mathbb{I}(\hat{S}_{j}^{c} \leq 0.5), \hspace{0.1cm} \tilde{Q}_{j}^{c} \big\}, \quad \mbox{for $c\in\{1,2\}$ and $j \in \mathcal{D}^{test}$}.
\end{equation}

In the third step, we compare the R-values against the designated level \(\alpha_c\):
\begin{equation}\label{R-rule}
\hat{Y}_{j} = \sum_{c \in \{1, 2\}} c \cdot \mathbb{I}(R^{c}_{j} \leq \alpha_c), \quad j \in \mathcal{D}^{test}.
\end{equation}

\begin{remark}\label{rem:R-value}\rm{
In binary classification, the \(j\)th individual is associated with two R-values, \(R_j^1\) and \(R_j^2\). Definition \eqref{R-value-max} provides a crucial adjustment to ensure that, in effect, only one R-value is used for decision-making. Specifically, if \(\hat S_j^c \leq 0.5\), we set \(R_j^c = 1\). Since \(0 < \alpha_c < 1\) is a small constant, this adjustment guarantees that any \(R_j^c = 1\) is effectively discarded, meaning the \(j\)th individual is never assigned to class \(c\) when \(\hat S_j^c \leq 0.5\). Moreover, because \(\hat{S}^1_j + \hat{S}^2_j = 1\), one of the two R-values in \(\{R_j^c : c = 1,2\}\) must equal 1, thereby preventing overlapping selections. Finally, if both R-values exceed \(\alpha_c\), we output an indecision, \(\hat Y_j = 0\). 
}
\end{remark}

The FASI algorithm, summarized in Algorithm \ref{FASI:alg}, offers several attractive properties. First, the R-value serves as an estimate of a proportion, making it easily interpretable and comparable across groups. Second, the FSR analysis based on R-values is straightforward: practitioners can directly make decisions by comparing the R-values with a user-specified FSR level. Third, fairness notion is integrated into the R-value. As demonstrated in Lemma \ref{lemma:score} in Section \ref{proof:thm2} of the Supplement, the decision rule in \eqref{R-rule} involves finding the smallest group-wise threshold for \(\hat S_j^c\) that satisfies \(\text{FSR}^{\{c\}}_a \leq \alpha_c\), thereby approximately aligning the group-wise FSR levels with the nominal level. Finally, FASI is model-free, providing a robust framework for FSR control, as discussed in the next subsection.

\begin{algorithm}[t!]
\caption{The FASI Algorithm}\label{FASI:alg} 
\hspace*{\algorithmicindent} 

\noindent\textbf{Input}: $\{(X_i, A_i, Y_i): i\in\mathcal{D}\}$, $\{(X_j, A_j): j\in\mathcal{D}^{test}\}$, FSR levels $\{\alpha_c: c=1, 2\}$. \\
\textbf{\;\;\;\;\; Output}: a selective classification rule $\{\hat Y_j\in\{0, 1, 2\}: j\in\mathcal D^{test}\}$.

\begin{algorithmic}[1]
    \State Randomly split $\mathcal{D}$ into $\mathcal{D}^{train}$ and $\mathcal{D}^{cal}$.
    \State Train a machine learning model on $\{(X_i, A_i, Y_i): i\in \mathcal{D}^{train}\}$. 
    \State Predict confidence scores $\hat S_i^c$ for $i\in\mathcal{D}^{cal}\cup\mathcal{D}^{test}$. 
    \State Compute the R-values $\{R_j^c: c=1, 2; j\in\mathcal D^{test}\}$ according to Equations \eqref{Q-kc} to \eqref{R-value-max}.
    \State Make decisions by comparing the R-values with $\alpha_c$ using \eqref{R-rule}, for all $j\in \mathcal{D}^{test}$.
\end{algorithmic}

\end{algorithm}

\begin{remark}\rm{
There are two potential strategies to achieve fairness across protected groups. The first strategy, as adopted in the FASI algorithm, involves modifying the current confidence scores to generate new scores (R-values) that are directly comparable across groups.  The second strategy, on the other hand, involves retaining the original confidence scores and implementing group-adjusted thresholds. As demonstrated in Lemma \ref{lemma:score} in Appendix \ref{proof:thm2}, this strategy is mathematically equivalent to the first. However, in practical applications, this approach may be seen as confusing or even controversial because it applies different thresholds to various protected groups. Such disparate treatment is difficult to interpret and could be perceived as introducing an alternative form of discrimination. In contrast, FASI employs a universal threshold for all individuals, with the R-value serving as a statistical wrapper that distills complex factors -- such as error rate control and fairness -- into a single, easy-to-use index.}
\end{remark}

\subsection{Why FASI works?}\label{why-fasi:sec}

We start by explaining why the R-value provides a sensible estimate of the FSR. To simplify the discussion, we focus on a specific group $A=a$ and consider a thresholding rule of the form $\{\mathbb I(\hat S_{j}^c\geq t): j \in \mathcal{D}^{test}_a\}$.  Consider the false selection proportion (FSP) process: 
\beq\label{FSP-1}
\mbox{FSP}_a^{\{c\}}(t)=\frac{\sum_{j\in \mathcal D^{test}_a} \mathbb I\left(\hat S_j^c\geq t, Y_j\neq c\right)}{\left\{\sum_{j\in \mathcal D^{test}_a}\mathbb I(\hat S_j^c\geq t)\right\} \vee 1}, 
\eeq
with $\mbox{FSP}(t)=0$ if no individual is selected. The FSP cannot be computed from data because we do not observe the true states $\{Y_j: j\in \mathcal D^{test}_a\}$. The effectiveness of the FASI algorithm relies on the following exchangeability condition:
\begin{assumption}\label{ex:as}
The data points $\{(X_i, Y_i): i\in \mD^{cal}_a \cup \mD^{test}_a\}$ are exchangeable for all $a\in\mathcal A$.
\end{assumption}
As FASI uses the same fitted model to compute the scores (Assumption \ref{ex:as}), the confidence scores are exchangeable. Consequently, the unobserved process 
 $\sum_{j\in \mathcal D^{test}_a} \mathbb I\left(\hat S_j^c\geq t, Y_j\neq c\right)$ is strongly resembled by its ``mirror process'' in the calibration data $\sum_{i\in \mathcal D^{cal}_a} \mathbb I\left(\hat S_i^c\geq t, Y_i\neq c\right)$. Constructing a mirror process and exploiting its symmetry for inference is a powerful idea that has been explored in recent works (cf. \citealp{BarCan15, weinstein17counting, lei18adapt, LeuSun22, Duetal21}). To account for the unequal sample sizes between $\mathcal D^{cal}_a$ and $\mathcal D^{test}_a$, we derive the mirror FSP process as follows:
\beq\label{FSP-est}
\widehat{\mbox{FSP}}^{\{c\}}_a(t)= \frac{\big\{\sum_{i\in \mathcal D^{cal}_a} \mathbb I(\hat S_i^c\geq t, Y_i\neq c)+1\big\}/(n_a+1)}{\big\{\sum_{j\in \mathcal{D}^{test}_a}\mathbb I(\hat S_j^c\geq t)\big\}/m_a}. 
\eeq
This provides insight into why \eqref{Q-kc} (and similarly \eqref{Q-kc-plus}) has been employed in constructing the R-value. When computing the R-values, \eqref{FSP-est} is only evaluated at values in $\{S_j^c: j\in\mathcal D^{test}\}$; hence the denominator is always greater or equal to 1. 

\begin{remark}\rm{
When inspecting the numerator in \eqref{FSP-est}, we observe that a ``+1'' has been included in $\sum_{j\in \mathcal D^{cal}_a} \mathbb I\bigl(\hat S_j^c \geq t, Y_j \neq c\bigr)$ and $n_a$. This technical adjustment has only a minor impact on the empirical performance of FASI but guarantees that \eqref{FSP-est} corresponds to a martingale, which is crucial for proving the theory.
}
\end{remark}

The FSP process \eqref{FSP-1} and its mirror process \eqref{FSP-est} together provide an intuitive interpretation of the R-value. Roughly speaking, the R-value represents the smallest estimated FSP at which the \(i^{\text{th}}\) individual is just selected. In other words, if we set the threshold at \(R=r\) and select all individuals with R-values less than or equal to \(r\) into class \(c\), then we expect that, for every group \(a \in \mathcal{A}\), approximately \(100r\%\) of the selections will be incorrect decisions. The fairness notion is inherently integrated into the R-value, enabling the calibration of a universal threshold to align all group-wise FSRs with the nominal level. Moreover, our interpretation resembles the q-value (\citealp{Sto03}) in FDR analysis; further details are provided in Section \ref{RQP:sec} of the Supplement. We emphasize that while Storey's q-value relies on the empirical distribution of p-values, our R-value is derived from calibration data through a carefully designed mirror process.

\subsection{Theory on FSR control}\label{subsec:FSR-theory}

We now present a theorem establishing the validity of FASI for FSR control. Our theory differs from existing work in that we make no assumptions about the accuracy of \(\hat S_i^c\). Instead, the accuracy of the scores influences only the power of FASI, leaving its validity (for FSR control) unaffected. Practical guidelines on constructing more accurate confidence scores (and, hence, effective R-values) are provided in Section \ref{subsec:tr} and Section \ref{opt:sec} of the Supplement.

\begin{theorem}\label{FSR:thm}
Define $\gamma_{c,a}= \mathbb E\left(p^{test,a}_{c,null} / p_{c, null}^{cal,a}\right)$, where $p^{test,a}_{c, null}$ and $p_{c,null}^{cal,a}$ are the empirical proportions of individuals in group $a$ that do not belong to class $c$ in the test and calibration data, respectively. Then under Assumption \ref{ex:as}, for all $a\in\mathcal A$, we have:
\begin{description}
\item (a). The FASI algorithm with R-value defined via \eqref{Q-kc}, \eqref{Q-value-mod} and \eqref{R-value-max}  satisfies $\mbox{FSR}_a^{\{c\}}\leq \gamma_{c,a}\alpha_c$. 

\item (b). The stable version of the FASI algorithm with R-value defined via \eqref{Q-kc-plus} --
\eqref{R-value-max}  satisfies 
\begin{equation}\label{equ:theory}
\mbox{FSR}_a^{\{c,*\}}\leq \gamma_{c,a}^\prime \alpha_c+\frac{\alpha}2\mathbb E\left| {\texttt{RES}}(\tau_a^c)-1 \right|,
\end{equation}
where $\gamma_{c,a}^\prime$ is a constant, $\tau_a^c$ is a stopping time, both defined in Appendix \ref{proof:thm1b}, 
\begin{eqnarray}
\mbox{FSR}^{\{c\},*}_a& = & \mathbb E\left[\frac{\sum_{j \in \mathcal{D}^{test}_a} \mathbb I(\hat Y_{j}=c, {Y}_{j}\neq c)}{\sum_{j \in \mathcal{D}^{test}_a} \mathbb I(\hat Y_{j}=c)+1}\right], \quad \mbox{and} \label{FSR-star}\\
{\texttt{RES}}(\tau_a^c) & = & \frac{\sum_{i\in\mathcal D^{cal}_a} \mathbb I(\hat S_i^c\geq \tau_a^c, Y_i=c) }{\sum_{j\in\mathcal D^{test}_a} \mathbb I(\hat S_j^c\geq \tau_a^c, Y_j=c)+1}\cdot\frac{\sum_{j\in\mathcal D^{test}_a} \mathbb I(\hat S_j^c\geq \tau_a^c, Y_j\neq c) }{\sum_{i\in\mathcal D^{cal}_a} \mathbb I(\hat S_i^c\geq \tau_a^c, Y_j\neq c)+1}.\label{equ:Res}    
\end{eqnarray}
\end{description}
\end{theorem}

Assumption \ref{ex:as} on exchangeability implies that \(\gamma_{c,a}\) is typically close to 1, allowing nearly exact control in Part (a), as confirmed by our numerical studies (Section \ref{gamma_sim_est:sec} of the Supplement). Moreover, \(\gamma_{c,a}^\prime\) in Part (b) is also close to 1 (cf. Remark \ref{rem:conv} in Appendix \ref{proof:thm1b}). Assumption \ref{ex:as} implies that both terms in the product on the right-hand side of \eqref{equ:Res} are stochastically close to 1. To provide a more rigorous characterization of the upper bound on the FSR level in Part (b), Section \ref{subsec:convergence-tau} of the Supplement presents an asymptotic analysis that specifies sufficient conditions for the strong convergence of \(\tau_a^c\). Specifically, let \(\tau^* \in (0,1)\) be a constant. If \(\tau_a^c \xrightarrow{a.s.} \tau^*\), one can show that  $ \lim_{(n_a, m_a)\rightarrow\infty} \mathbb{E}\left|\texttt{Res}(\tau_a^c) - 1\right| = 0. $ Hence, \eqref{equ:theory} indicates that the stable version of FASI controls the FSR at \(\alpha + o(1)\).  

\begin{remark}\rm{
Under the conditions outlined in Appendix \ref{subsec:convergence-tau}, we establish that group-wise FSR control asymptotically ensures overall FSR control (see Proposition \ref{prop:OFSR-theory} in Section \ref{sec:group-overall-FSR} of the Appendix). The proof of this proposition also provides heuristic insight into why obtaining finite-sample guarantees of overall FSR control may be inherently challenging. If asymptotic guarantees are sufficient, then group-wise FSR control constitutes a stricter requirement than overall FSR control; hence adding an overall FSR constraint is unnecessary. }
\end{remark}

\begin{remark}\rm{
In the modified FSR definition \eqref{FSR-star}, the ``+1'' adjustment is adopted, mirroring a technique used in Theorem 1 of \cite{BarCan15} though for distinct purposes. The difference between \(\text{FSR}^{\{c\},*}_a\) and \(\text{FSR}^{\{c\}}_a\) is typically negligible in practice. Furthermore, Section \ref{appendix:different_rval} includes a corollary showing that a conservative version of the R-value guarantees FSR control below \(\alpha\) without the \(\gamma_{c,a}\) term in the bound. However, because this conservative variant often leads to significant power loss, we recommend the more efficient R-value defined in \eqref{R-value-max}, which achieves nearly exact control empirically.}
\end{remark}

Three major challenges in proving Theorem \ref{FSR:thm} are (i) handling the dependence between the scores \(\hat S_i^c\) (since the same training data were used to compute \eqref{FSP-est}), and (ii) evaluating the FSR without any knowledge about the quality of the scores. Inspired by elegant ideas in the FDR literature \citep{Stoetal04, BarCan15}, we have carefully designed the R-values so that the corresponding FSP process \eqref{FSP-est} is stochastically dominated by a supermartingale. We then apply the optional stopping theorem and leverage the exchangeability assumption to establish an upper bound for the FSR. We stress that, in Theorem \ref{FSR:thm}, part (a) guarantees validity in finite samples, and both parts (a) and (b) do not rely on any assumptions regarding the underlying models or the quality of the scores. 

\subsection{Connections to existing work}\label{subsec:connections}

This section explores the connections and distinctions between FASI and existing methods developed under the sufficiency principle in fairness research. Additionally, we provide insights about recent developments in conformal inference relevant to FASI.

Our formulation in \eqref{FSR-a} is closely related to the sufficiency principle \eqref{suff-princ} in the fairness literature, but it overcomes several of its limitations. First, \eqref{FSR-a} operates within a selective classification framework by offering an indecision option for cases requiring further review, thereby enabling effective error rate control at user-specified levels. Second, we define the FSR notion to aggregate decision errors over m new individuals, which addresses the sufficiency principle’s limitation of only pertaining to the error rate of an individual decision. Lastly, many algorithms developed under the sufficiency principle are complex and computationally intensive, lacking finite sample guarantees when applied to outputs from black-box models. In contrast, the FASI algorithm effectively controls the FSR in finite samples without relying on assumptions about the underlying model, classification algorithm, or score accuracy. A detailed comparison with related works, including \cite{zeng2022fair} and \cite{Lee_21}, is provided in Section \ref{appendix:related_algorithms} of the Supplement.

The R-value can be interpreted within the conformal inference framework \citep{vovk05, lei14prediction}. In Section \ref{RQP:sec} of the Supplement, we show that a variant of our R-value coincides with the Benjamini–Hochberg (BH) adjusted q‑value applied to conformal p‑values \citep{mary22semi, Batetal21} in the one‑class classification setting \citep{Moyetal96, Khaetal09, Kemetal13}. {In contrast to our focus on selective inference, recent contributions such as \cite{Angelopoulos_25_book} and \cite{Gibbs_25} primarily address group-conditional coverage without providing guarantees in selective settings. As noted in \cite{benjamini2005false} and \cite{Gazetal25}, selective inference poses significantly greater statistical challenges than standard conditional coverage, corresponding to a distinct inferential objective. We expand on these points in Appendix \ref{subsec:RV-conformal} (Remark \ref{rem:conditional-cover}).

The theory presented in \cite{Batetal21} encounters a complication similar to ours, as the conformal p-values are dependent. To address this, \cite{Batetal21} first shows that the conformal p-values satisfy the condition of positive regression dependence on a subset (PRDS) and then applies the theory in \cite{BenYek01} to establish the validity of FDR control. While we conjecture that the PRDS approach may be relevant, its extension to our specific context is non-trivial because our R-values do not explicitly utilize conformal p-values under the binary classification setup. Therefore, our martingale-based theory appears to be a suitable and equally effective alternative. Moreover, incorporating conformal p-values—which rely on one-class classifiers—directly into our binary classification problem would entail discarding labeled outliers and consequently lead to information loss; this issue has been explored in a recent study by \cite{Liaetal22}.

Our mirror process leverages a calibration set containing data from both classes, unlike the counting knockoff approach (e.g., \citealp{weinstein17counting, Batetal21}), which relies solely on null training data (see Section \ref{RQP:sec} of the Supplement for further discussion). Using data from both classes eliminates the need for Storey's adjustment, which is required by both the counting knockoff and conformal BH methods \citep{Batetal21, JinCan23} to mitigate the conservativeness of the BH procedure. Additionally, our method addresses fairness in FSR control -- a topic that the aforementioned conformal methods have not explored.

\subsection{Theoretical R-value and optimality theory}\label{subsec:tr}

We briefly discuss the theoretical R-value and its optimality theory, which extends the work of \cite{SunCai07} and \cite{Caietal19} from multiple testing to selective binary classification. Details are deferred to Section \ref{theory:sec} of the Supplement due to space limitations. Despite being developed under an idealized setup, the theory offers practical insights for training score functions to construct more powerful R-values that aim to minimize the number of indecisions while controlling the FSR rates for all sensitive groups. We emphasize two key messages.

First, the choice of an optimal score function indicates that, during the training stage, we should utilize all features, including the sensitive attribute A, to best capture individual-level information. Scores trained without the sensitive attribute are suboptimal. Fairness adjustments should not be made during the training stage but rather in the calibration stage, where the fully informative scores can be converted into R-values to adjust the disparity in error rates across groups. This strategy shares the same spirit as the selection-by-prediction or learn-then-test framework advocated by \cite{JinCan23} and \cite{Angetal22}.

Second, the optimal selection rule equalizes group-wise error rates. To minimize the EPI, the pre-specified marginal FSR (mFSR, defined in the Appendix, Equation \ref{mfsr-ratio}) must be exhausted in every group, making the mFSRs equal to the nominal level. In other words, the constrained optimization formulation \eqref{FSR-a} leads to asymptotic equality of error rates. Our numerical studies support this claim, although a complete analysis is hindered by the dependence among scores, which we leave for future research.

\section{Simulations }\label{section:simulation}

This section presents two simulations under the binary classification setup. The objective is to compare the performance of FASI against the Full Covariate Classifier (FCC). We did not include the Restricted Covariate Classifier (RCC) in these simulations, as RCC has consistently demonstrated larger deviations from the target group-wise FSR levels. We demonstrate that both the oracle and data-driven versions of FASI can control the group-wise FSRs, while RCC fails to do so. The oracle versions of FASI and FCC use the exact {class probabilities}, defined in Equation \ref{prob-score1}, while the data-driven procedures employ the softmax scores via the GAM method \citep{hastie2009elements, James2013, chen_xgboost}. 

In all simulations, we set $|\mathcal D^{train}|=1{,}500$, $|\mathcal D^{cal}|=1{,}000$ and $|\mathcal D^{test}|=1{,}000$. Gender is our protected attribute taking two values $A=F$ (females) and $A=M$ (males). The feature vectors $\pmb X\in \mathbb{R}^3$ are simulated according to the following model:
\begin{equation}\label{rmix-model}
F(\cdot)=\pi_{M}\{\pi_{1|M} F_{1, M}(\cdot)+\pi_{2|M} F_{2, M}(\cdot)\}+\pi_{F}\{\pi_{1|F} F_{1, F}(\cdot)+\pi_{2|F} F_{2, F}(\cdot)\},
\end{equation}
where $\pi_a=\PP(A=a)$, $\pi_{c|a}=\PP(Y=c|A=a)$ and $F_{c, a}$ is the conditional distribution of $\pmb X$ given $Y=c$ and $A=a$. Let $\pi_M=0.5$, and $\pi_F=1-\pi_M=0.5$. The Supplement (Section \ref{sec:sim_overall_fsr}) includes a setup with markedly imbalanced group sizes (e.g., $\pi_M \gg \pi_F$). 
Although only the GAM method is employed in our simulation, we report that our findings remain consistent regardless of the specific learning algorithms utilized. For a comparison of different machine learning algorithms, please refer to Section \ref{sec:multiple_ML_FSR.plot} of the Supplement. We consider two scenarios.

\begin{figure}[t]
\centering
   \scalebox{1.1}
   { \input{Figures/sim_1_r4}}
    \caption{\label{fig:simulation_1_plot}\small Top row: the oracle procedure. Middle and Bottom rows: the data-driven procedure. Left and middle columns (excluding the bottom row): group-wise FSRs (i.e. $\mbox{FSR}^{1}_a$ and $\mbox{FSR}^{2}_a$), Right most column: the EPI levels. Bottom row: overall FSRs (i.e. $\mbox{FSR}^{1}$ and $\mbox{FSR}^{2}$). } 
\end{figure}

\begin{figure}[ht]
\centering
   \scalebox{1.1}
   { 
    \input{Figures/sim_2_r4}
    }
    \caption{\label{fig:simulation_2_plot} \small Comparable setup to Simulation 1 except that the female and male distributions now differ.}
\end{figure}

\begin{figure}[t]
    \centering 
    \scalebox{1.1}{%
\begin{tikzpicture}[x=1pt,y=1pt]
\definecolor{fillColor}{RGB}{255,255,255}
\path[use as bounding box,fill=fillColor,fill opacity=0.00] (0,0) rectangle (289.08,216.81);
\begin{scope}
\path[clip] (  0.00,  0.00) rectangle (289.08,216.81);
\definecolor{drawColor}{RGB}{255,255,255}
\definecolor{fillColor}{RGB}{255,255,255}

\path[draw=drawColor,line width= 0.4pt,line join=round,line cap=round,fill=fillColor] ( -0.00,  0.00) rectangle (289.08,216.81);
\end{scope}
\begin{scope}
\path[clip] ( 24.53, 19.53) rectangle (285.58,203.63);
\definecolor{fillColor}{RGB}{255,255,255}

\path[fill=fillColor] ( 24.53, 19.53) rectangle (285.58,203.63);
\definecolor{drawColor}{RGB}{204,121,167}

\path[draw=drawColor,line width= 1.1pt,dash pattern=on 2pt off 2pt on 6pt off 2pt ,line join=round] ( 36.40,109.84) --
	( 53.35,112.80) --
	( 70.30,113.30) --
	( 87.25,113.43) --
	(104.20,113.51) --
	(121.16,113.19) --
	(138.11,113.55) --
	(155.06,112.10) --
	(172.01,113.43) --
	(188.96,111.48) --
	(205.91,112.60) --
	(222.86,110.98) --
	(239.81,106.67) --
	(256.76,101.32) --
	(273.71, 87.12);
\definecolor{drawColor}{RGB}{0,114,178}

\path[draw=drawColor,line width= 1.1pt,line join=round] ( 36.40,113.21) --
	( 53.35,112.13) --
	( 70.30,110.91) --
	( 87.25,112.60) --
	(104.20,112.79) --
	(121.16,111.93) --
	(138.11,112.66) --
	(155.06,113.18) --
	(172.01,113.09) --
	(188.96,114.88) --
	(205.91,112.18) --
	(222.86,112.01) --
	(239.81,112.34) --
	(256.76,112.47) --
	(273.71,112.10);
\definecolor{drawColor}{RGB}{0,0,0}

\path[draw=drawColor,line width= 0.6pt,dash pattern=on 1pt off 3pt ,line join=round] ( 24.53,114.93) -- (285.58,114.93);
\definecolor{fillColor}{RGB}{248,118,109}

\path[fill=fillColor,fill opacity=0.10] ( 36.40,138.63) --
	( 53.35,142.67) --
	( 70.30,146.48) --
	( 87.25,150.65) --
	(104.20,147.52) --
	(121.16,154.28) --
	(138.11,158.26) --
	(155.06,156.84) --
	(172.01,164.51) --
	(188.96,166.95) --
	(205.91,174.10) --
	(222.86,179.00) --
	(239.81,181.98) --
	(256.76,195.26) --
	(273.71,194.58) --
	(273.71, 27.90) --
	(256.76, 27.90) --
	(239.81, 27.90) --
	(222.86, 48.59) --
	(205.91, 58.96) --
	(188.96, 57.91) --
	(172.01, 67.45) --
	(155.06, 69.05) --
	(138.11, 73.92) --
	(121.16, 75.11) --
	(104.20, 78.52) --
	( 87.25, 77.53) --
	( 70.30, 82.11) --
	( 53.35, 84.11) --
	( 36.40, 84.74) --
	cycle;

\path[] ( 36.40,138.63) --
	( 53.35,142.67) --
	( 70.30,146.48) --
	( 87.25,150.65) --
	(104.20,147.52) --
	(121.16,154.28) --
	(138.11,158.26) --
	(155.06,156.84) --
	(172.01,164.51) --
	(188.96,166.95) --
	(205.91,174.10) --
	(222.86,179.00) --
	(239.81,181.98) --
	(256.76,195.26) --
	(273.71,194.58);

\path[] (273.71, 27.90) --
	(256.76, 27.90) --
	(239.81, 27.90) --
	(222.86, 48.59) --
	(205.91, 58.96) --
	(188.96, 57.91) --
	(172.01, 67.45) --
	(155.06, 69.05) --
	(138.11, 73.92) --
	(121.16, 75.11) --
	(104.20, 78.52) --
	( 87.25, 77.53) --
	( 70.30, 82.11) --
	( 53.35, 84.11) --
	( 36.40, 84.74);
\definecolor{fillColor}{RGB}{0,191,196}

\path[fill=fillColor,fill opacity=0.10] ( 36.40,161.57) --
	( 53.35,157.74) --
	( 70.30,156.83) --
	( 87.25,158.25) --
	(104.20,154.91) --
	(121.16,160.21) --
	(138.11,157.29) --
	(155.06,160.42) --
	(172.01,154.42) --
	(188.96,162.04) --
	(205.91,158.07) --
	(222.86,156.24) --
	(239.81,159.35) --
	(256.76,156.70) --
	(273.71,161.27) --
	(273.71, 68.55) --
	(256.76, 72.71) --
	(239.81, 69.61) --
	(222.86, 66.79) --
	(205.91, 69.34) --
	(188.96, 69.55) --
	(172.01, 72.14) --
	(155.06, 69.54) --
	(138.11, 69.71) --
	(121.16, 68.90) --
	(104.20, 72.00) --
	( 87.25, 72.39) --
	( 70.30, 67.70) --
	( 53.35, 69.72) --
	( 36.40, 68.61) --
	cycle;

\path[] ( 36.40,161.57) --
	( 53.35,157.74) --
	( 70.30,156.83) --
	( 87.25,158.25) --
	(104.20,154.91) --
	(121.16,160.21) --
	(138.11,157.29) --
	(155.06,160.42) --
	(172.01,154.42) --
	(188.96,162.04) --
	(205.91,158.07) --
	(222.86,156.24) --
	(239.81,159.35) --
	(256.76,156.70) --
	(273.71,161.27);

\path[] (273.71, 68.55) --
	(256.76, 72.71) --
	(239.81, 69.61) --
	(222.86, 66.79) --
	(205.91, 69.34) --
	(188.96, 69.55) --
	(172.01, 72.14) --
	(155.06, 69.54) --
	(138.11, 69.71) --
	(121.16, 68.90) --
	(104.20, 72.00) --
	( 87.25, 72.39) --
	( 70.30, 67.70) --
	( 53.35, 69.72) --
	( 36.40, 68.61);
\definecolor{drawColor}{RGB}{204,121,167}

\path[draw=drawColor,line width= 0.6pt,dash pattern=on 4pt off 4pt ,line join=round] ( 36.40, 84.74) --
	( 53.35, 84.11) --
	( 70.30, 82.11) --
	( 87.25, 77.53) --
	(104.20, 78.52) --
	(121.16, 75.11) --
	(138.11, 73.92) --
	(155.06, 69.05) --
	(172.01, 67.45) --
	(188.96, 57.91) --
	(205.91, 58.96) --
	(222.86, 48.59) --
	(239.81, 27.90) --
	(256.76, 27.90) --
	(273.71, 27.90);
\definecolor{drawColor}{RGB}{0,114,178}

\path[draw=drawColor,line width= 0.6pt,dash pattern=on 4pt off 4pt ,line join=round] ( 36.40, 68.61) --
	( 53.35, 69.72) --
	( 70.30, 67.70) --
	( 87.25, 72.39) --
	(104.20, 72.00) --
	(121.16, 68.90) --
	(138.11, 69.71) --
	(155.06, 69.54) --
	(172.01, 72.14) --
	(188.96, 69.55) --
	(205.91, 69.34) --
	(222.86, 66.79) --
	(239.81, 69.61) --
	(256.76, 72.71) --
	(273.71, 68.55);
\definecolor{drawColor}{RGB}{204,121,167}

\path[draw=drawColor,line width= 0.6pt,dash pattern=on 4pt off 4pt ,line join=round] ( 36.40,138.63) --
	( 53.35,142.67) --
	( 70.30,146.48) --
	( 87.25,150.65) --
	(104.20,147.52) --
	(121.16,154.28) --
	(138.11,158.26) --
	(155.06,156.84) --
	(172.01,164.51) --
	(188.96,166.95) --
	(205.91,174.10) --
	(222.86,179.00) --
	(239.81,181.98) --
	(256.76,195.26) --
	(273.71,194.58);
\definecolor{drawColor}{RGB}{0,114,178}

\path[draw=drawColor,line width= 0.6pt,dash pattern=on 4pt off 4pt ,line join=round] ( 36.40,161.57) --
	( 53.35,157.74) --
	( 70.30,156.83) --
	( 87.25,158.25) --
	(104.20,154.91) --
	(121.16,160.21) --
	(138.11,157.29) --
	(155.06,160.42) --
	(172.01,154.42) --
	(188.96,162.04) --
	(205.91,158.07) --
	(222.86,156.24) --
	(239.81,159.35) --
	(256.76,156.70) --
	(273.71,161.27);
\end{scope}
\begin{scope}
\path[clip] (  0.00,  0.00) rectangle (289.08,216.81);
\definecolor{drawColor}{RGB}{0,0,0}

\path[draw=drawColor,line width= 0.4pt,line join=round] ( 24.53, 19.53) --
	( 24.53,203.63);
\end{scope}
\begin{scope}
\path[clip] (  0.00,  0.00) rectangle (289.08,216.81);
\definecolor{drawColor}{gray}{0.30}

\node[text=drawColor,anchor=base east,inner sep=0pt, outer sep=0pt, scale=  0.56] at ( 21.38, 25.97) {0.00};

\node[text=drawColor,anchor=base east,inner sep=0pt, outer sep=0pt, scale=  0.56] at ( 21.38, 69.48) {0.05};

\node[text=drawColor,anchor=base east,inner sep=0pt, outer sep=0pt, scale=  0.56] at ( 21.38,113.00) {0.10};

\node[text=drawColor,anchor=base east,inner sep=0pt, outer sep=0pt, scale=  0.56] at ( 21.38,156.51) {0.15};

\node[text=drawColor,anchor=base east,inner sep=0pt, outer sep=0pt, scale=  0.56] at ( 21.38,200.03) {0.20};
\end{scope}
\begin{scope}
\path[clip] (  0.00,  0.00) rectangle (289.08,216.81);
\definecolor{drawColor}{gray}{0.20}

\path[draw=drawColor,line width= 0.4pt,line join=round] ( 22.78, 27.90) --
	( 24.53, 27.90);

\path[draw=drawColor,line width= 0.4pt,line join=round] ( 22.78, 71.41) --
	( 24.53, 71.41);

\path[draw=drawColor,line width= 0.4pt,line join=round] ( 22.78,114.93) --
	( 24.53,114.93);

\path[draw=drawColor,line width= 0.4pt,line join=round] ( 22.78,158.44) --
	( 24.53,158.44);

\path[draw=drawColor,line width= 0.4pt,line join=round] ( 22.78,201.95) --
	( 24.53,201.95);
\end{scope}
\begin{scope}
\path[clip] (  0.00,  0.00) rectangle (289.08,216.81);
\definecolor{drawColor}{RGB}{0,0,0}

\path[draw=drawColor,line width= 0.4pt,line join=round] ( 24.53, 19.53) --
	(285.58, 19.53);
\end{scope}
\begin{scope}
\path[clip] (  0.00,  0.00) rectangle (289.08,216.81);
\definecolor{drawColor}{gray}{0.20}

\path[draw=drawColor,line width= 0.4pt,line join=round] ( 53.35, 17.78) --
	( 53.35, 19.53);

\path[draw=drawColor,line width= 0.4pt,line join=round] (121.16, 17.78) --
	(121.16, 19.53);

\path[draw=drawColor,line width= 0.4pt,line join=round] (188.96, 17.78) --
	(188.96, 19.53);

\path[draw=drawColor,line width= 0.4pt,line join=round] (256.76, 17.78) --
	(256.76, 19.53);
\end{scope}
\begin{scope}
\path[clip] (  0.00,  0.00) rectangle (289.08,216.81);
\definecolor{drawColor}{gray}{0.30}

\node[text=drawColor,anchor=base,inner sep=0pt, outer sep=0pt, scale=  0.56] at ( 53.35, 12.52) {0.2};

\node[text=drawColor,anchor=base,inner sep=0pt, outer sep=0pt, scale=  0.56] at (121.16, 12.52) {0.4};

\node[text=drawColor,anchor=base,inner sep=0pt, outer sep=0pt, scale=  0.56] at (188.96, 12.52) {0.6};

\node[text=drawColor,anchor=base,inner sep=0pt, outer sep=0pt, scale=  0.56] at (256.76, 12.52) {0.8};
\end{scope}
\begin{scope}
\path[clip] (  0.00,  0.00) rectangle (289.08,216.81);
\definecolor{drawColor}{RGB}{0,0,0}

\node[text=drawColor,anchor=base,inner sep=0pt, outer sep=0pt, scale=  0.70] at (155.06,  4.86) {$\pi_{2|F}$};
\end{scope}
\begin{scope}
\path[clip] (  0.00,  0.00) rectangle (289.08,216.81);
\definecolor{drawColor}{RGB}{0,0,0}

\node[text=drawColor,rotate= 90.00,anchor=base,inner sep=0pt, outer sep=0pt, scale=  0.70] at (  8.32,111.58) {$\mbox{FSR}^{1}_a$};
\end{scope}
\begin{scope}
\path[clip] (  0.00,  0.00) rectangle (289.08,216.81);
\definecolor{drawColor}{RGB}{0,0,0}

\node[text=drawColor,anchor=base,inner sep=0pt, outer sep=0pt, scale=  0.70] at (155.06,208.49) { };
\end{scope}
\end{tikzpicture}}
    	\caption{\label{quantiles_FSR_control.plot}\small $90\%$ quantiles of the FSPs. Red region: female group; Blue region: male group. }
\end{figure}

In the first scenario, the conditional distributions of $\pmb X$ given class $Y$ are assumed to be multivariate normal and are identical for males and females:
$$
F_{1, M}=F_{1, F} =\mathcal N(\pmb \mu_1, 2\cdot \mathbf I_3),
\quad F_{2, M}=F_{2, F} = \mathcal N(\pmb \mu_2, 2\cdot \mathbf I_3),
$$
where $\mathbf I_3$ is a $3\times 3$ identity matrix, $\pmb \mu_1=(0,1,6)^\top$ and $\pmb \mu_2= (2,3,7)^\top$. The only difference between the group-wise distributions lies in the conditional proportions: we fix $\pi_{2|M} = \mathbb{P}(Y=2|A=M) = 0.5$, while varying $\pi_{2|F} = \mathbb{P}(Y=2|A=F)$ from $0.15$ to $0.85$. We shall see that in the asymmetric situation (i.e., when $\pi_{2|F}$ is very large or small), the unadjusted FCC rule leads to unfair policies (i.e. we observe disparate FSRs across the male and female groups). 

We simulate 1,000 data sets and apply both the FCC and FASI methods at an FSR level of 0.1 to these simulated data sets. The FASI method is implemented with R-values defined in \eqref{Q-kc-plus}--\eqref{R-value-max}. For the FCC method, the protected attributes are ignored when computing the fractions in \eqref{Q-kc-plus}, and these fractions are denoted as \(Q_k^{c, \text{FCC}}\). Then, the \(Q_k^{c, \text{FCC}}\) values are adjusted according to \eqref{R-value-max} to obtain the R-values, denoted as \(R^{c, \text{FCC}}_{j}\). The corresponding selection rule is 
$\hat{Y}_j^{\text{FCC}}= \sum_{c \in \{1, 2\}} c \cdot \mathbb{I}\bigl(R^{c, \text{FCC}}_{j} \leq 0.1\bigr)$, for $j \in \mathcal{D}^{test}.$

The FSR levels are computed by averaging the respective false discovery proportions (FSPs) from $1{,}000$ replications. The simulation results are summarized in Figure~\ref{fig:simulation_1_plot}. The first and second rows respectively correspond to the oracle and data-driven versions of each method. The first two columns respectively plot the group-wise FSRs for class 1 and class 2 as functions of $\pi_{2|F}$. The final column plots the EPI \eqref{def:EPI}, obtained by averaging the results from 1,000 replications. The following patterns can be observed.

\begin{itemize}
      
 \item FCC fails to control the group-wise FSRs. As $\pi_{2|F}$ moves away from $\pi_{2|M}=0.5$, the gap between the FSR control for Females and Males dramatically widens due to the asymmetry in the proportions of the signals (true class 2 observations) in the male and female groups. 
    
    \item Both the oracle and data-driven FASI procedures consistently control the FDR at the nominal level. However, when $\pi_{2|F}$ is high, the number of selections decreases, resulting in a reduced total number of selections from both groups. Consequently, both methods exhibit increased conservativeness. This pattern can be attributed to the conservative nature of the R-value, which includes a ``+1'' adjustment and functions as an estimate of the true false selection proportion: the level of conservativeness becomes more pronounced as the proportion $\pi_{2|F}$ become close to either 0 or 1. 

    \item Both oracle and data-driven FASI algorithms are able to roughly equalize the group-wise FSRs between the Female and Male groups, while also controlling the overall FSR. The data-driven FASI is able to closely mirror the behavior of the oracle method. 
                 
    \item The parity in FSR control is achieved at the price of slightly higher EPI levels.
    
\end{itemize}

Our second simulation considers the setting where $F_{c, M}\neq F_{c, F}$. Denoting the mean for class c and protected attribute a as $\pmb \mu_{c,a}$, the data is generated from $F_{c, a} = \mathcal N(\pmb \mu_{c,a}, 2\cdot \mathbf I_3)$, with components $\pmb \mu_{1,M} = (0,1,6)^\top$, $\pmb \mu_{2,M} = (2,3,7)^\top$, $\pmb \mu_{1,F} = (1,2,7)^\top$ and $\pmb \mu_{2,F} = (3,4,8)^\top$. In all other respects Simulations 1 and 2 are identical. The results for the second simulation scenario are provided in Figure~\ref{fig:simulation_2_plot}. We notice very similar patterns to our first simulation setup. FASI controls the group-wise FSRs for all values of $\pi_{2|F}$ while the FCC fails to do so. The data-driven FASI closely emulates the oracle procedure, for both the FSR and EPI levels. 

Finally, we examine the variability of the false discovery proportions (FSP), which can fluctuate across replications. Specifically, the FSR is derived as the average of the FSPs. While our theory ensures that the FSR can be controlled under the nominal level $\alpha$, it is important to note that the FSP may deviate significantly from $\alpha$. To investigate this variability, we focus on the same experimental setting in Simulation 1 used to generate Figure \ref{fig:simulation_1_plot}, and present the $90\%$ quantiles of the group-wise FSPs. The summarized results are depicted in Figure \ref{quantiles_FSR_control.plot}.

The group-wise FSRs, represented by solid blue and dot-dashed red lines, are effectively controlled at the desired $10\%$ level. The quantiles are visually depicted by blue/red regions, corresponding to the male/female  groups, respectively. For the male group, where $\pi_{2|M}$ remains constant, the $90\%$ quantiles range between $5\%$ and $15\%$. In contrast, the FSP variability for the female group is more pronounced, with greater variability when $\pi_{2|F}$ is larger, as few selections are made from the female group.

\section{Real Data Examples}
\label{section:real_data}

This section demonstrates the application of FASI on two real data sets. Sections~\ref{compas_application_section} and ~\ref{census.sec} respectively analyze the COMPAS data \citep{Angetal16-COMPAS1} and US census data \citep{Dua:2019}. For the COMPAS and census data, we have employed GAM and Adaboost models, respectively, to construct confidence scores. It is important to note that users have the flexibility to choose the best model for their specific application by utilizing their own training data. To facilitate the implementation of FASI with user-specified models, the R package \texttt{fasi} has been developed and is readily available on CRAN.  
  
\subsection{COMPAS data analysis}
\label{compas_application_section}

In 2016, ProPublica's investigative journalists curated a data set of 6,172 individuals, where 3,175 were Black and the remaining 2,997 belonged to other racial categories, who had been arrested in Broward County, Florida. These racial categories, Black and Other, serve as our protected attributes in this study. Within the data set, the ``Black'' group consisted of 1,773 individuals who were identified as having recidivated within the 2-year time frame considered in the study, while the ``Other'' group consisted of 1,217 individuals who also recidivated during this period. This 2-year window was chosen as a proxy for the true label of identifying recidivists. 

All individuals were assigned a risk score by the COMPAS algorithm (a whole number between 1 and 10) developed by NorthPointe Inc. This score was used to inform the judge of each person's risk of recidivating during their bail hearing. 
The data set contains demographic information about each person including their race, age, number of previous offenses, sex, number of prior offenses, and their assigned COMPAS risk score. 

In this analysis, our objective is to utilize FASI to address potential disparities in FSRs among different racial groups. The literature has extensively examined various fairness notions, such as disparate treatment \citep{Zafar_disparate_treatment}, as well as studies specifically related to the COMPAS data set \citep{Angetal16-COMPAS1, Dieetal16-COMPAS2}. It is crucial to carefully evaluate and scrutinize the societal trade-offs associated with different definitions of fairness. 

\begin{figure}[t]
   \scalebox{1.1}{
    \centering 
    \input{Figures/compas_analysis_plt}
    } 
    	\caption{\label{compas_application.plot}\small COMPAS data analysis for predicting recidivists. Left and Middle: False Selection Rate minus the desired control level for varying levels of $\alpha$ for the FCC and FASI method respectively. Right: The EPI for both the FCC and FASI method.    }
\end{figure}

We performed 100 random splits of the data set, where for each protected group and class label $Y$ (our proxy for recidivism), 90\% of the data was assigned to $\mathcal{D}$ and the remaining 10\% to $\mathcal{D}^{test}$. Furthermore, we evenly split $\mathcal{D}$ into $\mathcal{D}^{train}$ and $\mathcal{D}^{cal}$. To assess the performance, we present the results across a range of $\alpha$ values from 0.15 to 0.30. The first two columns in Figure~\ref{compas_application.plot} illustrate the difference between the true and target FSRs for the FCC and FASI algorithms, respectively. The last column of the figure plots the EPI levels.

While the FCC approach effectively controls the overall FSR, it falls short in controlling the FSRs across different racial groups. In the left panel of Figure~\ref{compas_application.plot}, we can observe that the race-wise FSRs deviate substantially from the nominal level, and the FSR levels for the Black group are significantly lower compared to those of the Other group. This discrepancy persists consistently across all values of $\alpha$. In contrast, the middle panel of Figure~\ref{compas_application.plot} demonstrates that by employing the FASI algorithm, the race-wise and overall FSR levels are effectively controlled below the nominal level and are approximately equalized across the sensitive groups. Moreover, the right panel illustrates that FASI achieves a nearly identical EPI level as the FCC approach.

\subsection{1994 census income data analysis}
\label{census.sec}

The US census is a primary source of information for generating data concerning the American population. Consequently, the data they collect plays a direct role in informing future policy decisions, such as allocating resources for programs that offer economic assistance to vulnerable populations. These resources encompass necessities such as food, healthcare, job training, housing, and other forms of economic aid, which rely on accurate estimates of income levels within the population. 
The potential consequences of making unfair decisions when predicting income levels can be significant, as these predictions contribute to determining how hundreds of billions of dollars in federal funding will be allocated over the next decade. In this case study, we utilize the 1994 US Census Data set from the UCI Machine Learning Repository to predict whether an individual earns above or below \$50,000 per year, with Class 1 representing individuals earning less than \$50,000 and Class 2 representing those earning more than \$50,000. To avoid overlapping selections, we utilize the two-stage procedure described in Section \ref{subsec:R-multiclass} of the Supplement.  

The data set in this study comprises 32,561 observations on 14 variables, predominantly demographic factors such as education level, age, and hours worked per week, among others. The protected attributes under consideration are Female and Male. Specifically, the Female group consists of a total of 10,771 observations, with 1,179 individuals earning over \$50,000 per year. Similarly, the Male attribute encompasses the remaining 21,790 observations, with 6,662 individuals earning over \$50,000 per year.

We applied the FCC and FASI algorithms at different FSR levels, ranging from 0.05\% to 10\%. We performed 100 random splits of the data set, where for each gender and class label, 70\% of the data was randomly assigned to $\mathcal{D}$, and the remaining 30\% was assigned to $\mathcal{D}^{test}$. Furthermore, $\mathcal{D}$ was evenly divided into $\mathcal{D}^{train}$ and $\mathcal{D}^{cal}$. The left and middle panels of Figure~\ref{adult_application.plot} in Appendix \ref{subsec:income-plot} respectively show the FSR levels for both the FCC and FASI.

From the left column, we can observe that the group-wise FSR levels of FCC consistently deviate from the nominal level $\alpha$, resulting in unfair decisions for the sensitive groups. This pattern is observed in both Class 1 and Class 2, although in opposite directions. The disparity in group-wise FSR levels becomes more pronounced as $\alpha$ increases. 
In contrast, the middle column demonstrates that for Class 1, the group-wise and overall FSR levels of FASI remain close to $\alpha$. For Class 2, the group-wise and overall FSR levels of FASI exhibit conservativeness but are roughly equalized across the two sensitive groups. The conservativeness can be attributed to the R-value, which provides a conservative estimate of the true FSP. 
Furthermore, the right column highlights that FASI effectively achieves approximate parity, ensuring that the burden is roughly equally shared across the two genders, with only a slight increase in the EPI level.

\section{Discussion}\label{si-general:sec}

This section concludes the article by discussing additional fairness notions, highlighting limitations in existing research and suggesting future directions.

Fairness in machine learning presents a complex challenge. Multiple studies focus on addressing representation or sampling bias, which arises when data are collected in a non-representative fashion \citep{Mehrabi_fairness_review}. By contrast, algorithmic bias emerges when the model itself introduces bias beyond the inherent biases in the input data. This article addresses the issue of algorithmic bias, with the objective of ensuring an equitable distribution of erroneous decisions across different groups. FASI is model-free, allowing for deployment with any user-specified model. It achieves fairness by aligning group-wise FSRs to the same designated level, requiring only mild conditions on data exchangeability.

In addition to the sufficiency principle, the \emph{separation principle} \citep{Bar17} has been widely used. It requires that 
$P(Y\neq \hat Y|Y=c, A=a)$ are the same for all $a\in\mathcal A$. This principle differs from the sufficiency principle \eqref{suff-princ}, whereby $\hat{Y}$ and $Y$ interchange positions in the conditional probability expression. A third notion on fairness, in the context of prediction intervals, has been considered in \cite{Rometal20-Eq}. Rather than conditioning on either $Y$ or $\hat Y$, this fairness criterion is concerned with the joint probabilities of $(\hat Y, Y)$, requiring that the misclassification rates are equalized across all protected groups $P(Y\neq \hat Y|A=a)$ are the same for all $a\in\mathcal A$. 
The fourth notion, known as \emph{demographic parity} \citep{Jiang_2020} requires that
$P(\hat Y \neq c|A=a)$ are the same for all $a \in \mathcal A$. Other popular fairness notions include \emph{equalized odds} (\citealp{Haretal16, Rometal20-Odds}) and \emph{equalized risks} (\citealp{Coretal18}).

 \cite{Zafar_disparate_treatment} proposed the use of cost-sensitive classifiers with group-specific costs \citep{Menon_cost_fairness} to address a fairness issue comparable to our work. However, their technique forces a decision to be made on all individuals, whereas our approach is a selective inference procedure that only makes confident judgments on a subset of subjects. Given human intervention, FASI can achieve higher accuracy than cost-sensitive classifiers, as practitioners are aware of the undecided cases that merit additional scrutiny, ultimately reducing erroneous decisions with potentially extensive societal costs.

Our fairness criterion, as described in Equation \ref{FSR-a}, constitutes a group fairness notion that presupposes full knowledge of the protected groups. This approach is widely adopted in the literature and finds applications across diverse domains, including medicine and the criminal justice system \citep{Manrai_medicine_disparities, Angetal16-COMPAS1}, often facilitated by specialized software tools \citep{IBM_fair_software, fairness_audit_toolkit}. However, situations may arise where the protected groups lack clear delineation, such as when the sensitive attribute pertains to age or income. New ideas, such as individual fairness and counterfactual fairness, provide useful alternatives. Specifically, individual fairness aims to ensure that comparable individuals receive commensurate outcomes \citep{Mukherjee_individual_fairness}, while counterfactual fairness posits that fairness should not be exclusively contingent on observable attributes but should also consider potential counterfactual factors. Given the substantial  complexities associated with individual and counterfactual fairness algorithms, we leave exploration of this promising avenue in future research.

A highly contentious matter is that disparate fairness criteria often yield distinct algorithms and different decisions in practice. For instance, the sufficiency and separation principles can be incompatible with one another (\citealp{Kleetal16, Frietal21}), and classification parity or group calibration can potentially harm the very groups that these algorithms are intended to protect (\citealp{Coretal18}). Despite growing awareness of fairness concerns in decision-making, a consensus is yet to be reached on the best approaches for achieving fairness in machine learning. While we do not claim that FASI is ubiquitously superior to competing approaches, adjusting group-wise FSRs appears to be an effective and suitable fairness criterion for high-stake applications, overcoming several limitations of the widely used sufficiency principle. Much research is still needed for understanding the trade-offs and applicability of different fairness notions across diverse contexts and applications. 

{\singlespace
\small
\bibliographystyle{plainnat}
\bibliography{myrefs.bib}
}
\newpage

\setcounter{page}{1}

\appendix 

\renewcommand{\theequation}{\thesection.\arabic{equation}}
\setcounter{equation}{0}

\begin{center}
\Large Online Supplementary Material for ``A Burden Shared is a Burden Halved: A Fairness-Adjusted Approach to Classification'' 
\end{center}

\medskip
\medskip
\medskip

\label{appendix:different_rval}
This supplement provides a comparison of various R-value notions (Section \ref{appendix:different_rval}), additional technical details of the methodology (Sections \ref{theory:sec}-\ref{RQP:sec}), technical proofs (Sections \ref{proof-thm1:sec}-\ref{proof-thm3:sec}), discussion of related fairness algorithms and possible extensions (Sections \ref{appendix:related_algorithms} and \ref{subsec:R-multiclass}), and supplementary numerical results (Section \ref{RvRp:sec}).

\section{Variants of the (empirical) R-value}\label{appendix:different_rval}

The R-value \eqref{R-value-max} has been proposed as the basic operational unit of our FASI algorithm; we discuss its empirical variants in this section and its theoretical version in Section \ref{theory:sec}.

First, while including both $\mathcal{D}^{cal}$ and $\mathcal{D}^{test}$ in the denominator of \eqref{Q-kc-plus} enhances the algorithm's stability, the resulting FASI algorithm can only control a modified version of the FSR [cf. \eqref{FSR-star}] asymptotically. The simpler version, which only includes $\mathcal{D}^{test}$ in the denominator of \eqref{Q-kc}, is particularly relevant for readers who prefer a validity theory in finite samples. 

Secondly, one consideration, pertaining to the multiplicative factor $\gamma_{c,a} = \mathbb E\left(p^{test,a}_{c,null} / p_{c, null}^{cal,a}\right)$ in the theorem, is that FASI fails to provide precise FSR control due to the possible fluctuations in $\gamma_{c,a}$. While this issue is also minor (as under Assumption \ref{ex:as}, this constant is approximately 1 and numerically negligible, cf. Section \ref{gamma_sim_est:sec} of this Supplement), we present a conservative version of the R-value next. We demonstrate that this multiplicative factor $\gamma_{c,a}$ can be eliminated from the theory when the conservative version is employed. It is important to note that the conservative R-value is primarily of theoretical interest, as it leads to a substantial loss in power in many practical scenarios.

We summarize related results in the subsequent corollary. The proof of the corollary follows directly from the proof of Theorem \ref{FSR:thm} and is therefore omitted.

\begin{corollary}\label{modified-R-col}
Suppose we apply the FASI algorithm with the conservative R-value: 
\beq\label{conservative-R}
{R}_{j}^{c, *}=\max\left\{ \II(\hat S_j^c\leq 0.5), \frac{n_{a}^{cal}+1}{ n^{cal,c}_{a, null}+1} \tilde Q_{j}^{c}\right\} 
\eeq
for $j \in \mathcal{D}^{test}$, where $\tilde Q_{j}^{c}$ is defined via \eqref{Q-kc} and \eqref{Q-value-mod}, $n^{cal}_{a}=\sum_{i\in \mathcal D^{cal}}\II(A_i=a)$ and $n^{cal,c}_{a,null}=\sum_{i\in \mathcal D^{cal}}\II(A_i=a, Y_i\neq c)$. Further define $n^{test}_{a}=\sum_{j\in \mathcal D^{test}}\II(A_j=a)$ and $n^{test,c}_{a,null}=\sum_{j\in \mathcal D^{test}}\II(A_j=a, Y_j\neq c)$.
Then we have, for all $a\in\mathcal A$, 
$
\mbox{FSR}_a^{\{c\}}\leq \EE\left({n_{a,null}^{test,c}}/{n_{a}^{test}} \right)\alpha\leq \alpha. 
$ 
\end{corollary}

The ratio ${n_{a,null}^{test,c}}/n_{a}^{test}$, in Corollary \ref{modified-R-col} is referred to as the null proportion in multiple testing, also appears in the classical Benjamini-Hochberg (BH) procedure for FDR control. In Section \ref{subsec:RV-BH} of this supplement, we will elaborate the connection between the FASI algorithm and the BH algorithm implemented with conformal p-values.   

It is expected that the FASI algorithm with conservative R-values \eqref{conservative-R} can be enhanced by incorporating the unknown ratio ${n_{a,null}^{test}}/n_{a}^{test}$ into the analysis. This approach has been successfully adopted in various works, such as \cite{BenHoc00} and \cite{Sto02}, to boost the power of the conservative BH algorithm in the context of FDR control. The FASI algorithm with R-value defined in \eqref{R-value-max} can be roughly regarded as such an approach. Specifically, the unknown ratio ${n_{a,null}^{test,c}}/n_{a}^{test}$ is initially estimated as $(n_{a,null}^{cal,c}+1)/(n_{a}^{cal}+1)$. This estimated ratio is then incorporated into the FASI algorithm by utilizing the conservative version of FASI at the modified level of $(n_{a}^{cal}+1)/(n_{a,null}^{cal,c}+1)\alpha$. This practice leads to improved power at the expense of the additional factor $\gamma_{c,a}$ in Theorem \ref{FSR:thm}.

\setcounter{equation}{0}

\section{Theoretical R-value and Optimality Theory}\label{theory:sec}

In this section, we introduce the theoretical R-value and derive the optimal score function under a simplified setup. Our subsequent discussions are purely theoretical, where we assume an oracle with access to all distributional information and make several simplifying assumptions. Our primary goal is to develop a theoretical version of the R-value and an optimality theory for FSR control. This theoretical framework serves as a foundation for our practical algorithm and provides valuable insights into the properties of the R-value.
Our theory provides practical insights for practitioners on how to train score functions to construct informative R-values. 
 
\subsection{The mixture model under an oracle setting}\label{thR:sec}

Denote $\pi_a=P(A_j=a)$ and $\pi_{c|a}=P(Y_j=c|A_j=a)$. We assume that $(X_j,A_j)$ are independent observations obeying the following random mixture model:
\beq\label{rmix1}
F(x,A_j)= \sum_{a\in \mathcal A} \II(A_j=a) \cdot \left\{\pi_{1|a} F_{1|a}(x)+ \pi_{2|a} F_{2|a}(x)\right\},
\eeq
where $F_{1|a}(x)$ and $F_{2|a}(x)$ are the conditional CDFs of $X_j$ coming from classes $1$ and $2$ given that $A_j=a$, respectively. Let $f_{c|a}(x)$ be the corresponding density function.  For our analysis, we consider a class of oracle rules of the form 
\begin{equation}\label{select-rule-oracle}
\hat{Y}_j\coloneqq \hat Y_j(X_j, A_j)~=~\sum_{a\in\mathcal A} \mathbb I(A_j=a)\cdot \left\{\sum_{c \in \{1, 2\}} c \cdot \mathbb{I}(S_j^c > t_a^c)\right\}, \quad \text{for } j \in \mathcal{D}^{test},
\end{equation}
where the thresholds satisfy \( t_a^c > 0.5 \) to avoid overlapping selections.
We assume that an oracle has knowledge of the conditional probabilities and conditional density functions defined above.

\subsection{The conversion algorithm}\label{conversion:subsec}

In this section, we present a systematic approach for converting an arbitrary score \( S^c(x,a) \) into a fair score \( R^c(S^c) \), which we refer to as the theoretical R-value. Although the discussion is theoretical in nature, it highlights the existence of a fair score corresponding to every confidence score. This algorithm can be regarded as a method of \emph{calibration by group}, a widely used technique in the fairness literature (see \cite{Bar17} for an example). Our discussion assumes that (a) the score function is known and (b) the distributional information of the scores is available; this is referred to as the oracle setup, which does not involve utilizing labeled training and calibration data \( \mathcal D^{train} \) and \( \mathcal D^{cal} \). 

The conversion algorithm consists of three steps. In Step 1, we define and derive important quantities for \( S^c_j \coloneqq S^c(X_j, A_j) \), where \( j \in \mathcal D^{test} \). Under the random mixture model \eqref{rmix1} (for the observed data points), the scores \( S^c_j \) obey the following mixture model:
\begin{equation}\label{model:rmix-yns}
G^c(s)~=~\sum_{a\in\mathcal A} \II(A_j=a)\cdot G^c_a(s) ~=~\sum_{a\in\mathcal A}\II(A_j=a)\cdot\left\{ \pi_{1|a} G_{1|a}^c(s) + \pi_{2|a} G_{2|a}^c(s) \right\},
\end{equation}
where \( G_{c^\prime|a}^c(s) \) denotes the conditional CDF of \( S^c \) given \( A = a \) and \( Y = c^\prime \), and \( \pi_{c^\prime|a} = \PP(Y_i = c | A_i = a) \) are the conditional probabilities for \( c^\prime = 1, 2 \). 

Under the oracle setup, the conditional probabilities and conditional CDFs defined above are assumed to be known. In Step 2, we compute the conditional error probabilities for individuals from group \( a \) when the threshold for \( S_j^c \)  is \( t_a^c\) using decision rule \eqref{select-rule-oracle}: 
\begin{eqnarray*}
\texttt{err}^1_a (t_a^1) & = & \PP(Y =2 | S_j^1 > t_a^1, A = a) =\frac{\pi_{2|a}\left\{1-G_{2|a}(t_a^1)\right\}}{1-G_a^1(t_a^1)}; \\
\texttt{err}^2_a (t_a^2) & = & \PP(Y =1 | S_j^2 > t_a^2, A = a) =\frac{\pi_{1|a}\left\{1-G_{1|a}(t_a^2)\right\}}{1-G_a^2(t_a^2)}.
\end{eqnarray*}

Finally, in Step 3 we compute a pair of fair scores for every individual from group $a$ with observed scores $(S^1_j=s_j^1, S^2_j=s_j^2)$: 
\beq\label{TR-th}
\texttt{TQ}_j^c~\equiv~\texttt{TQ}_j^c(s^c_j)~=~\sum_{a\in\mathcal A}\II(A_j=a)\cdot\inf_{t\leq s^c_j}\left\{\texttt{err}^c_a (t)~=~\PP(Y_j\neq c | s_j^c> t, A = a)\right\}, 
\eeq
for $c\in\{1,2\}$ and $j\in\mathcal D^{test}$. If the confidence score satisfies the monotone likelihood ration condition (MLRC, \citealp{SunCai07}), then the infimum is achieved at $s_j^c$ exactly. To avoid overlapping selections, we define the theoretical R-values as 
\begin{equation}\label{TR-value}
\texttt{TR}_j^{c} = \max\left\{\mathbb{I}({S}_{j}^{c} \leq 0.5), \hspace{0.1cm} \texttt{TQ}_j^c \right\}, \quad c\in\{1,2\}, \; j \in \mathcal{D}^{test}.
\end{equation}

\subsection{Theoretical R-value and fairness}\label{thR:sec}

Consider random mixture model \eqref{rmix1}. Let \( \{S^c_j \coloneqq S^c(X_j, A_j): c\in\{1,2\}, j \in \mathcal D^{test}\} \) be the confidence scores and $\{\texttt{TR}_j^{c}: c\in\{1,2\}, j\in\mathcal D^{test}\}$ denote the corresponding theoretical R-values. The goal is to assign labels ``0'', ``1'' and ``2'' to new instances {$\{(X_{j}, A_{j}): j\in \mathcal{D}^{test}\}$}. Consider the following classification rule: 
\begin{equation}\label{TR-rule} 
\hat Y_j=\sum_{c\in\{1,2\}} c\cdot \II(\texttt{TR}_j^c\leq\alpha_c).
\end{equation}

Define the marginal FSR
\beq\label{mfsr-ratio}
\mbox{mFSR}^{\{c\}}_a=\frac{\mathbb E\left\{\sum_{j \in \mathcal{D}^{test}} \mathbb I(\hat Y_{j}=c, {Y}_{j}\neq c, : A_{j}=a)\right\}}{\mathbb E\left\{\sum_{j \in \mathcal{D}^{test}} \mathbb I(\hat Y_{j}=c, A_{j}=a)\right\}}.
\eeq
We assume that the instances {$(X_j, A_j)$} are independent draws from an underlying random mixture model \eqref{rmix1}. It can be shown that, following arguments in \cite{Sto03} for FDR analysis,
\beq\label{mfsr-ca}
\mbox{mFSR}^{\{c\}}_a=\PP(Y_j\neq c|\hat Y_j = c, A=a),
\eeq
which is the conditional probability required in the sufficiency principle [cf. Equation \eqref{suff-princ} in the main text].

Based on the work of \cite{Caietal19}, we can similarly show that under mild conditions, 
\beq\label{fsr-ca}
\mbox{FSR}^{\{c\}}_a=\mbox{mFSR}^{\{c\}}_a+o(1), \mbox{ when } m_a\coloneqq |\{j \in \mathcal{D}^{test}: A_{j}=a\}|\rightarrow \infty.
\eeq

The next proposition shows that thresholding the theoretical R-value leads to a fair selective inference procedure.

\begin{proposition}\label{theoretical-R:thm}\rm{
Consider the classification rule \eqref{TR-rule}. Then we have
\beq\label{suf:con}
\mbox{$\PP(Y_j\neq c| \hat Y_j = c, A=a)\leq \alpha_c$ for $c\in\{1,2\}$ and $a\in \mathcal A$.}   
\eeq
}
\end{proposition}

We would like to make two important remarks. Firstly, the theoretical R-value, which may be viewed as the counterpart of the data-driven R-value, represents the minimum conditional probability required to ensure that an individual with score $S_j^c=s^c$ is selected into class $c$. Secondly, the theoretical R-value is a fundamental quantity that is closely linked to the sufficiency principle in the fairness literature. Proposition \ref{theoretical-R:thm} highlights that by setting thresholds for the theoretical R-values, the thresholding procedure fulfills the sufficiency principle and controls the group-wise error rates. 

\subsection{Oracle theoretical R-value and the optimality theory}\label{opt:sec}

We present and prove an intuitive result that shows the class probability 
$$
S_{OR}^{c,j}(x,a)=\PP(Y_j=c|X_j=x, A_j=a)
$$ 
is the optimal choice of confidence score for calibrating the theoretical R-value. To simplify the arguments, we develop our optimality theory based on the mFSR, an asymptotically equivalent variation of the FSR. The relationship between the mFSR and FSR has been established in Equation \eqref{fsr-ca}. 

We aim to construct a selection rule under the binary classification setting that solves the following constrained optimization problem:
\beq\label{prob-form}
\text{Minimize the EPI, subject to $\mbox{mFSR}^{c}_a\leq\alpha_c$, for $c\in\{1, 2\}$ and $a\in\mathcal A$.} 
\eeq

Our strategy is to convert the oracle scores $S^{c,j}_{OR}$ to oracle theoretical R-values
$$
\left\{\texttt{TR}_{OR}^{c,j}: c\in\{1,2\}, j\in\mathcal D^{test}\right\}.
$$ 
The process of conversion follows the general strategy outlined in Section \ref{conversion:subsec}, and is described in more detail in the proof of Theorem \ref{optimality:thm} below.

Consider the selective classification problem outlined in \eqref{prob-form}. Define the oracle procedure $\pmb\delta_{OR}=\{\delta_{OR}^j: j \in \mathcal{D}^{test}\}$, where
\beq\label{oracle-proc}
\delta_{OR}^j=\sum_{c\in\{1,2\}} c\cdot \mathbb I(\texttt{TR}_{OR}^{c,j}\leq\alpha_c). 
\eeq
The optimality of the oracle procedure is established in the next theorem. 
\begin{theorem}\label{optimality:thm}
Consider random mixture model \eqref{rmix1}. Let $\mathcal D_{\alpha_1, \alpha_2}$ denote the collection of selection rules of the form \eqref{TR-rule} that satisfy $\mbox{mFSR}^{c}_a\leq\alpha_c$ for $c=1, 2$ and all $a\in \mathcal A$. Let $\mbox{EPI}_{\pmb\delta}$ denote the EPI of an arbitrary decision rule $\pmb\delta$ without overlapping selections. Then the oracle procedure \eqref{oracle-proc} is optimal in the sense that $\mbox{EPI}_{\pmb\delta_{OR}}\leq \mbox{EPI}_{\pmb\delta}$ for any $\pmb\delta\in\mathcal D_{\alpha_1, \alpha_2}$. 
\end{theorem}

The optimality theory indicates that, during the training stage, we should utilize \emph{all features}, including the sensitive attribute $A$, to best capture individual level information. 

\subsection{R-value and Storey's Q-value}\label{convert:sec}

Excluding the sensitive attribute $A$ in our analysis, the theoretical R-value is closely connected to the q-value, a useful tool in large-scale testing due to its intuitive interpretation and ease of use, as described in \cite{Sto03}. 

To test hypotheses {$\{H_j: j\in \mathcal{D}^{test}\}$} with associated p-values {$\{p_j: j\in \mathcal{D}^{test}\}$}, let $\pi$ be the proportion of non-nulls and $G(t)$ the alternative distribution of p-values. The q-value for hypothesis {$H_j$} is defined as 
$$
\inf_{t\geq p_j}\left\{\mbox{pFDR}(t)\coloneqq \frac{(1-\pi)t}{(1-\pi)t+\pi G(t)}\right\},
$$ 
which roughly measures the fraction of false discoveries when {$H_j$} is rejected. 

The q-value and R-value algorithms operate in the same manner. Conducting an FDR analysis at a given level $\alpha$ entails obtaining the q-value for hypothesis $j$ and rejecting it if the q-value is less than or equal to $\alpha$. Likewise, conducting an FSR analysis at level $\alpha$ involves obtaining the R-value for individual $j$ and selecting it if the R-value is less than or equal to $\alpha$.

\setcounter{equation}{0}

\section{R-value and Conformal P-value}\label{RQP:sec}

In this section, we adopt a multiple testing perspective to analyze the R-value. Although motivated differently, we show that the R-value is equivalent to the (BH) q-value of the conformal p-values \citep{Batetal21} in a one-class classification scenario. For comparability considerations, we exclude the sensitive attribute \(A\) in the following discussions.

\subsection{Selective inference for one-class classification: a multiple testing perspective}\label{subsec:OCC-MT}

The selective inference perspective described in Section \ref{si-binary:sec} provides a flexible framework accommodating various types of classification rules. For instance, if the interest lies solely in pinpointing high-risk individuals, the action space is defined as \( \Lambda = \{0, 2\} \), and one may employ the following rule for screening: 
\begin{equation}\label{select-rule2}
\hat{Y}_j = 2 \cdot \mathbb{I}(S_j^2 > t_2), \quad t_2 > 0.5, \quad \text{for } j \in \mathcal{D}^{test}.
\end{equation}
This setup is closely related to semi-supervised multiple testing or outlier detection in conformal inference (\citealp{mary22semi, Batetal21, Liaetal22}). 

Next, we explain the connection between the FSR and FDR from a multiple testing perspective. Consider testing \( m \) hypotheses:
\beq\label{outlier-detection}
H_{j0}:\; Y_j=1 \quad \text{vs.} \quad H_{j1}:\; Y_j\neq 1 \; (\text{i.e.,}\; Y_j=2), \quad \text{for }j \in \mathcal{D}^{test}. 
\eeq
In the context of the selective inference framework of Section~\ref{si-binary:sec}, this multiple testing problem has the state space \(\mathcal{C} = \{1, 2\}\). A multiple testing procedure represented by \(\hat{Y}_{j}\in\{0,2\}, \, j\in\mathcal{D}^{test}\) corresponds to the selection rule in \eqref{select-rule2}. Here, the action space \(\Lambda=\{0, 2\}\) differs from the state space \(\mathcal{C} = \{1, 2\}\), with \(\hat{Y}_{j}=2\) indicating that \(H_{j0}\) is rejected, and \(\hat{Y}_{j}=0\) signifying insufficient evidence to reject \(H_{j0}\). Consequently, \(\mbox{FSR}^{\{2\}}\), as defined in \eqref{FSR-c}, is equivalent to the widely used FDR, i.e., the expected proportion of false rejections among all rejections.

\subsection{A brief review of conformal p-values}

The problem of one-class classification, also known as outlier detection or out-of-distribution testing in conformal inference, can be formulated within the framework of selective inference. Consider the observed data $\{(X_i, Y_i): i \in \mathcal{D}\}$ originating from two classes, $Y_i=1$ and $Y_i=2$. We divide the set $\mathcal{D}$ into two subsets, $\mathcal{D}^c=\{i: Y_i=c\}$, $c=1, 2$, where $\mathcal{D}^1$ and $\mathcal{D}^2$ represent the index sets of inliers and outliers, respectively.

In the context of one-class classification, the objective is to accurately identify outliers (individuals with label $Y=2$) in a set of unlabeled test data $\{X_j: j\in \mathcal{D}^{test}\}$, while maintaining strict control over the error rate. By considering individuals in class ``1'' as the null cases, we can formulate an equivalent multiple testing problem \eqref{outlier-detection}.

\begin{remark}\rm{
In the context of outlier detection, the standard practice is to only consider the labeled inliers $\mathcal{D}^1$ when computing conformal p-values. It is worth noting that recent research by \cite{Liaetal22} has revealed that this approach may result in potential efficiency loss. Nevertheless, for the sake of comparability, we adhere to the conventional practice and exclude $\mathcal{D}^2$ in our investigation.}
\end{remark}

The construction of split-conformal p-values \citep{Batetal21} involves partitioning $\mathcal{D}^1$ into two subsets: $\mathcal{D}^{train}$ for training a score function and $\mathcal{D}^{cal}$ for calibrating a significance index. Treating $\hat{S}^2(\cdot)$ as a conformity score function that indicates the likelihood of belonging to class $2$, the conformal p-value for testing $H_{j0}$ can be expressed using our notation as:
\beq\label{conform-p}
\hat{u}_j \equiv \hat{u}(X_j) = \frac{\sum_{i\in \mathcal{D}^{cal}} \mathbb{I}\{\hat{S}^2(X_i) \geq \hat{S}^2(X_j)\} + 1}{n^{cal} + 1}.
\eeq

\begin{remark}\rm{
To avoid confusion, note that in our framework a higher score indicates a higher likelihood of being an outlier. This is opposite to the convention in \cite{Batetal21}, where a lower score reflects stronger evidence. To align the definitions, we have replaced the expression ``\(S \leq t\)'' in the conformal p-value definition of \cite{Batetal21} with ``\(S \geq t\)'' in our formulation \eqref{conform-p}. This adjustment ensures that both formulations are equivalent.}
\end{remark}

\subsection{R-value is the BH q-value of conformal p-values (for outlier detection problems)}\label{subsec:RV-BH}

For the outlier detection problem \eqref{outlier-detection}, we consider using the thresholding rule \eqref{select-rule2} instead of the selection rule \eqref{R-rule} as we are only interested in selecting the high-risk class ($Y=2$). 

To see the connection of our R-value to the conformal p-value \eqref{conform-p}, recall the definition of Storey's q-value 
$$
\hat q^{ST}\left\{\hat u(t)\right\}={(1-\pi)\hat u(t)}/{G\{\hat u(t)\}}, 
$$
where $\pi$ is the proportion of non-null cases in $\mathcal D^{test}$ and $G(\cdot)$ is the cumulative distribution function (CDF) of the p-values. Now recall $m=|\mathcal D^{test}|$, let $\hat G(t)$ denote the empirical process of the scores  $\{\hat S_j^2: j\in \mathcal D^{test}\}$: 
\beq\label{hat-G}
\hat G(t)=\frac{1}{m}\sum_{j\in \mathcal D^{test}} \mathbb I\left\{\hat u(\hat S_j^2)\leq \hat u(t)\right\} = \frac{1}{m} \sum_{j\in \mathcal D^{test}}\mathbb I\left(\hat S_j^2\geq t\right),
\eeq 
where the last equality holds because, by \eqref{conform-p}, a larger score corresponds to a smaller conformal p-value. Next we consider a modification of Storey's q-value, referred to as the BH q-value, which ignores the $(1-\pi)$ term and substitutes $\hat G$ in place of $G$ in Storey's q-value:
\beq\label{q-BH}
\hat q^{BH}_{i}=\frac{\hat u(\hat S_{i}^2)}{\hat G(\hat S_{i}^2)} \quad \mbox{for } i \in \mathcal{D}^{cal}\cup\mathcal D^{test}.
\eeq

Combining \eqref{conform-p} -- \eqref{q-BH}, we have
\begin{eqnarray}\label{q-BH2}
\hat q^{BH}(t) & = & \frac{m}{n^{cal}+1}\cdot \frac{\sum_{i\in \mathcal D^{cal}} \mathbb I\left(\hat S_i^2\geq t \right)+1}{\sum_{j\in \mathcal D^{test}}\mathbb I\left(\hat S_j^2\geq t\right)} \nonumber \\ & =& \frac{m}{n^{cal}+1}\cdot \frac{\sum_{i\in \mathcal D^{cal}} \mathbb I\left(\hat S_i^2\geq t, Y_i=1\right)+1}{\sum_{j\in \mathcal D^{test}}\mathbb I\left(\hat S_j^2\geq t\right)}. 
\end{eqnarray}
The last equality holds because under the one-class classification setup, $\mathcal D^{cal}$ is a ``pure'' training set in which all observations are from the null class ``1''. Let $t$ take values in $\{S_k: k\in\mathcal D^{cal}\cup\mathcal D^{test}\}$ and denote the corresponding values $\{\hat q^{BH}_k: k\in\mathcal D^{cal}\cup\mathcal D^{test}\}$. 

We also need to apply a monotonicity adjustment to ensure that the q-value function is non-decreasing in the conformity score. Let
\begin{equation}\label{conf-bhqv} 
   q^{BH}_{j} = \min_{k\in\mathcal D^{cal}\cup\mathcal D^{test}: \hat S_{k}^2 \leq \hat S^2_{j}} \hat q^{BH}_{k}, \quad\text{for } j\in\mathcal D^{test}.
\end{equation}
Since there is no overlapping selection, the adjustment in \eqref{R-value-max} is unnecessary. This precisely recovers the R-value defined by \eqref{Q-kc} and \eqref{Q-value-mod} (excluding the sensitive attribute).

\subsection{Discussion}\label{subsec:RV-conformal}

We emphasize that the fundamental connection between the R-value and conformal q-values only holds under the one-class classification setup. The BH q-value \eqref{q-BH2} will be different from the R-value \eqref{FSP-est} under the binary classification setup that we have considered in this article. Specifically, the cardinalities of the calibration sets will be different under the two setups, and the equality \eqref{q-BH2} does not hold. Our R-value does not explicitly utilize conformal p-values under the binary classification setup.

The conformal p-value approach by \cite{Batetal21} remains applicable for selective inference in the binary classification setup, specifically for the selection of cases from class 2. Nevertheless, it is noteworthy that the conformal p-value method utilizes a smaller data set, as the data set $\mathcal D^2$ is discarded, in comparison to our R-value approach. Consequently, this may lead to suboptimal information utilization and a reduction in statistical power. In addition, it is worth noting that the FASI algorithm may not be well-suited for the outlier detection problem, as it presumes that the test data and calibration data are exchangeable, which is unlikely to hold in practical scenarios. Therefore, both the conformal p-value and FASI approaches would require modification to address the outlier detection problem with labeled outliers. Related issues have gone beyond the scope of this study and will be pursued in future research.

\begin{remark}\label{rem:conditional-cover}\rm{

The recent contributions such as \cite{Angelopoulos_25_book} and \cite{Gibbs_25} addressed group-conditional coverage without providing guarantees in selective settings. Specifically, conditional coverage methods -- regardless of whether they incorporate protected groups -- seek to ensure that prediction sets achieve the target coverage rate on average over the entire test dataset. In contrast, selective inference is concerned with controlling error rates within a data-dependent subset of observations. As noted in \cite{benjamini2005false} and \cite{Gazetal25}, selective inference poses significantly greater statistical challenges. This distinction becomes especially clear in binary classification, where prediction sets are limited to the forms \(\{1\}\), \(\{2\}\), or \(\{1,2\}\). Because the set \(\{1,2\}\) is uninformative, one might wish to restrict evaluation only to those observations assigned \(\{1\}\) or \(\{2\}\). However, narrowing the focus to such selected cases leads to an inflated miscoverage rate (exceeding the nominal level $\alpha$): all misclassification errors necessarily occur among the singleton sets, since any observation assigned \(\{1,2\}\) trivially contains the true label with probability one. This observation underscores the necessity of developing methods explicitly tailored to selective inference. }
\end{remark}

\setcounter{equation}{0}

\section{Proof of Theorem \ref{FSR:thm}}\label{proof-thm1:sec}

We begin by presenting the proof of part (a) of Theorem \ref{FSR:thm} in Section \ref{proof:thm2}, followed by the more involved proof of part (b) and its corresponding lemmas in Section \ref{proof:thm1b}. Since the non-asymptotic theory in Theorem \ref{FSR:thm}(b) provides only an upper bound on the FSR, we also offer an asymptotic analysis in Section \ref{subsec:convergence-tau} to demonstrate that, under the standard regularity conditions commonly used in multiple testing, the upper bound from Theorem \ref{FSR:thm}(b) converges to 0, thereby establishing the asymptotic validity of the stable version of FASI algorithm.

\subsection{Proof of part (a)}\label{proof:thm2}

\subsubsection{An equivalent expression of the FASI algorithm} 

Consider a class of decision rules that select subjects into class $c$ if the confidence scores exceed a threshold $t$. For the $a$th group, $a\in\mathcal A$, the estimated false discovery proportion (FSP), as a function of $t$, can be described as the following empirical process:  
\beq\label{Qc-t}
 \widehat{\mbox{FSP}}^{c}_a(t)= \frac{\big\{\sum_{i\in \mD^{cal}_a}\II(\hat S_i^c\geq t, Y_i\neq c)+1\big\}/(n_a+1)}{\big\{\sum_{j\in \mD^{test}_a} \II(\hat S_{j}^c\geq t) \vee 1 \big\}/m_a}.
\eeq 
Let $\mathbf S_a^c=\{\hat S_{i}^{c}: i\in\mathcal D^{cal}_a\cup\mathcal D^{test}_a\}$ for $a\in\mathcal A$.

Consider a selection procedure represented by the process given in \eqref{Qc-t}. We  aim to find the smallest threshold, denoted as $\tau_a^c$, for which the estimated FSP is less than $\alpha$:
\beq\label{st}
\tau_a^c=\left[\min\left\{t\in\mathbf S_a^c: \widehat{\mbox{FSP}}^{c}_a(t)\leq \alpha_c\right\}\right]\vee 0.5.
\eeq
The adjustment ``$(\cdot) \vee 0.5$'', which shares similar ideas to the adjustment in \eqref{R-value-max}, indicates that we never assign the \(j\)th individual to class \(c\) when the confidence score \(\hat S_j^c\leq 0.5\) [cf. the equivalent rule based on $\hat S_j^c$ given in Equation \eqref{FASI:alg2} below]; this also effectively avoids overlapping selections; see Remark \ref{rem:R-value} for related discussions. Note that while the thresholds for the R-values from different groups are identical at \(\alpha_c\), the thresholds for the scores \( \hat{S}_j^c \), denoted $\{\tau_a^c: a\in\mathcal A\}$, vary depending on $A_j=a$, the group membership of the \( j \)-th subject. 

The R-value defined in \eqref{R-value-max} of the main text can be written as:
$$
\tilde Q_{j}^{c}=\sum_{a\in\mathcal A} \mathbb I(A_j=a)\cdot \min_{\{t\in\mathbf S_a^c: t \leq \hat S_{j}^c\}} \widehat{\mbox{FSP}}^{c}_a(t), \quad R_j^{c}=\max\left\{ \II(\hat S_j^c\leq 0.5), \tilde Q_j^{c}\right\}, \quad \mbox{for $j\in \mathcal D^{test}$}.  
$$

The following lemma shows that the decision rule based on thresholding the R-value can be equivalently represented using a decision rule based on thresholding the scores $\{\hat S_j^c, j\in\mathcal D^{test}\}$.  

\begin{lemma}\label{lemma:score}
Consider $\tau_a^{c}$ defined in \eqref{st}. Then the following two rules are equivalent:  
\begin{equation}\label{FASI:alg2}
\delta_j=\II(R_{j}^{c}\leq \alpha_c) \quad \Longleftrightarrow \quad \delta_j^\prime=\sum_{a\in\mathcal A} \mathbb I(A_j=a)\cdot \II(\hat S_{j}^c\geq \tau_a^c),\quad j\in\mathcal D^{test}.
\end{equation}
\end{lemma}

The proof of this lemma is provided in Section \ref{proof:lem1}. It follows that 
$$
\hat Y_j=\sum_{c\in\{1,2\}}c\cdot\mathbb I(R_j^c\leq\alpha_c)=\sum_{c\in\{1,2\}}c\cdot \left\{\sum_{a\in\mathcal A} \mathbb I(A_j=a)\cdot \II(\hat S_{j}^c\geq \tau_a^c)\right\}. 
$$
We note that the thresholding rule for the R-value incorporates the sensitive attribute, whereas the thresholding rule based on confidence scores does not; see Lemma \ref{lem:score} in Section \ref{proof:lem1} for further discussions.

\subsubsection{Upper bounding the FSP process by martingales} 

We now describe the true FSP process of the FASI algorithm using the confidence scores, where the FSP process is outlined in \eqref{Qc-t} and the algorithm is given by the second representation in  \eqref{FASI:alg2}. Suppose our selection procedure chooses threshold $t$ in the test set $\mD^{test}_a$. Let 
\begin{eqnarray*}
V^{t}_a(t) & = & \sum_{j\in \mD^{test}_a} \II(\hat S_{j}^c\geq t, Y_j\neq c), \quad 
W^{t}_a(t) = \sum_{j\in \mD^{test}_a} \II(\hat S_{j}^c\geq t, Y_j=c) \quad \mbox{and}\\
K^{t}_a(t) & = & \sum_{j\in \mD^{test}_a} \II(\hat S_{j}^c\geq t) ~=~ V^{t}(t)+W^{t}(t)
\end{eqnarray*}
denote the counts of false selections, correct selections and total selections, respectively.  For the calibration set $\mD^{cal}_a$, define 
\begin{eqnarray*}
V^{c}_a(t) & = & \sum_{i\in \mD^{cal}_a} \II(\hat S_{i}^c\geq t, Y_i\neq c), \quad  
W^{c}_a(t) = \sum_{i\in \mD^{cal}_a} \II(\hat S_{i}^c\geq t, Y_i=c),\\
K^{c}_a(t) & = & \sum_{i\in \mD^{cal}_a} \II(\hat S_{i}^c\geq t)
~=~V^{c}_a(t)+W^{c}_a(t)
\end{eqnarray*}
as the corresponding counts of selections. Consider the data-driven thresholds $\tau_a^c$ defined in \eqref{st}. Then the group-wise FSPs of the proposed FASI algorithm, as defined by the second representation in \eqref{FASI:alg2}, can be computed as
$$
\mbox{FSP}^{\{c\}}_a(\tau_a^c)=\frac{V^{t}_a(\tau_a^c)}{K^{t}_a(\tau_a^c)\vee 1}, \quad \mbox{ for $a\in\mathcal A$\; and \; $c\in\{1,2\}$}.
$$
The operation of the FASI algorithm implies that 
\begin{eqnarray}
\mbox{FSP}^{\{c\}}_a(\tau_a^c)& = & \frac{V^{t}(\tau_a^c)}{V^{c}(\tau_a^c)+1}\cdot \frac{V^{c}(\tau_a^c)+1}{K^{t}(\tau_a^c)\vee 1} \nonumber
          \\ & = &  \widehat{\mbox{FSP}}^{c}_a(\tau_a^c)\cdot \frac{|\mD^{cal}_a|+1}{|\mD^{test}_a|}\cdot \frac{V^{t}(\tau_a^c)}{V^{c}(\tau_a^c)+1} \nonumber
          \\ & \leq & \alpha \cdot \frac{|\mD^{cal}_a|+1}{|\mD^{test}_a|} \cdot \frac{V^{t}(\tau_a^c)}{V^{c}(\tau_a^c)+1}, \label{ratio-mart}
\end{eqnarray}
where the last two steps utilize definitions \eqref{Qc-t} and \eqref{st}, respectively.

\subsubsection{Martingale arguments}\label{sec:mart-proof}

The ratio appearing in \eqref{ratio-mart} motivates us to consider the following process
\beq\label{Ratio1}
{V^{t}_a(t)}/\big\{V^{c}_a(t)+1\big\},
\eeq
which we show is a martingale. We start with the following continuous-time filtration:
\begin{eqnarray*}
\mathcal F_t^{a} & = & \sigma \{V^{t}_a(s), V^{c}_a(s), W^{t}_a(s), W^{c}_a(s): t_a^l\leq s\leq t \} \\
 & = & \sigma \{V^{t}_a(s), V^{c}_a(s), K^{t}_a(s), K^{c}_a(s): t_a^l\leq s\leq t \},
\end{eqnarray*}
where $t_a^l$ represents the lower limit of the threshold. That is, if $t_a^l$ is employed, then all subjects in group $a$ are classified into class $c$. 

In our proof, it is sufficient to consider a discrete-time filtration since FASI only selects thresholds from $\mathbf S_a^c$. Let $m^*_a=|\mathcal D^{cal}_a|+|\mathcal D^{test}_a|$ denote the total number of selections in both $\mathcal D^{cal}_a$ and $\mathcal D^{test}_a$ when the threshold is $t_a^l$. We consider a $\sigma$-field that contains all  information of the entire selection process. Specifically, let $\{s_k: k=m_a^*, \cdots, 1\}$ denote a sequence of thresholds (times), where $s_k$ is the threshold when exactly $k$ subjects, including those from both $\mathcal D^{cal}$ and $\mathcal D^{test}$, are selected into class $c$, and $k$ takes values in the order of $m^*_a, m^*_a-1, \cdots, 1$ (backward in time). This leads to the following discrete-time filtration: 
\begin{equation}\label{fil-k2}
\mathcal F_k^{a}=\sigma \left\{V^{t}_a(s_j), V^{c}_a(s_j), W^{t}_a(s_j), W^{c}_a(s_j): j=m^*_a, m^*_a-1, \ldots, k\right\}.
\end{equation}
We can see that $\mathcal F_k^{a}$ is a backward-running filtration as for $k_1<k_2$, $\mathcal F_{k_2}^{a}\subset \mathcal F_{k_1}^{a}$. Note that at time $s_k$, only one of the four following events is possible: 
{\small
\begin{eqnarray*}
A_1 & = & \{ V^{t}_a(s_{k-1})=V^{t}_a(s_k)-1, V^{c}_a(s_{k-1})=V^{c}_a(s_k), W^{t}_a(s_{k-1})= W^{t}_a(s_{k}), W^{c}_a(s_{k-1})= W^{c}_a(s_{k})\}, \\
A_2 & = & \{ V^{t}_a(s_{k-1})=V^t(s_k), V^{c}_a(s_{k-1})=V^{c}_a(s_k)-1, W^{t}_a(s_{k-1})= W^{t}_a(s_{k}), W^{c}_a(s_{k-1})= W^{c}_a(s_{k})\}, \\
A_3 & = & \{ V^{t}_a(s_{k-1})=V^{t}_a(s_k), V^{c}_a(s_{k-1})=V^{c}_a(s_k), W^{t}_a(s_{k-1})= W^{t}_a(s_{k})-1, W^{c}_a(s_{k-1})= W^{c}_a(s_{k})\},\\
A_4 & = & \{ V^{t}_a(s_{k-1})=V^{t}_a(s_k), V^{c}_a(s_{k-1})=V^{c}_a(s_k), W^{t}_a(s_{k-1})= W^{t}_a(s_{k}), W^{c}_a(s_{k-1})= W^{c}_a(s_{k})-1\}. 
\end{eqnarray*}
}

According to Assumption \ref{ex:as}, and the fact that FASI uses same fitted model to compute the scores, we have 
${\PP(A_1|\mathcal F_k^{a})}/{\PP(A_2|\mathcal F_k^{a})}={V^{t}_a(s_k)}/{V^{c}_a(s_k)}.
$
Moreover, we must have
$\PP(A_1|\mathcal F_k^{a})+\PP(A_2|\mathcal F_k^{a})+\PP(A_3|\mathcal F_k^{a})+\PP(A_4|\mathcal F_k^{a})=1. $
It follows that there exists a $\gamma_k$, such that
\begin{eqnarray*}
\PP(A_1|\mathcal F_k^{a})=\gamma_k\cdot \frac{V^{t}_a(s_k)}{V^{t}_a(s_k)+V^{c}_a(s_k)}, \quad
\PP(A_2|\mathcal F_k^{a})=\gamma_k\cdot \frac{V^{c}_a(s_k)}{V^{t}_a(s_k)+V^{c}_a(s_k)},
\end{eqnarray*}
and $\PP(A_3|\mathcal F_k^{a})+\PP(A_4|\mathcal F_k^{a})=1-\gamma_k$. It will soon become evident that the value of $\gamma_k$ does not matter in the theory, as it will be canceled out in the calculations.

To see why \eqref{Ratio1} is a martingale wrt $\mathcal F_k^{a}$, note that
\begin{eqnarray*}
\EE\left\{ \frac{V^{t}_a(s_{k-1})}{V^{c}_a(s_{k-1})+1} | \mathcal F_k^{a} \right\}
&  = &
\frac{V^{t}_a(s_k)-1}{V^{c}_a(s_k)+1}\cdot \gamma_k\cdot \frac{V^{t}_a(s_k)}{V^{t}_a(s_k)+V^{c}_a(s_k)}\\ && +
\frac{V^{t}_a(s_k)}{V^{c}_a{s_k}}\cdot\gamma_k\cdot \frac{V^{c}_a(s_k)}{V^{t}_a(s_k)+V^{c}_a(s_k)}
+\frac{V^{t}_a(s_k)}{V^{c}_a(s_k)+1}(1-\gamma_k)   
 \\ & = & \frac{V^{t}_a(s_{k})}{V^{c}_a(s_{k})+1}\cdot(\gamma_k+1-\gamma_k) ~=~ \frac{V^{t}_a(s_{k})}{V^{c}_a(s_{k})+1}.
\end{eqnarray*}

\subsubsection{FSR Control}
\label{fsr_control_pf1}

The threshold $\tau_a^c$ defined by \eqref{st} is a stopping time with respect to the filtration $\mathcal F_k^{a}$ since $\{\tau_a^c\leq s_k\}\in \mathcal F_k^{a}$. In other words, the event whether the $k$th selection occurs completely depends on the information prior to time $s_k$ (including $s_k$). 

Let $\mD^{test,0}_a$ and $\mD^{cal,0}_a$ be the index sets for subjects in  $\mD^{test}_a$ and $\mD^{cal}_a$ that do not belong to class $c$, respectively. In the final step of our proof, we shall apply the optional stopping theorem to the filtration $\{\mathcal F_k^{a}\}$. Recall that  $t_l^a$ is lower limit of the threshold, and $m^*_a$ is the total number of misclassifications in both $\mD^{cal}_a$ and $\mD^{test}_a$ when the threshold is $t_l^a$. The group-wise FSR is 
\begin{eqnarray}
\mbox{FSR}^{\{c\}}_a  & = & \EE \{\mbox{FSP}^{\{c\}}_a(\tau_a^c)\} \label{FSR-FSP}
\\ & \leq & \alpha \cdot \EE\left[\cdot \EE \left\{\frac{|\mD^{cal}_a|+1}{|\mD^{test}_a|}\frac{V^{t}_a(\tau_a^c)}{V^{c}_a(\tau_a^c)+1}|\mathcal F_{m^*_a}\right\}\right]\nonumber
\\ & = & \alpha \cdot \EE\left[\frac{|\mD^{cal}_a|+1}{|\mD^{test}_a|}\cdot \frac{V^{t}_a(t_l)}{V^{c}_a(t_l)+1}\right]\nonumber
\\ & = & \alpha \cdot \EE\left\{\frac{|\mD^{cal}_a|+1}{|\mD^{test}_a|}\cdot \frac{|\mD^{test,0}_a|}{|\mD^{cal,0}_a|+1}\right\}\label{qq2}
\\ & \leq & \gamma_{c,a}\alpha, \nonumber
\end{eqnarray}
To get Equation \eqref{qq2} we have used the fact that when $t_l^a$ is used then all subjects are classified to class $c$. This completes the proof.

\begin{remark}\rm{
We provide a remark to explain the $\EE$ operator in \eqref{FSR-FSP}. As indicated by \eqref{FASI:alg2}, the FASI algorithm is equivalent to a thresholding rule based on $\hat S_i^c$. The data-driven threshold (or stopping time), $\tau_a^c$, is a random variable that varies across different realizations or data sets. The FSP, denoted as $\mbox{FSP}_a^{\{c\}}(\tau_a^c)$, is a random variable that differs across data sets. The FSR, defined as the expectation of the FSP, integrates the randomness across the training, calibration and test data.}
\end{remark}

\subsubsection{Proof of Lemma~\ref{lemma:score}}\label{proof:lem1}

  First, it is easy to see that the two decision rules are equivalent (i.e. $\delta_j=\delta_j^\prime=0$) if $\hat S_j^c\leq 0.5$. We only consider the situation where $\hat S_j^c>0.5$. 
    
    Next, suppose that $\delta_j^\prime=1$ holds for some $j \in \mathcal{D}^{test}$. Without loss of generality, assume that $A_j=a$. It follows that $\II(\hat S_{j}^c>\tau_a^c)=1$, indicating that the stopping time $\tau_a^c$ must satisfy 
\begin{equation}\label{eqn:tau-ac}
    \tau_a^c\in \{t\in\mathbf S_a^c: t \leq \hat S_{j}^c\}.
\end{equation}     
We conclude that
$$
R^c_{j}=\tilde Q_j^{c}=\min_{\{t\in\mathbf S_a^c: t \leq \hat S_{j}^c\}} \widehat{\mbox{FSP}}^{c}_a(t) \leq \widehat{\mbox{FSP}}^{c}_a(\tau_c) \leq \alpha. 
$$
where the first two equalities are due to the definition of R-value and the monotonicity adjustment, whereas the first inequality follows from \eqref{eqn:tau-ac} and the second inequality follows from the definition of $\tau_a^c$. 
Hence $\II\{R^{c}_{j}\leq\alpha \}=1$, proving the first direction of the equivalence.  
    
    Conversely, suppose that $\II\{R^{c}_{j}\leq\alpha\}=1$. Without loss of generality, assume that $A_j=a$. By the definition of the R-value, we have 
   $$
  R^c_{j}=\tilde Q_j^{c}=\min_{\{t\in\mathbf S_a^c: t \leq \hat S_{j}^c\}}  \widehat{\mbox{FSP}}^{c}_a(t) \leq\alpha. 
   $$
   That is, there exists a threshold $t\leq \hat S_j^c$ in $\mathbf S_a^c$ such that $\widehat{\mbox{FSP}}^{c}_a(t) \leq\alpha$. It follows that
    $$
    \tau_a^c=\min_{t\in\mathbf S_a^c}\left\{t: \widehat{\mbox{FSP}}^{c}_a(t)\leq\alpha\right\}\leq \min_{t\in\mathbf S_a^c, t\leq \hat S_j^c}\left\{t: \widehat{\mbox{FSP}}^{c}_a(t)\leq\alpha\right\}\leq \hat S^c_{j},
    $$
    implying that $\delta_j^\prime=\sum_{\kappa\in\mathcal A} \mathbb I(A_j=\kappa)\cdot \II(\hat S_{j}^c\geq \tau_\kappa^c)=\mathbb I(A_j=a)\II(\hat S_{j}^c\geq \tau_a^c)=1$.

Combining the two arguments above, we have
$
\delta_j=1  \Longleftrightarrow  \delta_j^\prime=1.
$
We can similarly show that     
$\delta_j=0  \Longleftrightarrow  \delta_j^\prime=0$, establishing \eqref{FASI:alg2}.

\begin{remark}\label{lem:score}\rm{
In sensitive applications, there may be reservations about procedures that employ different thresholds based on membership in a protected group, even if the resulting fairness guarantees are equivalent. Therefore, it is crucial to communicate our fairness guarantee in a manner that avoids misinterpretation associated with the use of multiple thresholds. A key advantage of the FASI procedure based on the R-value [the first thresholding rule in \eqref{FASI:alg2}] is its simplicity in conveying fairness while remaining user-friendly in operation. By employing a single universal threshold, we provide a tool that is more interpretable from a fairness perspective. For practitioners, this approach is significantly easier to understand compared to managing multiple thresholds (when multiple thresholds are involved, it can become challenging to explain to the public how the resulting classification algorithm is fair).  }

\end{remark}

\subsection{Proof of part (b)}\label{proof:thm1b}

The proof is more complicated but follows essentially the same strategy of the proof for part (a).
Details are provided for new arguments and omitted for repeated arguments.

\subsubsection{Preliminaries and notations}

Consider the R-value \eqref{R-value-max} that utilizes both \(\mathcal{D}^{cal}_a \cup \mathcal{D}^{test}_a\) via \eqref{Q-kc-plus}. The estimated FSP in group $a$ for a given threshold $t$ is: 
\begin{equation}\label{FSP-new-est} 
 \widehat{\mbox{FSP}}^{c}_a(t)= \frac{\big\{\sum_{i\in \mD^{cal}_a}\II(\hat S_i^c\geq t, Y_i\neq c)+1\big\}/(n_a+1)} {\big\{\sum_{i\in \mD^{test}_a\cup \mD^{cal}_a} \II(\hat S_i^c\geq t)+1\big\}/(n_a+m_a+1)}.  
\end{equation}
Now employing the stable version of the FSP estimate \eqref{FSP-new-est}, define
$$
\tau_a^c=\big[\min \big\{t\in\mathbf S_a^c:  \widehat{\mbox{FSP}}^{c}_a(t)\leq \alpha\big\}\big]\vee 0.5,  
$$
Similar to the previous proof, Lemma \ref{lemma:score} indicates that the FASI algorithm is equivalent to the following thresholding rule based on confidence scores:
$$
\textstyle \hat Y_j=\sum_{a\in\mathcal A} \mathbb I(A_j=a) \left\{\sum_{c\in\{1,2\}} c\cdot\II(\hat S_{j}^c\geq \tau_a^c)\right\},  j\in\mathcal D^{test}.
$$

Consider the modified FSP definition in Theorem \ref{FSR:thm}: $\mbox{FSR}^{\{c\},*}_a=\mathbb E\left[\mbox{FSP}^{\{c\},*}_a(\tau_a^c)\right]$, where 
$$
\mbox{FSP}^{\{c\},*}_a(\tau_a^c)=\frac{\sum_{j \in \mathcal{D}^{test}_a} \mathbb I(\hat S_i^c\geq \tau_a^c, {Y}_{j}\neq c)}{\sum_{j \in \mathcal{D}^{test}_a} \mathbb I(\hat S_i^c\geq \tau_a^c)+1}. 
$$
It follows from the definition \eqref{FSP-new-est} and the operation of FASI algorithm that
\begin{eqnarray*}
\mbox{FSP}^{\{c\},*}_a(\tau_a^c) \leq  \alpha \cdot  \frac{n_a+1}{n_a+m_a+1}
\cdot \frac{\sum_{i\in \mD^{test}_a\cup \mD^{cal}_a} \II(\hat S_i^c\geq \tau_a^c)+1}{\sum_{j \in \mathcal{D}^{test}_a} \mathbb I(\hat S_i^c\geq \tau_a^c, {Y}_{j}\neq c)+1} \cdot \frac{\sum_{j \in \mathcal{D}^{test}_a} \mathbb I(\hat S_i^c\geq \tau_a^c, {Y}_{j}\neq c)}{\sum_{j \in \mathcal{D}^{test}_a} \mathbb I(\hat S_i^c\geq \tau_a^c)+1}.
\end{eqnarray*}
The product of the last two terms can be reorganized as
\begin{eqnarray}\label{key-inequ}
&& \left\{1+ \frac{\sum_{i\in \mD^{cal}_a} \II(\hat S_i^c\geq \tau_a)}{\sum_{j \in \mathcal{D}^{test}_a} \mathbb I(\hat S_i^c\geq \tau_a)+1}\right\} \cdot \frac{\sum_{j \in \mathcal{D}^{test}_a} \mathbb I(\hat S_i^c\geq \tau_a, {Y}_{j}\neq c)}{\sum_{j \in \mathcal{D}^{cal}_a} \mathbb I(\hat S_i^c\geq \tau_a, {Y}_{j}\neq c)+1} \nonumber
\\ & = & 
 \left\{1+ \frac{\sum_{i\in \mD^{cal}_a} \II(\hat S_i^c\geq \tau_a, Y_i=c)+\sum_{i\in \mD^{cal}_a} \II(\hat S_i^c\geq \tau_a, Y_i\neq c)}{\sum_{j \in \mathcal{D}^{test}_a} \mathbb I(\hat S_i^c\geq \tau_a, Y_i=c)+\sum_{j \in \mathcal{D}^{test}_a} \mathbb I(\hat S_i^c\geq \tau_a, Y_i\neq c)+1}\right\} \cdot M_1(\tau_a^c) \nonumber
\\ & \leq & M_1(\tau_a^c) + \max\left\{M_2(\tau_a^c), M_3(\tau_a^c)\right\},
\end{eqnarray}
where in the above equation, we have three martingales respectively defined as:
\begin{eqnarray*}
M_1(\tau_a^c) & = & \frac{\sum_{j \in \mathcal{D}^{test}_a} \mathbb I(\hat S_i^c\geq \tau_a^c, {Y}_{j}\neq c)}{\sum_{j \in \mathcal{D}^{cal}_a} \mathbb I(\hat S_i^c\geq \tau_a^c, {Y}_{j}\neq c)+1}; \\
M_2(\tau_a^c) & = & \frac{\sum_{i\in\mathcal D^{cal}_a}\mathbb I(\hat S_i^c\geq \tau_a^c, Y_i\neq c) }{ \sum_{i\in\mathcal D^{cal}_a}\mathbb I(\hat S_i^c\geq \tau_a^c, Y_i\neq c)+1};
\\
M_3(\tau_a^c) & = & \frac{\sum_{i\in\mathcal D^{cal}_a}\mathbb I(\hat S_i^c\geq \tau_a^c, Y_i= c)}{\sum_{i\in\mathcal D^{test}_a}\mathbb I(\hat S_i^c\geq \tau_a^c, Y_i= c)+1} \cdot \frac{\sum_{i\in\mathcal D^{test}_a}\mathbb I(\hat S_i^c\geq \tau_a^c, Y_i\neq c)}{\sum_{i\in\mathcal D^{cal}_a}\mathbb I(\hat S_i^c\geq \tau_a^c, Y_i\neq c)+1}.
\end{eqnarray*}
In the derivation of \eqref{key-inequ}, we have used the following inequality: 
$$
\frac{x+y}{z+(t+1)}\cdot \frac{z}{x+1} \leq \max\left\{\frac{x}{x+1}\; , \; \frac{yz}{(t+1)(x+1)}\right\}
$$
for any positive integers $x$, $y$, $z$, and $t$. The proof for this inequality is elementary and hence omitted. 

\subsubsection{The main proof}\label{subsec:main-proof}

Noting that $M_2(\tau_a^c)\leq 1$ holds trivially true, and utilizing the fact $\max(x, y)=\frac{(x+y)}2+\frac{|x-y|}{2},$ we can easily derive the following upper bound:
\begin{equation}\label{equ:FSP-star1}
\mathbb E\left[\mbox{FSP}^{\{c\},*}_a(\tau_a^c)\right] \leq  
 \mathbb E\left[\frac{\alpha(n_a+1)}{n_a+m_a+1}\cdot \left\{M_1(\tau_a^c)+ \frac{M_3(\tau_a^c)+1}2+\frac{|M_3(\tau_a^c)-1|}{2} \right\}\right]. 
\end{equation}

To characterize the FSP process, recall the counts in Section \ref{sec:mart-proof}: 
\begin{eqnarray*}
V^{t}_a(t) & = & \sum_{j\in \mD^{test}_a} \II(\hat S_{j}^c\geq t, Y_j\neq c), \; W^{t}_a(t) = \sum_{j\in \mD^{test}_a} \II(\hat S_{j}^c\geq t, Y_j=c), \; K^{t}_a(t) = \sum_{j\in \mD^{test}_a} \II(\hat S_{j}^c\geq t), \\
V^{c}_a(t) & = & \sum_{i\in \mD^{cal}_a} \II(\hat S_{i}^c\geq t, Y_i\neq c),\; W^{c}_a(t) = \sum_{i\in \mD^{cal}_a} \II(\hat S_{i}^c\geq t, Y_i=c), \; K^{c}_a(t) =\sum_{i\in \mD^{cal}_a} \II(\hat S_{i}^c\geq t),
\end{eqnarray*}
defined respectively for the test set and calibration set. Moreover, we employ the same discrete-time filtration as in Section \ref{sec:mart-proof}:
$$
\mathcal F_k^{a}=\sigma \left\{V^{t}_a(s_j), V^{c}_a(s_j), W^{t}_a(s_j), W^{c}_a(s_j): j=m^*_a, m^*_a-1, \ldots, k\right\}.
$$ 
Again, the threshold $\tau_a^c$ of the FASI algorithm with modified mirror process \eqref{FSP-new-est} is  a stopping time conditional on $\mathcal F_k^{a}$. 

In the next subsections (Sections \ref{subsec:M1-mart} and \ref{subsec:M3-mart}), we demonstrate that both $M_1(t)$ and $M_3(t)$ are super-martingales adapted to  $\mathcal F_k^{a}$. By applying the optional stopping theorem, we obtain that
\begin{equation}\label{equ:M1-M3}
\mathbb E\big\{M_1(\tau_a^c)\big\}  \leq  \frac{m_a^0}{n_a^0+1} \quad \mbox{and} \quad
\mathbb E\big\{M_3(\tau_a^c)\big\}  \leq  \frac{n_a^1m_a^0}{(m_a^1+1)(n_a^0+1)},
\end{equation}
where $m_a^0=\sum_{j\in\mathcal D^{test}_a} \mathbb I(Y_j\neq c)$, $m_a^1=\sum_{j\in\mathcal D^{test}_a} \mathbb I(Y_j= c)$, $n_a^0=\sum_{i\in\mathcal D^{cal}_a} \mathbb I(Y_j\neq c)$, and $n_a^1=\sum_{i\in\mathcal D^{cal}_a} \mathbb I(Y_j=c)$. As we have focused on selection individuals from class $c$, for simplicity, we have suppressed $c$ in the above notations. It follows from \eqref{equ:FSP-star1} and \eqref{equ:M1-M3} that
\begin{eqnarray*}
\mathbb E\left[\mbox{FSP}^{\{c\},*}_a(\tau_a^c)\right] & = &
\mathbb E\left[ \frac{\alpha(n_a+1)}{n_a+m_a+1} \cdot\left\{ \frac{m_a^0}{n_a^0+1}+ \frac 1 2+\frac{n_a^1\cdot m_a^0}{2(m_a^1+1)(n_a^0+1)}+\frac{|M_3(\tau_a^c)-1|}{2} \right\} \right]
\\&\leq& \alpha_c\gamma_{c,a}^\prime+\frac{\alpha_c}2 \mathbb E|M_3(\tau_a^c)-1|,
\end{eqnarray*}
proving the desired result. \qed
\begin{remark}\label{rem:conv}
\rm{
Under the exchangeability condition, some simple calculations (considering only the leading terms) show that
\begin{eqnarray}
\gamma_{c,a}^\prime & = & \mathbb E\left[\frac{(n_a+1)}{n_a+m_a+1}\cdot \left\{\frac{m_a^0}{n_a^0+1}+ \frac 1 2+\frac{n_a^1\cdot m_a^0}{2(m_a^1+1)(n_a^0+1)}\right\}\right]  \nonumber \\
 & = &\frac 1 2 \mathbb E\left\{\frac{n_a(m_a^0+n_a^0)}{n_a^0(m_a+n_a)}+\frac{n_am_a^0}{n_a^0m_a}\cdot\frac{m_a(m_a^1+n_a^1)}{m_a^1(m_a+n_a)}\right\}+o(1). \label{gamma-ca}
\end{eqnarray}
We can see that \(\gamma_{c,a}^\prime\) is very similar to \(\gamma_{c,a}\) (defined in Part (a) of the theorem), as both are essentially related to the empirical proportions. Moreover, under the exchangeability condition, both \(\gamma_{c,a}^\prime\) and \(\gamma_{c,a}\) are very close to 1.

Moreover, in Section \ref{subsec:convergence-tau}, we discuss sufficient conditions under which the data-driven threshold satisfies the almost sure convergence $\tau_a^c \xrightarrow{a.s.} \tau^*$ for some constant $\tau^* \in (0,1)$. In the asymptotic regime, we assume that $m_a^0 \asymp m_a^1 \asymp m_a$ and $n_a^0 \asymp n_a^1 \asymp n_a$, with all of these quantities diverging to infinity. Specifically, we establish the strong convergence of the data-driven threshold $\tau_a^c$. It follows that
\begin{eqnarray*}
\lim_{n_a, m_a\rightarrow\infty}\mathbb E|M_3(\tau_a^c)-1| = \left| \frac{\PP(S\geq\tau^*, Y=c)}{\PP(S\geq\tau^*, Y\neq c)}\cdot \frac{\PP(S\geq\tau^*, Y\neq c)}{\PP(S\geq\tau^*, Y=c)}-1 \right|=0. 
\end{eqnarray*}
Hence the stable version of the FASI algorithm controls the FSR asymptotically: 
$$
\mathbb E\left[\mbox{FSP}^{\{c\},*}_a(\tau_a^c)\right]\leq \alpha_c+o(1).
$$
}
\end{remark}

\subsubsection{Martingale arguments for $M_1(t)$}\label{subsec:M1-mart}

Consider the events $A_1$ to $A_4$ defined in Section \ref{sec:mart-proof}. The conditional probabilities of these events along the filtration $\mathcal F_k^a$ are given by 
\begin{eqnarray*}
\PP(A_1|\mathcal F_k^{a}) & = & \gamma_k\cdot \frac{V^{t}_a(s_k)}{V^{t}_a(s_k)+V^{c}_a(s_k)}, \quad
\PP(A_2|\mathcal F_k^{a})=\gamma_k\cdot \frac{V^{c}_a(s_k)}{V^{t}_a(s_k)+V^{c}_a(s_k)}, \\
\PP(A_3|\mathcal F_k^{a}) & = &(1-\gamma_k)\cdot \frac{W^{t}_a(s_k)}{W^{t}_a(s_k)+W^{c}_a(s_k)}, \quad 
\PP(A_4|\mathcal F_k^{a})=(1-\gamma_k)\cdot \frac{W^{c}_a(s_k)}{W^{t}_a(s_k)+W^{c}_a(s_k)}.
\end{eqnarray*}
It is easy to verify that $M_1(t)=\frac{V^t_a(t)}{V^c_a(t)+1}$ is a backward-running martingale by noting that:
\begin{eqnarray*}
&& \EE\left\{M_1(s_{k-1})|\mathcal F_k^{a} \right\} \\
& = & 
\frac{V^t_a(s_k)-1}{V^c_a(s_k)+1}\cdot \gamma_k\cdot \frac{V^{t}_a(s_k)}{V^{t}_a(s_k)+V^{c}_a(s_k)}
+
\frac{V^t_a(s_k)}{V^c_a(s_k)}\cdot \gamma_k\cdot \frac{V^{c}_a(s_k)}{V^{t}_a(s_k)+V^{c}_a(s_k)}
+
\frac{V^t_a(s_k)}{V^c_a(s_k)+1}(1-\gamma_k)
\\ & = & \frac{V^t_a(s_k)}{V^c_a(s_k)+1}~=~M_1(s_k). \qed
\end{eqnarray*}

\subsubsection{Martingale arguments for $M_3(t)$}\label{subsec:M3-mart}

We consider the same events, same probabilities and same filtration as before. Write
$$
M_3(t)=\frac{W^c_a(t)}{W^t_a(t)+1}\cdot\frac{V^t_a(t)}{V^c_a(t)+1}. 
$$ 
To see why $M_3(t)$ is a martingale, note that
\begin{eqnarray*}
&& \EE\left\{M_3(s_{k-1})|\mathcal F_k^{a} \right\} \\
& = & 
\frac{W^c_a(s_k)}{W^t_a(s_k)+1}\cdot \frac{V^t_a(s_k)-1}{V^c_a(s_k)+1}
\cdot\gamma_k\cdot \frac{V^{t}_a(s_k)}{V^{t}_a(s_k)+V^{c}_a(s_k)}
+
\frac{W^c_a(s_k)}{W^t_a(s_k)+1}\cdot \frac{V^t_a(s_k)}{V^c_a(s_k)}
\cdot\gamma_k\cdot \frac{V^{c}_a(s_k)}{V^{t}_a(s_k)+V^{c}_a(s_k)}
\\ & & +
\frac{W^c_a(s_k)}{W^t_a(s_k)}\cdot \frac{V^t_a(s_k)}{V^c_a(s_k)+1}
\cdot(1-\gamma_k)\cdot \frac{W^{t}_a(s_k)}{W^{t}_a(s_k)+W^{c}_a(s_k)}
\\ && +
\frac{W^c_a(s_k)-1}{W^t_a(s_k)+1}\cdot \frac{V^t_a(s_k)}{V^c_a(s_k)+1}
\cdot(1-\gamma_k)\cdot \frac{W^{c}_a(s_k)}{W^{t}_a(s_k)+W^{c}_a(s_k)}.
\end{eqnarray*} 
Some simple calculations yield
$$
\EE\left\{M_3(s_{k-1})|\mathcal F_k^{a} \right\} = \frac{W^c_a(s_k)}{W^t_a(s_k)+1}\cdot\frac{V^t_a(s_k)}{V^c_a(s_k)+1}~=~M_3(s_k). \qed
$$

\subsection{Asymptotic analysis of upper bounds}\label{subsec:convergence-tau} 
 
To rigorously establish the asymptotic validity of the stable version of FASI, we conduct an asymptotic analysis of the residual term from Theorem \ref{FSR:thm}(b). This analysis precisely shows that 
\[
\lim_{(n, m) \rightarrow \infty} \mathbb{E}\left|\texttt{Res}(\tau) - 1\right| = 0,
\]
thereby corroborating our numerical results and formally confirming our intuition that FASI effectively controls the FSR. In contrast to the finite-sample theory presented in the main text, this component of our theory utilizes classical limit theorems, which require standard regularity conditions commonly employed in statistics. It is important to emphasize that these conditions are necessitated by the available theoretical tools and do not imply that the FASI method is inherently dependent on them.

\subsubsection{Preliminaries and assumptions}\label{subsec:assumptions}

Since our focus is on group-wise FSR control and given that the theoretical analysis can be applied to each group to establish the properties of the FASI algorithm, we restrict our analysis to a particular group. The asymptotic regime assumes that $m_a^0 \asymp m_a^1 \asymp m_a$ and $n_a^0 \asymp n_a^1 \asymp n_a$, with all of these quantities diverging to infinity. In what follows, we omit the notation \(a\) for group membership in order to reduce notational complexity. 

Suppose our objective is to select cases with \(Y_j=2\). To simplify the discussion, we assume that the scores are defined as \(S_i \coloneqq 1 - \hat{S}^{c=2}(X_i, A_i) \in (0, 1)\). Consequently, we select individuals when their scores are small. This slightly modified notation system aligns more closely with the framework of multiple testing while utilizing the same decision rule as before.

Denote the estimated FSP as
$$
\widehat{\mbox{FSP}}(t)= \frac{\big\{\sum_{i\in \mD^{cal}}\II(S_i<t, Y_i\neq c)+1\big\}/(n+1)} {\big\{\sum_{i\in \mD^{test}\cup \mD^{cal}} \II(S_i<t)+1\big\}/(n+m+1)}.   
$$
Then the data-driven threshold of the FASI algorithm is given by
\beq\label{def:tau}
\tau=\big\{t\in[0,1]:  \widehat{\mbox{FSP}}(t)\leq\alpha\big\}.
\eeq 

\begin{remark}\label{iid-assumption}\rm{
Our analysis focuses on the scenario where \( S_i \) are continuous random variables, such as softmax score outputs from machine learning algorithms.  The data points \((S_i, Y_i)\) are assumed to be i.i.d., which is reasonable because: (i) the score functions are trained using an independent data set, and (ii) the De Finetti Theorem can be employed to identify a latent variable, enabling analysis conditional on that latent variable. This assumption leads to the random mixture model defined in \eqref{rmix1}, as considered in Section \ref{thR:sec}. }
\end{remark}

Omitting the group membership \( a \) notation, the CDF of the scores (within a specific group) is given by:
\beq\label{two-group} 
F(t) = \PP(S_i < t, Y = 1) + \PP(S_i < t, Y = 2) = \pi_1 F_1(t) + \pi_2 F_2(t), \quad i \in \mathcal{D}^{\text{cal}} \cup \mathcal{D}^{\text{test}},
\eeq 
where \(\pi_c = \PP(Y_i = c)\) for \(c \in \{1, 2\}\). We assume that the conditional CDFs \(F_1(t)\) and \(F_2(t)\) are strictly greater than 0 on the interval \((0,1)\).

Define the marginal false selection rate as \(\mbox{mFSR}(t) = {\pi_1 F_1(t)}/{F(t)}\). The oracle threshold \(\tau^*\) is defined as:
\beq\label{def:tau-star}
\tau^* = \sup\big\{t \in (0,1) : \mbox{mFSR}(t) \leq \alpha\big\}.
\eeq
The marginal FSR is analogous to the marginal false discovery rate (mFDR) in multiple testing \citep{GenWas02, SunCai07}. 

We introduce the following assumption, which plays a key role in our asymptotic analysis to ensure the existence and convergence of the data-driven threshold. This critical assumption was also utilized in \cite{JinCan23}, albeit in a slightly different context. 

\begin{assumption}\label{as:MLRC}
Let \( t \in [0,1] \) be a threshold for the scores \( S_j \). Denote \(\alpha\) as the nominal FSR level, and \(\mbox{mFSR}(t)\) as the marginal FSR of the thresholding rule \(\{I(S_i < t) : i \in [m]\}\). For any $\varepsilon>0$, we have
\begin{eqnarray}\label{cons:1}
\mbox{ $\exists~~\tau^*-\varepsilon<\tau^\prime<\tau^*$, such that $\mbox{mFSR}(\tau^\prime)<\alpha$.} 
\end{eqnarray} 
\end{assumption}

\begin{remark}\rm{
Assumption \ref{as:MLRC} can be derived from the Monotone Likelihood Ratio Condition (MLRC, \citealp{SunCai07}). Under this condition, it can be demonstrated that \(\mbox{mFSR}(t)\) is monotonically increasing in \(t\) \citep{Caoetal13}. For i.i.d. scores \(S_i\) following the random mixture model \eqref{two-group}, the MLRC naturally implies \eqref{cons:1}. The MLRC is a common (and often implicit) condition in the FDR literature. For instance, it simplifies to the assumption of concavity of the p-value CDF (cf. \citealp{GenWas02, Sto02}). Therefore, while Assumption \ref{as:MLRC} might appear complex, it is actually less restrictive than many standard assumptions commonly used in the FDR literature.
}
\end{remark}

\subsubsection{Key lemmas}

We first state three lemmas. The first two, which are similar to the large-sample theories in \cite{Stoetal04}, are consequences of the well-known Glivenko-Cantelli Theorem, and are therefore stated without proofs. 

Again, we omit the group notation \( a \); let \( n = |\mathcal{D}^{\text{cal}}| \) and \( m = |\mathcal{D}^{\text{test}}| \).

\begin{lemma}\label{lem:CDF-conv} 
Consider the random mixture model \eqref{two-group}. The empirical CDFs obey the following strong uniform convergence:
\begin{eqnarray*}
\sup_{t\in(0,1)}\left| \frac{1+\sum_{i\in\mathcal D^{cal}}\mathbb I(S_i<t)}{n+1}-F(t) \right|  \xrightarrow{a.s.}  0; \quad \sup_{t\in(0,1)}\left| \frac{1+\sum_{j\in\mathcal D^{test}}\mathbb I(S_j<t)}{m+1}-F(t) \right|  \xrightarrow{a.s.} 0.
\end{eqnarray*}
\end{lemma}

\begin{lemma}\label{lem:cond-CDF-conv}
Consider the random mixture model \eqref{two-group}. The conditional empirical CDFs obey the following strong uniform convergence: for $c\in\{1, 2\}$,
\begin{eqnarray*}
\sup_{t\in(0,1)}\left| \frac{1+\sum_{i\in\mathcal D^{cal}}\mathbb I(S_i<t, Y_i=c)}{n+1}-\pi_cF_c(t) \right| & \xrightarrow{a.s.} & 0; \\
\sup_{t\in(0,1)}\left| \frac{1+\sum_{i\in\mathcal D^{test}}\mathbb I(S_j<t, Y_j=c)}{m+1}-\pi_cF_c(t) \right| & \xrightarrow{a.s.} & 0. 
\end{eqnarray*}
\end{lemma}

The final lemma, which is integral to establishing the strong convergence of the data-driven threshold, follows from standard $\epsilon$-$N$ arguments; we provide a proof in Section \ref{subsec:proof-lemma4} for completeness.

\begin{lemma}\label{lem:convergence}
Let $\{f_n\}_{n\geq 1}: [0,1]\rightarrow (0,\infty)$ be a sequence of functions and $f: [0,1]\rightarrow (0,\infty)$ be another function. For a given constant $\alpha$, define 
$$
\tau_n=\sup\{t\in[0,1]: f_n(t)\leq\alpha\}\quad \mbox{and}\quad \tau^*=\sup\{t\in[0,1]: f(t)\leq\alpha\}.
$$
Assume the following conditions hold: 
\begin{description}
\item (i) For any $\epsilon>0$, there exists some $t\in[\tau^*-\epsilon,\tau^*)$ such that $f(t)<\alpha$; 
\item (ii) There exists a $\delta\in(0,\tau^*)$ satisfying $f(\delta)<\alpha$ such that 
$\sup_{t\in[\delta,1)}|f_n(t)-f(t)|\xrightarrow{a.s.} 0$. 
\end{description}
Then we have $\tau_n\xrightarrow{a.s.} \tau^*$. 

\end{lemma}

\subsubsection{Uniform convergence of the threshold}\label{subsec:uni-conv}

Now we establish the strong convergence of $\tau$ as mentioned in Remark \ref{rem:conv} in Appendix \ref{subsec:main-proof}. The next proposition establishes the strong convergence of the data-driven threshold.
\begin{proposition}\label{prop:as-conv}
Let $\tau$ [defined by \eqref{def:tau}] and $\tau^*$ [defined by \eqref{def:tau-star}] denote the data-driven and oracle thresholds, respectively. Under Model \eqref{two-group} and Assumption \ref{as:MLRC}, we have $\tau\xrightarrow{a.s.} \tau^*$.
\end{proposition}

\noindent\textbf{Proof.} To prove the desired result, we apply Lemma \ref{lem:convergence} by substituting $f(t)$ and $f_n(t)$ with $\mbox{mFSR}(t)$ and $\widehat{\mbox{\rm FSP}}(t)$, respectively. Our primary task is to ensure that both Conditions (i) and (ii) of the lemma are met. Condition (i) is satisfied as a result of Condition \eqref{cons:1} from Assumption \ref{as:MLRC}, combined with the equation $\mbox{mFSR}(\tau^*) = \alpha$. Further, according to Lemmas \ref{lem:CDF-conv} and \ref{lem:cond-CDF-conv}, we can therefore find a $\kappa > 0$ such that the subsequent uniform strong convergence holds:
\[
\sup_{t \in [\kappa, 1)} \left| \widehat{\mbox{\rm FSP}}(t) - \mbox{\rm mFSR}(t) \right| \xrightarrow{a.s.} 0.
\]
Thus, Condition (ii) is fulfilled. Recognizing the equal roles of \(\tau\) and \(\tau^*\) in both Lemma \ref{lem:convergence} and our proposition, by applying Lemma \ref{lem:convergence} we conclude that \(\tau \xrightarrow{a.s.} \tau^*\). \qed

\subsubsection{Asymptotic FSR control}

To demonstrate that \(\lim_{(n, m) \rightarrow \infty} \mathbb{E}\left|\texttt{Res}(\tau) - 1\right| = 0\), it suffices to prove the following lemma.

\begin{lemma}
Consider Model \eqref{two-group}. Suppose \(\tau \xrightarrow{a.s.} \tau^* \in (0,1)\). Then, for \(c \in \{1,2\}\), we have
\begin{eqnarray}\label{equ:strong-conv1}
\PP\left\{\lim_{n,m\rightarrow\infty}\frac{1+\sum_{i\in\mathcal D^{cal}}\mathbb I(S_i\leq\tau, Y_i=c)}{n+1}=\pi_cF_c(\tau^*) \right\} & = & 1; \\ \label{equ:strong-conv2}
\PP\left\{\lim_{n,m\rightarrow\infty}\frac{1+\sum_{j\in\mathcal D^{test}}\mathbb I(S_j\leq\tau, Y_j=c)}{m+1}=\pi_cF_c(\tau^*) \right\} & = & 1.
\end{eqnarray}
\end{lemma}

\noindent\textbf{Proof.} We  only prove the first equality in \eqref{equ:strong-conv1}, as the second can be established in a similar manner. The proof involves two simple decompositions. The first decomposition is
\begin{eqnarray*}
&& \frac{1+\sum_{i\in\mathcal D^{cal}}\mathbb I(S_i\leq\tau, Y_i=c)}{n+1}-\pi_cF_c(\tau^*) 
\\ & = & 
\left[\frac{1+\sum_{i\in\mathcal D^{cal}}\mathbb I(S_i\leq\tau, Y_i=c)}{n+1}-\frac{1+\sum_{i\in\mathcal D^{cal}}\mathbb I(S_i\leq\tau^*, Y_i=c)}{n+1}\right]\\ 
&&+\left[\lim_{n,m\rightarrow\infty}\frac{1+\sum_{i\in\mathcal D^{cal}}\mathbb I(S_i\leq\tau^*, Y_i=c)}{n+1}-\pi_cF_c(\tau^*)\right]
\\ & = & \mbox{I}+ \mbox{II}. 
\end{eqnarray*}

Term II converges to 0 almost surely due to the uniform strong convergence of the CDFs. To address Term I, we employ a second decomposition:
\begin{eqnarray}
&& \frac{1+\sum_{i\in\mathcal D^{cal}}\mathbb I(S_i\leq\tau, Y_i=c)}{n+1}-\frac{1+\sum_{i\in\mathcal D^{cal}}\mathbb I(S_i\leq\tau^*, Y_i=c)}{n+1} \nonumber
\\ & = & 
\frac{1+\sum_{i\in\mathcal D^{cal}}\mathbb I(\tau^*<S_i\leq\tau, Y_i=c)}{n+1}-\frac{1+\sum_{i\in\mathcal D^{cal}}\mathbb I(\tau<S_i\leq\tau^*, Y_i=c)}{n+1}. \label{equ:conv3}
\end{eqnarray}
We can see that \eqref{equ:conv3} converges to 0 almost surely if \(S_i\) is continuous and \(\tau \xrightarrow{a.s.} \tau^*\). This completes the proof of the lemma, thereby establishing asymptotic FSR control. \qed

\subsubsection{Proof of Lemma \ref{lem:convergence}}\label{subsec:proof-lemma4}

The lemma can be proven by combining the results from two directions.  

\noindent\textbf{Direction 1 (lower bound).} We show that $\liminf_{n\to\infty}\tau_n\geq\tau^*$ almost surely. Let $\epsilon>0$ be an arbitrarily small constant. Condition (i) implies that there exists $t_\epsilon\in[\tau^*-\epsilon, \tau^*)$ such that $f(t_\epsilon)<\alpha$. Since $t_\epsilon\geq \delta$ for sufficiently small $\epsilon$, the uniform convergence in Condition (ii) applies on $[\delta,1)$. Hence we can find $N_1$ such that for all $n\geq N_1$, $|f_n(t_\epsilon)-f(t_\epsilon)|<\alpha-f(t_\epsilon)$. This implies 
$$
f_n(t_\epsilon)<f(t_\epsilon)+\alpha-f(t_\epsilon)=\alpha. 
$$
Since $f_n(t_\epsilon)<\alpha$, we have
$\tau_n\geq t_\epsilon\geq \tau^*-\epsilon$ for all $n\geq N_1$. As $\epsilon>0$ is arbitrary, we conclude that $\liminf_{n\to\infty}\tau_n\geq\tau^*$ almost surely. 

\noindent\textbf{Direction 2 (upper bound).} We now show the more complicated direction:
\[
\mbox{$\limsup_{n\to\infty}\tau_n\leq\tau^*$ almost surely.}
\]
We argue by contradiction. If the upper bound does not hold, then we can find a subsequence \(\{n^\prime\}\) and \(\epsilon>0\) such that \(\tau_{n^\prime}\geq\tau^*+\epsilon\) holds with positive probability. Consider an arbitrary \(t_0\in[\tau^*,\tau^*+\epsilon)\). By the definition of \(\tau^*\), we must have \(f(t_0)>\alpha\). Note that if we choose \(\epsilon>0\) sufficiently small, then \(t_0\geq \delta\), so we can apply the uniform convergence in Condition (ii). Specifically, we can find \(N_2\) such that for all \(n^\prime\geq N_2\),
$
|f_{n^\prime}(t_0)-f(t_0)|<\frac{f(t_0)-\alpha}{2}.
$
Hence,
\[
f_{n^\prime}(t_0)>f(t_0)-\frac{f(t_0)-\alpha}{2}>\alpha.
\]
This implies that for the subsequence \(\{n^\prime\}\), we must have \(\tau_{n^\prime} < t_0\) (strict inequality), as otherwise it would contradict the definition of \(\tau_{n^\prime}\). However, since we have chosen \(t_0 \in [\tau^*, \tau^* + \epsilon)\) and \(\tau_{n^\prime} \ge \tau^* + \epsilon\) holds with positive probability, we obtain a contradiction. Hence such a subsequence cannot exist, and we conclude that  
$
\limsup_{n \to \infty} \tau_n \le \tau^*
$  
almost surely.
\qed

\setcounter{equation}{0}

\section{Asymptotic Guarantees for Overall FSR Control}\label{sec:group-overall-FSR}

This section develops the asymptotic theory underpinning the FASI algorithm introduced in Remark \ref{rem:asymptotic-theory}. To simplify notation, all discussion here focuses on the selection of cases into a pre-specified class \(c\), subject to a group-wise FSR bound of \(\alpha\). 

We establish that, under slightly stronger yet standard conformal assumptions, FASI guarantees asymptotic overall FSR control at level $\alpha$ -- complementing its finite-sample group-wise guarantees (Theorem \ref{FSR:thm}). These theoretical findings are consistently supported by our numerical experiments. A key practical implication is that, asymptotically, group-wise FSR control is stricter than overall control, making an additional overall FSR constraint unnecessary.

The theoretical analysis in this section assumes that the pairs of scores $S_j^c$ are continuous, and \((S^c_j, Y_j)\) are i.i.d. and follow the mixture model \eqref{model:rmix-yns}, rewritten here for convenience:
\begin{equation}\label{model:rmix-yns2}
G^c(s)~=~\sum_{a\in\mathcal A} \II(A_j=a)\cdot G^c_a(s) ~=~\sum_{a\in\mathcal A}\II(A_j=a)\cdot\left\{ \pi_{1|a} G_{1|a}^c(s) + \pi_{2|a} G_{2|a}^c(s) \right\},
\end{equation}
where \(G_{c^\prime|a}^c(s)\) denotes the conditional cumulative distribution function of \(S^c\) given \(A=a\) and \(Y=c^\prime\), and \(\pi_{c^\prime|a} = \PP(Y_i = c \mid A_i = a)\) are the corresponding conditional probabilities for \(c^\prime = 1, 2\). Let $\pi_a=\PP(A_j=a)$ denotes the expected fraction of individuals belonging to group $a$. 

The i.i.d. assumption, detailed in Remark \ref{iid-assumption}, is adopted here only to simplify the theoretical exposition. In practice, our method requires only the weaker exchangeability condition and remains model-free -- no distributional specification of \(G^c_a(s)\) is needed for implementation.

\begin{proposition}\label{prop:OFSR-theory}
Consider $(Y_i, S_i)$ from the mixture model \eqref{model:rmix-yns2}. Consider the FASI algorithm with R-value defined via \eqref{Q-kc}, \eqref{Q-value-mod} and \eqref{R-value-max} and the stable version of FASI with R-value defined via \eqref{Q-kc-plus}-\eqref{R-value-max}. Then under Model \eqref{model:rmix-yns2} and Assumption \ref{as:MLRC} in Section \ref{subsec:assumptions}, we have the overall FSR level [defined in Eq~\eqref{FSR-c}] is given by $\alpha+o(1)$. 
\end{proposition}

\subsection{Proof of Proposition \ref{prop:OFSR-theory}}

We prove the result for the more complex stable version of FASI, which uses the R-value defined in equations \eqref{Q-kc-plus}–\eqref{R-value-max}. The analogous result for the simpler version of FASI follows by a similar argument. In Theorem \ref{FSR:thm}(b), we have shown that 
$$
\mbox{FSR}_a^{\{c,*\}}\leq \gamma_{c,a}^\prime \alpha_c+\frac{\alpha}2\mathbb E\left| {\texttt{RES}}(\tau_a^c)-1 \right|.
$$

Consider $\tau_a^c$ and $\tau_a^{c,*}$ defined in \eqref{def:tau} and \eqref{def:tau-star}, respectively. Proposition \ref{prop:as-conv} in Section \ref{subsec:uni-conv}
implies that $\tau_a^c\xrightarrow{a.s.} \tau_a^{c,*}$. Invoking \eqref{equ:strong-conv1} and \eqref{equ:strong-conv2}, it follows that 
\begin{equation}\label{conv:res}
\lim_{n,m\rightarrow\infty}\EE\left| {\texttt{RES}}(\tau_a^c)-1 \right|=0.
\end{equation}
From Eq~\eqref{gamma-ca}, it is easy to show that 
$
\lim_{n,m\rightarrow\infty} \gamma_{c,a}=1. 
$
Combining with \eqref{conv:res}, we have
\begin{equation}\label{control:FSR-ac}
\mbox{FSR}_a^{\{c\}}\leq\alpha_c+o(1). 
\end{equation}

Next, we present a lemma to establish the asymptotic equivalence between the marginal false selection rate (mFSR) and the FSR. For clarity, we first recall the definition of the mFSR in the context of the FASI algorithm: 
\begin{eqnarray*}
\mbox{mFSR}^{\{c\}}_a & = & \frac{\mathbb E\left\{\sum_{j \in \mathcal{D}^{test}_a} \mathbb I(\hat S_j^c\geq\tau_a, {Y}_{j}\neq c)\right\}}{\mathbb E\left\{\sum_{j \in \mathcal{D}^{test}_a} \mathbb I(\hat S_j^c\geq\tau_a)\right\}}, \quad \mbox{for $a\in\mathcal A$}, \\
\mbox{mFSR}^{\{c\}} & = & \frac{\mathbb E\left\{\sum_{a\in\mathcal A}\sum_{j \in \mathcal{D}^{test}_a} \mathbb I(\hat S_j^c\geq\tau_a, {Y}_{j}\neq c)\right\}}{\mathbb E\left\{\sum_{a\in\mathcal A}\sum_{j \in \mathcal{D}^{test}_a} \mathbb I(\hat S_j^c\geq\tau_a)\right\}}
\end{eqnarray*}
The term \(\text{mFSR}^{\{c\}}\) (without group indicator \(a\)) represents the overall FSR. It is obtained by applying FASI group-wise, pooling all rejections, and calculating the ratio.

\begin{lemma}\label{lem:mFSR}
Under Model \eqref{two-group} and Assumption \ref{as:MLRC}, we have
\begin{equation}
\mbox{mFSR}^{\{c\}}_a=\mbox{FSR}^{\{c\}}_a+o(1)\; \mbox{for $a\in\mathcal A$, and }\; \mbox{mFSR}^{\{c\}}=\mbox{FSR}^{\{c\}}+o(1).
\end{equation}
\end{lemma}
\emph{Proof sketch for Lemma \ref{lem:mFSR}.} The lemma can be established by arguments analogous to those used in the proof of Lemma 7 in \cite{Caietal19}; we therefore omit the details. We only note that the two conditions required in Lemma 7 of \cite{Caietal19} are readily verified under Model \eqref{two-group} and Assumption \ref{as:MLRC}. Specifically, the convergence of the thresholds to fixed limits \(\tau^*\) and \(\tau^*_a\) ensures the lower‑bound condition, while the i.i.d. assumption and an application of the Cauchy–Schwarz inequality—as in their original proof—suffice to complete the verification.  \qed

Now we continue our proof of the proposition. Consider oracle thresholds $\tau^{*}$ and $\tau_a^{*}$. Define
\begin{eqnarray*}
\mbox{mFSR}^{\{c\}}_a (\tau_a^{*})& = & \frac{\mathbb E\left\{\sum_{j \in \mathcal{D}^{test}_a} \mathbb I(\hat S_j^c\geq\tau_a^*, {Y}_{j}\neq c)\right\}}{\mathbb E\left\{\sum_{j \in \mathcal{D}^{test}_a} \mathbb I(\hat S_j^c\geq\tau_a^*)\right\}}, \quad \mbox{for $a\in\mathcal A$}, \\
\mbox{mFSR}^{\{c\}}(\tau^{*}_a; a\in\mathcal A) & = & \frac{\mathbb E\left\{\sum_{a\in\mathcal A}\sum_{j \in \mathcal{D}^{test}_a} \mathbb I(\hat S_j^c\geq\tau_a^*, {Y}_{j}\neq c)\right\}}{\mathbb E\left\{\sum_{a\in\mathcal A}\sum_{j \in \mathcal{D}^{test}_a} \mathbb I(\hat S_j^c\geq\tau_a^*)\right\}}.
\end{eqnarray*}

Invoking the lemma and note that $\tau_a\xrightarrow{a.s.} \tau_a^{*}$,  we have
\begin{eqnarray}\label{FSR-mFSR2}
\lim_{n,m\rightarrow\infty}\mbox{FSR}^{\{c\}}_a(\tau_a)=\lim_{n,m\rightarrow\infty}\mbox{mFSR}^{\{c\}}_a(\tau_a)
=\lim_{n,m\rightarrow\infty}\mbox{mFSR}^{\{c\}}_a(\tau^{*}_a)=\mbox{mFSR}^{\{c\}}_a(\tau^{*}_a).
\end{eqnarray}
It is important to note that the last equality does not involve a limit, as the quantity is constant under the random mixture model. Concretely, for group-specific mFSRs with oracle threshold $\tau_a^{*}$, some calculations reveal that
$$
\mbox{mFSR}^{\{c\}}_a(\tau^{*}_a)=\frac{G_a^c(\tau_a^*)-\pi_{c|a}G_{c|a}(\tau_a^*)}{G_a^c(\tau_a^*)}.
$$
According to \eqref{control:FSR-ac}, $\mbox{mFSR}^{\{c\}}_a(\tau^{*}_a)=\lim_{n,m\rightarrow\infty}\mbox{FSR}^{\{c\}}_a(\tau^c_a)\leq\alpha$. It follows from \eqref{FSR-mFSR2} that 
\begin{equation}\label{ineq1}
G_a^c(\tau_a^*)-\pi_{c|a}G_{c|a}(\tau_a^*)\leq \alpha G_a^c(\tau_a^*), \;\mbox{for $a\in\mathcal A$.} 
\end{equation}
This indicates that the overall mFSR can be controlled asymptotically:
\begin{eqnarray}
\mbox{mFSR}^{{c}}(\tau_a^*; a\in\mathcal A) & = & 
\frac{\sum_{a\in\mathcal A}\left\{ G_a^c(\tau_a^*)-\pi_{c|a}G_{c|a}(\tau_a^*)\right\} }{ \sum_{a\in\mathcal A} G_a^c(\tau_a^*) }\nonumber \\
& \leq & \frac{\alpha\cdot \{\sum_{a\in\mathcal A} G_a^c(\tau_a^*)\}}{\sum_{a\in\mathcal A} G_a^c(\tau_a^*)}=\alpha,\label{inequality2}
\end{eqnarray} 
where the last inequality is due to \eqref{ineq1}. Finally, we note that
$$
\lim_{n,m\rightarrow\infty}\mbox{FSR}^{\{c\}}(\tau^a; a\in\mathcal A)=\lim_{n,m\rightarrow\infty}\mbox{FSR}^{\{c\}}(\tau^{*}_a; a\in\mathcal A)
=\mbox{mFSR}^{\{c\}}(\tau^{*}_a; a\in\mathcal A)\leq\alpha.
$$
We conclude that the FASI algorithm controls the overall FSR at level $\alpha+o(1)$. \qed

\begin{remark}\rm{
The proof relies on the asymptotic equivalence between the mFSR and the FSR. The mFSR formulation offers a key analytical advantage: within this framework, group-wise mFSR control directly implies overall mFSR control via inequality \eqref{inequality2}. This implication, however, does not extend to the original FSR, which helps explain why deriving finite-sample guarantees for overall FSR control remains challenging.}
\end{remark}

\setcounter{equation}{0}

\section{Proof of Theorem \ref{optimality:thm}}\label{proof-thm3:sec}

The theorem implies that the optimal confidence score for constructing R-values should be $S_{OR}^c(x,a)=\PP(Y=c|X=x, A=a)$. A similar optimality theory has been developed in the context of multiple testing with groups \citep{CaiSun09}. However, the proof for the binary classification setup with the indecision option is much more complicated; we provide the proof here for completeness. We first establish an essential monotonicity property in Section \ref{monotone:sec}, then prove the optimality theory in Section \ref{proof-thm2-part2:sec}. 

\subsection{A monotonicity property}\label{monotone:sec}

The oracle rule employs $\{S_{OR}^{c,j}: c\in\{1,2\}, j \in \mathcal{D}^{test}\}$ as the confidence scores. The corresponding oracle theoretical R-values $\{\texttt{TR}_{OR}^{c,j}: c\in\{1,2\}, j\in\mathcal D^{test}\}$ can be obtained via the conversion algorithm in Appendix \ref{convert:sec} [cf. Equation \ref{TR-value}]. Let 
$$
\pmb t=\big\{t_a^c: t_a^c\in[0.5, 1], c\in\{1,2\}, a\in\mathcal A\big\}
$$ 
be a collection of eligible thresholds.

Consider a class of thresholding rules of the form:
$$
d_{OR}^j(\pmb t)=\sum_{c\in\{1,2\}}c\cdot\left\{ \sum_{a\in\mathcal A}\II(A_j=a) \II(S_{OR}^{c,j}>t_a^c)\right\}, \quad j\in\mathcal D^{test}.
$$
Denote the mFSR level in group $a$ for selecting class $c$ as $Q^c_{OR, a}(t_a^c)$. The next proposition characterizes the relationship between $Q^c_{OR,a}(t_a^c)$ and $t_a^c$. 
\begin{proposition}\label{monotone:prop}
$Q_{OR,a}^c(t_a^c)$ is monotonically decreasing in $t_a^c$.  
\end{proposition}

\noindent\textbf{Proof of Proposition \ref{monotone:prop}}.  Define $\tilde Q_{OR,a}^c(t_a^c)=1-Q_{OR,a}^c(t_a^c)$. We only need to show that $Q_{OR,a}^c(t_a^c)$ is monotonically increasing in $t_a^c$. Let $\mathcal M_a=\{j \in \mathcal{D}^{test}: A_j=a\}$. According to the definition of the mFSR and the definition of $S_{OR,j}^c$, we have
\beq\label{FSR-identity}
\EE \left\{ \sum_{j\in\mathcal M_a}\left\{ S_{OR,j}^c -\tilde Q_a^c(t_a^c) \right\} \II(S_{OR,j}^c>t_a^c) \right\}=0,
\eeq
where the expectation is taken over $\mathcal D_a^{test}$. It is important to note that the oracle procedure, which assumes that all distributional information is known, does not utilize $\mathcal D^{train}$ and $\mathcal D^{cal}$. It is easy to see from Equation \eqref{FSR-identity} that $\tilde Q_a^c(t)>t_a^c$ otherwise the summation on the LHS must be positive, leading to a contradiction.

Next we show that $t_1<t_2$ implies $\tilde Q_{OR, a}^c(t_1) \leq \tilde Q_{OR,a}^c(t_2)$. We argue by contradiction. Assume instead that $\tilde Q_{OR,a}^c(t_1)> \tilde Q_{OR,a}^c(t_2)$, then we have
\begin{eqnarray*}
 && \sum_{j\in\mathcal M_a}\{S_{OR}^{c,j}-\tilde Q_{OR,a}^c(t_1)\}\II(S_{OR}^{c,j}>t_1) \\ & = & \sum_{j\in\mathcal M_a} \{S_{OR}^{c,j}-\tilde Q_{OR,a}^c(t_2)+\tilde Q_{OR,a}^c(t_2)-\tilde Q_{OR,a}^c(t_1)\}\II(S_{OR}^{c,j}>t_1)\\ 
&=& \sum_{j\in\mathcal M_a} \{S_{OR}^{c,j}-\tilde Q_{OR,a}^c(t_2)\}\II(S_{OR}^{c,j}>t_2) + \sum_{j\in\mathcal M_a} \{S_{OR}^{c,j}-\tilde Q_{OR,a}^c(t_2)\}\II(t_1\leq S_{OR}^{c,j}\leq t_2)\\ && +\sum_{j\in\mathcal M_a} \left\{\tilde Q_{OR,a}^c(t_2)-\tilde Q_{OR,a}^c(t_1)\right\}\II(S_{OR}^{c,j}>t_1)  =  I+II+III. 
\end{eqnarray*}
Taking expectations on both sides, it is easy to see that the LHS is zero. However, the RHS is strictly greater than zero. For term I, we have $\EE(I)=0$ according to the definition of mFSR. For term II, we have $\EE(II)<0$ as we always have $\tilde Q_{OR,a}^c(t)>t$. For term III, we have $\EE(III)<0$ since we assume $\tilde Q_{OR,a}^c(t_1)> \tilde Q_{OR,a}^c(t_2)$. It follows that the assumption $\tilde Q_{OR,a}^c(t_1)> \tilde Q_{OR,a}^c(t_2)$ cannot be true, and the proposition is proved.  \qed

\begin{remark}\rm{
The proposition is essential for expressing the oracle procedure as a thresholding rule based on $S_{OR}^{c,j}$. Specifically, denote $Q_{OR,a}^{c,-1}(\cdot)$ the inverse of $Q_{OR,a}^c(\cdot)$. The monotonicity of $Q_{OR,a}^c(t)$ and the definition of the theoretical R-value together imply that for all $S_{OR}^{c,j}>0.5$, we have for $j\in\mathcal D^{test}$,
$
S_{OR}^{c,j}=\sum_{a\in\mathcal A}\II(A_j=a)\cdot Q_{OR,a}^{c,-1}(\texttt{TR}_{OR}^{c,j}).
$
For notational convenience, let $T_{j}=\PP(Y_{j}=2|X_{j}=x, A_{j}=a)$. Then $S_{OR}^{1,j}=1-T_{j}$ and $S_{OR}^{2,j}=T_{j}$. Let $t_{OR}^{c,a}=\max\left\{0.5, (Q_{OR}^{c,a})^{-1}(\alpha_c)\right\}$. The oracle rule can be written as, for $j \in \mathcal{D}^{test}$, 
\begin{eqnarray*}
\delta_{OR}^{j} (X_j, A_j)& = & \textstyle\sum_{c\in\{0,1\}}c\cdot \II(\texttt{TR}_{OR}^{c,j}\leq\alpha_c) \\
& = & \textstyle\sum_{c\in\{1,2\}}c\cdot\left\{\sum_{a\in\mathcal A}\II(A_j=a) \II \left( S_{OR}^{c,j}\geq t_{OR}^{c,a}\right)\right\}
\\ & = & \textstyle\sum_{a\in\mathcal A}\II(A_j=a)\left\{\mathbb I\left(T_{j}\leq 1-t_{OR}^{c,a}\right)+2\mathbb I \left(T_{j}\geq t_{OR}^{c,a}\right)\right\}.
\end{eqnarray*}
}
\end{remark}

\subsection{Proof of the theorem}\label{proof-thm2-part2:sec}
Define the expected number of true selections $\mbox{ETS}=\sum_{j \in \mathcal{D}^{test}} \mathbb I(Y_{j}=c, \hat Y_{j}=c)$. Then it can be shown that minimizing the EPI subject to the FSR constraint is equivalent to maximizing the ETS subject to the same constraint. 

According to Proposition \ref{monotone:prop}, the oracle rule can be written as 
$$
\delta_{OR}^{j}~\coloneqq~\delta_{OR}^{j} (X_j, A_j)=\textstyle\sum_{a\in\mathcal A}\II(A_j=a)\left\{\mathbb I\left(T_{j}\leq 1-t_{OR}^{c,a}\right)+2\mathbb I \left(T_{j}\geq t_{OR}^{c,a}\right)\right\}.
$$
The group-wise mFSR constraints for the oracle rule imply that, for all $a\in\mathcal A$:
\beq\label{FSR-or-prop} 
  \EE \left\{ \sum_{j\in\mathcal M_a} (T_j-\alpha_1)\II(\delta_{OR}^j=1)\right\} =  0, \; 
  \EE \left\{\sum_{j\in\mathcal M_a} (1-T_j-\alpha_2)\II(\delta_{OR}^j=2)\right\}  =  0.  
\eeq
Let $\pmb\delta \in\{0, 1, 2\}^m$ be a general selection rule in $\mathcal D_{\alpha_1, \alpha_2}$. Then the mFSR constraints for $\pmb\delta$ implies that, for all $a\in\mathcal A$,
\beq\label{FSR-delta-prop}
  \EE \left\{ \sum_{j\in\mathcal M_a} (T_j-\alpha_1)\II(\delta_j=1)\right\}  \leq  0, \;
  \EE \left\{\sum_{j\in\mathcal M_a} (1-T_j-\alpha_2)\II(\delta_j=2)\right\} \leq  0. 
\eeq
The ETS of $\pmb\delta=\{\delta_j: j\in \mathcal{D}^{test}\}$ is given by
\begin{eqnarray*}
\mbox{ETS}_{\pmb\delta}=\EE \left[\sum_{a\in\mathcal A}\sum_{j\in\mathcal M_a}\left\{\II(\delta_j=1)(1-T_j)+\II(\delta_j=2)T_j \right\}\right] = \sum_{a\in\mathcal A}(\mbox{ETS}_{\pmb\delta}^{1,a}+\mbox{ETS}_{\pmb\delta}^{2,a}).
\end{eqnarray*}

The goal is to show that $\mbox{ETS}(\pmb\delta^{OR})\geq \mbox{ETS}(\pmb\delta)$.  We only need to show 
 $
 \mbox{ETS}_{\pmb\delta^{OR}}^{c,a}\geq  \mbox{ETS}_{\pmb\delta}^{c,a}
 $
 for all $c$ and $a$. We will show $
 \mbox{ETS}_{\pmb\delta^{OR}}^{1,a}\geq  \mbox{ETS}_{\pmb\delta}^{1,a}
 $
 for a given $a$. The remaining inequalities follow similar arguments. According to \eqref{FSR-or-prop} and \eqref{FSR-delta-prop}, we have
\beq\label{ee-ineq1}
 \EE \left[ \sum_{j\in\mathcal M_a} (T_j-\alpha_1)\left\{\II(\delta_{OR}^j=1)-\II(\delta_j=1) \right\}\right] \geq  0.
\eeq
Let $\lambda_{1,a}=(1-t_{OR}^{c,a}-\alpha_1)/t_{OR}^{c,a}$. It can be shown that $\lambda_{1,a}>0$. For $i\in\mathcal M_a$, we claim that the oracle rule can be equivalently written as
$$
\delta_{OR}^j=\II\left\{\frac{T_j-\alpha_1}{1-T_j}<\lambda_{1,a}\right\}.
$$  
Using the previous expression and techniques similar to the Neyman-Pearson lemma, we claim that the following result holds for all $j\in\mathcal M_a$:
$$
\left\{\II(\delta_{OR}^j=1)-\II(\delta_j=1) \right\} \left\{ T_j-\alpha_1-\lambda_{1,a}(1-T_j)\right\}\leq 0.
$$
It follows that 
\beq\label{ee-ineq2}
\EE \left[ \sum_{j\in\mathcal M_a}\left\{\II(\delta_{OR}^j=1)-\II(\delta_j=1) \right\}   \left\{ T_j-\alpha_1-\lambda_{1,a}(1-T_j)\right\}\right]\leq 0.
\eeq
According to \eqref{ee-ineq1} and \eqref{ee-ineq2}, we have
$$
\lambda_{1,a} \EE \sum_{j\in\mathcal M_a} (1-T_j)\left\{\II(\delta_{OR}^j=1)-\II(\delta_j=1) \right\}=\lambda_{1,a}\left(\mbox{ETS}_{\pmb\delta^{OR}}^{1,a}-\mbox{ETS}_{\pmb\delta}^{1,a}\right) \geq 0.
$$
Note that $\lambda_{1,a}>0$, the desired result follows. The theorem is proved by combining the results from all groups $a\in\mathcal A$.

\setcounter{equation}{0}

\section{Related Fairness Algorithms}\label{appendix:related_algorithms}

We discuss two closely related works developed based on the sufficiency principle, in order to emphasize the advantages of FASI. 

\cite{zeng2022fair} presents a group-wise thresholding rule that maximizes the classifier's power subject to the constraints imposed by the sufficiency principle. However, this method does not allow for indecisions, thereby rendering it impossible to control the error rate at user-specified levels. 
In contrast, \cite{Lee_21} proposes a selective classification procedure that satisfies the sufficiency principle and allows for indecisions. This enables fair decision-making with error rate control. However, the approach by \cite{Lee_21} relies on complex fitting algorithms and imposes stringent assumptions for theoretical development, which lacks reliable theoretical guarantees regarding output reliability in practical scenarios. Moreover, both methods fail to address the issue of inflated decision errors that arise when classifying multiple individuals simultaneously.

We emphasize that the choice of fairness definition should be contextual and informed by the specific automated decision-making scenario. FASI offers several advantages that make it a more practical choice for practitioners. Firstly, in high-stakes scenarios, the proposed selective inference framework with an indecision option effectively handles situations where the consequences of incorrect decisions are significant. This approach provides practitioners with guidance on which observations require further attention, rather than automatically making decisions when the accuracy may not be sufficient. Secondly, when multiple individuals need to be classified simultaneously, it is crucial to employ a suitable  error criterion that can aggregate cumulative errors and control for multiplicity. FASI addresses this concern by providing the FSR, which generalizes the powerful and practical FDR criterion in large-scale testing problems. Lastly, in scenarios where complex or blackbox machine learning models are utilized, having a model-free algorithm like FASI becomes essential. This allows for the deployment of user-specified blackbox models while simultaneously ensuring provable validity in controlling the associated risks without imposing strong model assumptions.

\begin{table}[ht]
\caption{Comparison of algorithms developed to fulfill the sufficiency principle.}
\label{table:compare_algorithms}
\centering
\scalebox{0.7}{
\begin{tabular}{c||c|c|c}
\textbf{} &  \textbf{\begin{tabular}[c]{@{}c@{}}User Specified Error Rate \end{tabular}} & \textbf{Finite Sample Theory} & \textbf{\begin{tabular}[c]{@{}c@{}} No Assumptions on Model Accuracy\end{tabular}} \\
\hline\hline
\textbf{\begin{tabular}[c]{@{}c@{}}Zeng et al. (2022): \\ FairBayes-DPP\end{tabular}} &  No & No & No \\
\hline
\textbf{\begin{tabular}[c]{@{}c@{}} Lee et al. (2021): \\ Fair Selective Classification \\ Via Sufficiency  \end{tabular}} &  No & No & No \\
\hline
\textbf{FASI} &  Yes & Yes & Yes \\
\end{tabular}
}
\end{table}

\section{The setup of multinomial classification}\label{subsec:R-multiclass}

Let $\mathcal C^\prime\subset \mathcal C$ represent the set of classes to be selected. With indecisions being allowed, the action space is $\Lambda=\{0, \mathcal C^\prime\}$. We denote the selection rule for $m$ individuals in the test set as $\{\hat Y_{j}: j\in \mathcal{D}^{test}\}\in \Lambda^m$. 

The FSR can be defined in two ways with respect to the subset $\mathcal C^\prime$. The first definition evaluates the fraction of incorrect selections for each individual class separately:
$$
\mbox{FSR}^{\{c\}}_a=\mathbb E\left[\frac{\sum_{j \in \mathcal{D}^{test}} \mathbb I(\hat Y_{j}=c, {Y}_{j}\neq c, A_{j}=a)}{\left\{ \sum_{j \in \mathcal{D}^{test}} \mathbb I(\hat Y_{j}=c, A_{j}=a)\right\}\vee 1}  \right], \quad  \mbox{for all $a\in \mathcal A$ and $c\in \mathcal C^\prime$.}
$$
By contrast, the second definition calculates an overall error rate by combining selections from all classes in $\mathcal C^\prime$: 
$$
\mbox{FSR}^{\mathcal C^\prime}_a=\mathbb E\Bigg[ \frac{\sum_{j \in \mathcal{D}^{test}} \II(\hat Y_{j}\in \mathcal C', \hat{Y}_{j} \neq Y_{j}, A_j=a) }{ \big\{ \sum_{j \in \mathcal{D}^{test}} \mathbb I(\hat Y_{j}\in \mathcal C', A_j=a)\big\} \vee 1 }  \Bigg], \quad  \mbox{for all $a\in \mathcal A$.}
$$

The second definition of FSR introduces several complicated issues. Firstly, it requires the employment of a new score function to achieve optimality under the oracle setting. Secondly, substantial adjustments must be made to the mirror process described in Section \ref{why-fasi:sec}. Thirdly, the development of martingale theories becomes notably more intricate. Finally, when dealing with scenarios involving more than two classes, an additional layer of complexity arises. These various issues offer intriguing and crucial avenues for future exploration and research.

\section{Additional Numerical Results}\label{RvRp:sec}

\subsection{The plot of the census income data analysis}\label{subsec:income-plot}

We present the supplementary figure for the the census income data analysis. 

\begin{figure}[ht]
    \scalebox{0.9}{
    \centering
    \begin{subfigure}{0.75\textwidth}
        \centering
        \input{Figures/adult_plts}
    \end{subfigure}
    \begin{subfigure}{0.24\textwidth}
        \centering
\begin{tikzpicture}[x=1pt,y=1pt]
\definecolor{fillColor}{RGB}{255,255,255}
\path[use as bounding box,fill=fillColor,fill opacity=0.00] (0,0) rectangle (158.99,144.54);
\begin{scope}
\path[clip] (  0.00,  0.00) rectangle (158.99,144.54);
\definecolor{drawColor}{RGB}{255,255,255}
\definecolor{fillColor}{RGB}{255,255,255}

\path[draw=drawColor,line width= 0.4pt,line join=round,line cap=round,fill=fillColor] (  0.00,  0.00) rectangle (158.99,144.54);
\end{scope}
\begin{scope}
\path[clip] ( 21.92, 19.95) rectangle (155.49,131.17);
\definecolor{fillColor}{RGB}{255,255,255}

\path[fill=fillColor] ( 21.92, 19.95) rectangle (155.49,131.17);
\definecolor{drawColor}{RGB}{0,158,115}

\path[draw=drawColor,line width= 1.1pt,line join=round] ( 27.99,125.21) --
	( 34.38,104.18) --
	( 40.77, 89.73) --
	( 47.17, 79.40) --
	( 53.56, 72.24) --
	( 59.95, 67.66) --
	( 66.34, 62.31) --
	( 72.73, 58.77) --
	( 79.12, 54.92) --
	( 85.51, 51.08) --
	( 91.90, 48.14) --
	( 98.29, 45.12) --
	(104.68, 42.62) --
	(111.08, 40.08) --
	(117.47, 37.52) --
	(123.86, 35.06) --
	(130.25, 32.76) --
	(136.64, 30.84) --
	(143.03, 28.97) --
	(149.42, 26.97);

\path[draw=drawColor,line width= 1.1pt,dash pattern=on 1pt off 3pt on 4pt off 3pt ,line join=round] ( 27.99,122.48) --
	( 34.38,101.73) --
	( 40.77, 87.72) --
	( 47.17, 77.43) --
	( 53.56, 70.26) --
	( 59.95, 65.65) --
	( 66.34, 60.33) --
	( 72.73, 56.52) --
	( 79.12, 52.52) --
	( 85.51, 48.64) --
	( 91.90, 45.92) --
	( 98.29, 42.85) --
	(104.68, 40.12) --
	(111.08, 37.51) --
	(117.47, 34.59) --
	(123.86, 31.96) --
	(130.25, 29.09) --
	(136.64, 26.50);
\definecolor{fillColor}{RGB}{0,158,115}

\path[fill=fillColor] ( 27.99,125.21) circle (  1.96);

\path[fill=fillColor] ( 34.38,104.18) circle (  1.96);

\path[fill=fillColor] ( 40.77, 89.73) circle (  1.96);

\path[fill=fillColor] ( 47.17, 79.40) circle (  1.96);

\path[fill=fillColor] ( 53.56, 72.24) circle (  1.96);

\path[fill=fillColor] ( 59.95, 67.66) circle (  1.96);

\path[fill=fillColor] ( 66.34, 62.31) circle (  1.96);

\path[fill=fillColor] ( 72.73, 58.77) circle (  1.96);

\path[fill=fillColor] ( 79.12, 54.92) circle (  1.96);

\path[fill=fillColor] ( 85.51, 51.08) circle (  1.96);

\path[fill=fillColor] ( 91.90, 48.14) circle (  1.96);

\path[fill=fillColor] ( 98.29, 45.12) circle (  1.96);

\path[fill=fillColor] (104.68, 42.62) circle (  1.96);

\path[fill=fillColor] (111.08, 40.08) circle (  1.96);

\path[fill=fillColor] (117.47, 37.52) circle (  1.96);

\path[fill=fillColor] (123.86, 35.06) circle (  1.96);

\path[fill=fillColor] (130.25, 32.76) circle (  1.96);

\path[fill=fillColor] (136.64, 30.84) circle (  1.96);

\path[fill=fillColor] (143.03, 28.97) circle (  1.96);

\path[fill=fillColor] (149.42, 26.97) circle (  1.96);

\path[fill=fillColor] ( 27.99,122.48) circle (  1.96);

\path[fill=fillColor] ( 34.38,101.73) circle (  1.96);

\path[fill=fillColor] ( 40.77, 87.72) circle (  1.96);

\path[fill=fillColor] ( 47.17, 77.43) circle (  1.96);

\path[fill=fillColor] ( 53.56, 70.26) circle (  1.96);

\path[fill=fillColor] ( 59.95, 65.65) circle (  1.96);

\path[fill=fillColor] ( 66.34, 60.33) circle (  1.96);

\path[fill=fillColor] ( 72.73, 56.52) circle (  1.96);

\path[fill=fillColor] ( 79.12, 52.52) circle (  1.96);

\path[fill=fillColor] ( 85.51, 48.64) circle (  1.96);

\path[fill=fillColor] ( 91.90, 45.92) circle (  1.96);

\path[fill=fillColor] ( 98.29, 42.85) circle (  1.96);

\path[fill=fillColor] (104.68, 40.12) circle (  1.96);

\path[fill=fillColor] (111.08, 37.51) circle (  1.96);

\path[fill=fillColor] (117.47, 34.59) circle (  1.96);

\path[fill=fillColor] (123.86, 31.96) circle (  1.96);

\path[fill=fillColor] (130.25, 29.09) circle (  1.96);

\path[fill=fillColor] (136.64, 26.50) circle (  1.96);
\end{scope}
\begin{scope}
\path[clip] (  0.00,  0.00) rectangle (158.99,144.54);
\definecolor{drawColor}{RGB}{0,0,0}

\path[draw=drawColor,line width= 0.4pt,line join=round] ( 21.92, 19.95) --
	( 21.92,131.17);
\end{scope}
\begin{scope}
\path[clip] (  0.00,  0.00) rectangle (158.99,144.54);
\definecolor{drawColor}{gray}{0.30}

\node[text=drawColor,anchor=base east,inner sep=0pt, outer sep=0pt, scale=  0.56] at ( 18.77, 30.62) {0.2};

\node[text=drawColor,anchor=base east,inner sep=0pt, outer sep=0pt, scale=  0.56] at ( 18.77, 60.80) {0.4};

\node[text=drawColor,anchor=base east,inner sep=0pt, outer sep=0pt, scale=  0.56] at ( 19.77, 90.99) {0.6};

\node[text=drawColor,anchor=base east,inner sep=0pt, outer sep=0pt, scale=  0.56] at ( 18.77,121.17) {0.8};
\end{scope}
\begin{scope}
\path[clip] (  0.00,  0.00) rectangle (158.99,144.54);
\definecolor{drawColor}{gray}{0.20}

\path[draw=drawColor,line width= 0.4pt,line join=round] ( 20.17, 32.55) --
	( 21.92, 32.55);

\path[draw=drawColor,line width= 0.4pt,line join=round] ( 20.17, 62.73) --
	( 21.92, 62.73);

\path[draw=drawColor,line width= 0.4pt,line join=round] ( 20.17, 92.92) --
	( 21.92, 92.92);

\path[draw=drawColor,line width= 0.4pt,line join=round] ( 20.17,123.10) --
	( 21.92,123.10);
\end{scope}
\begin{scope}
\path[clip] (  0.00,  0.00) rectangle (158.99,144.54);
\definecolor{drawColor}{RGB}{0,0,0}

\path[draw=drawColor,line width= 0.4pt,line join=round] ( 21.92, 19.95) --
	(155.49, 19.95);
\end{scope}
\begin{scope}
\path[clip] (  0.00,  0.00) rectangle (158.99,144.54);
\definecolor{drawColor}{gray}{0.20}

\path[draw=drawColor,line width= 0.4pt,line join=round] ( 53.56, 18.20) --
	( 53.56, 19.95);

\path[draw=drawColor,line width= 0.4pt,line join=round] ( 85.51, 18.20) --
	( 85.51, 19.95);

\path[draw=drawColor,line width= 0.4pt,line join=round] (117.47, 18.20) --
	(117.47, 19.95);

\path[draw=drawColor,line width= 0.4pt,line join=round] (149.42, 18.20) --
	(149.42, 19.95);
\end{scope}
\begin{scope}
\path[clip] (  0.00,  0.00) rectangle (158.99,144.54);
\definecolor{drawColor}{gray}{0.30}

\node[text=drawColor,anchor=base,inner sep=0pt, outer sep=0pt, scale=  0.56] at ( 53.56, 12.94) {0.025};

\node[text=drawColor,anchor=base,inner sep=0pt, outer sep=0pt, scale=  0.56] at ( 85.51, 12.94) {0.050};

\node[text=drawColor,anchor=base,inner sep=0pt, outer sep=0pt, scale=  0.56] at (117.47, 12.94) {0.075};

\node[text=drawColor,anchor=base,inner sep=0pt, outer sep=0pt, scale=  0.56] at (149.42, 12.94) {0.100};
\end{scope}
\begin{scope}
\path[clip] (  0.00,  0.00) rectangle (158.99,144.54);
\definecolor{drawColor}{RGB}{0,0,0}

\node[text=drawColor,anchor=base,inner sep=0pt, outer sep=0pt, scale=  0.70] at ( 88.71,  5.05) {$\alpha$};
\end{scope}
\begin{scope}
\path[clip] (  0.00,  0.00) rectangle (158.99,144.54);
\definecolor{drawColor}{RGB}{0,0,0}

\node[text=drawColor,rotate= 90.00,anchor=base,inner sep=0pt, outer sep=0pt, scale=  0.70] at (  9.32, 75.56) {$\widehat{EPI}$};
\end{scope}
\begin{scope}
\path[clip] (  0.00,  0.00) rectangle (158.99,144.54);
\definecolor{drawColor}{RGB}{0,0,0}

\node[text=drawColor,anchor=base,inner sep=0pt, outer sep=0pt, scale=  0.70] at ( 88.71,136.22) {Indecisions};
\end{scope}
\end{tikzpicture}
    \end{subfigure}
    }
    \caption{\label{adult_application.plot}\small Census income prediction. Class 1 (top row) comprises individuals earning \emph{less than} \$50,000 per year, while Class 2 (bottom row) includes individuals earning \emph{more than} \$50,000 per year. Left and Middle: False Selection Rate minus varying levels of $\alpha$. Right: The EPI levels. }
\end{figure}

\subsection{The stable version of the R-value}\label{rval_compare:sec}

In this section, we present simulation results to demonstrate that when $|\mathcal{D}^{test}|$ is small, the stable version of the R-value [defined via \eqref{Q-kc-plus}, \eqref{Q-value-mod}, and \eqref{R-value-max}] exhibits lower variability than the R-value defined via \eqref{Q-kc}, \eqref{Q-value-mod}, and \eqref{R-value-max}. The only difference is that the stable version employs both test and calibration data to stabilize the denominator of the FSP estimate. For clarity and easy presentation, we refer to the version using more data as the R-value and the version using fewer data points as the $R^\chi$-value in this subsection. However, in the main text, we do not give different names to these two R-values because their basic construction steps and underlying ideas are identical. 

To illustrate important patterns in variability, we examine the distribution of the R-value corresponding to a fixed confidence score of $s(x,a)=0.9$.

We consider the setting described in Section \ref{section:simulation} with $
F_{1, M}=F_{1, F} =\mathcal N(\pmb \mu_1, 2\cdot \mathbf I_3)$ and $F_{2, M}=F_{2, F} = \mathcal N(\pmb \mu_2, 2\cdot \mathbf I_3).$ We set  $\pi_{2|F}=\pi_{2|M}=0.8$, $\pmb \mu_1=(1,1,1)^\top$ and $\pmb \mu_2= (2,2,2)^\top$. The confidence scores are constructed as the oracle class probabilities $P(Y=c|X, A)$.  

In Figure \ref{rscore-compare.plot}, we compute $1,000$ $R^\chi$-values and R-values for a fixed score of $s=0.9$ based on randomly generated $\mathcal{D}^{cal}$ and $\mathcal{D}^{test}$. The size of the calibration set is fixed at $|\mathcal{D}^{cal}|=1,000$ and the test set has sizes $|\mathcal{D}^{test}|\in \{5, 50, 200\}$. The columns of Figure \ref{rscore-compare.plot} show the histograms the $R^\chi$-values (left) and R-values (right) with $\mathcal{D}^{test}$ increasing from $5$ (first row) to $200$ (last row).

\begin{figure}[htp]
    \centering 
    \scalebox{1.2}{%
    \input{Figures/r_score_comparison}}
    	\caption{\label{gamma_est_plt}\small  The comparison between the $R^\chi$-value and R-value for varying sizes of the test data set. The left column shows the histograms of the $R^\chi$-value (orange) and the right column shows the histograms of the R-value (green). The $R^\chi$-values and R-values are computed for a fixed confidence score of $s(x,a)=0.9$ based on $1,000$ randomly generated data sets. }
\end{figure}

When $|\mathcal{D}^{test}|=5$ , we notice that the $R^\chi$-value has much more variability than the R-value. This is because the denominator of the $R^\chi$-value only utilizes $5$ observations when computing the total number of selections. By contrast, the R-value uses $1,005$ observations since it has access to data from both $\mathcal{D}^{cal}$ and $\mathcal{D}^{test}$. Moving further down the rows of Figure \ref{rscore-compare.plot}, the advantage of the R-value slowly disappears as $|\mathcal{D}^{test}|$ increases. This causes the variability of both $R^\chi$-value and R-value to become almost identical. 

We conclude from this small simulation that the R-value [defined via \eqref{Q-kc-plus}] is more desirable in settings where $|\mathcal{D}^{test}|$ is small since it can use more data to decrease its variability. However, while the $R^\chi$-value [defined via \eqref{Q-kc}] has more variability for small $|\mathcal{D}^{test}|$, this disadvantage can be quickly overcome through the introduction of a reasonably sized test set. 

On the other hand, the $R^\chi$-value defined via \eqref{Q-kc} offers finite-sample guarantees for FSR control, whereas the R-value defined via \eqref{Q-kc-plus} controls the FSR only asymptotically. In practice, however, the differences in FSR levels between the two versions of FASI are negligible.

\subsection{Imbalanced group sizes}\label{sec:sim_overall_fsr}

In this section, we revisit Simulations 1 and 2 presented in Section \ref{section:simulation} to examine the impact of imbalanced group sizes on FASI's performance. In addition to evaluating group-wise FSRs, we also consider the overall FSR levels, as defined in Equation \eqref{FSR-c}. 

\begin{figure}[htp]
    \centering
    \begin{minipage}{\textwidth}
        \centering
        \scalebox{1}
        { \input{Figures/sim_1_appendix_r4}}
        \caption{\small A similar setup to Simulation 1 (described in Section 4) but with $\pi_{2|F}=\pi_{2|M}=0.5$, and $\pi_M$ ranging from $0.05$ to $0.95$. Overall FSR is controlled for all values of $\pi_M$.}
        \label{fig:first_figure}
    \end{minipage}

    \vspace{1em}

    \begin{minipage}{\textwidth}
        \centering
        \scalebox{1}
        { \input{Figures/sim_2_appendix_r4}}
        \caption{\small A similar setup to Figure \ref{fig:first_figure} above in this response file and Simulation 1 (described in Section 4). However, now with $\pi_{2|F}=0.5, \pi_{2|M}=0.2$, and $\pi_M$ ranging from $0.05$ to $0.95$. Overall FSR is controlled for all values of $\pi_M$.}
        \label{fig:second_figure}
    \end{minipage}
\end{figure}

Our simulation setups are similar to those in Section \ref{section:simulation}, except that we now vary $\pi_M$, the proportion of the Male protected group, from $0.05$ to $0.95$, rather than fixing $\pi_M=\pi_F=0.5$. We consider two settings: (i) $\pi_{2|F}=\pi_{2|M}=0.5$, and (ii) $\pi_{2|F}=0.5$ with $\pi_{2|M}=0.2$. The results from these settings are illustrated in Figures \ref{fig:first_figure} and \ref{fig:second_figure}, respectively. The following observations can be made:

\begin{itemize}
  \item The FASI method controls both the group-wise and overall FSR at the nominal level across all values of $\pi_M$. However, when $\pi_M$ is very small, the Male group-wise FSR control tends to be conservative due to the small sample size.
  \item When the conditional proportions are similar (e.g., $\pi_{2|F}=\pi_{2|M}=0.5$), indicating minimal disparity between male and female distributions, the FCC method performs well in terms of FSR control. In contrast, when heterogeneity is more pronounced [i.e., Setting (ii) with $\pi_{2|F}>\pi_{2|M}$], the FCC method only controls the overall FSR but fails to control the group-wise FSRs. 
\end{itemize}

\subsection{Numerical investigations of the factor $\gamma_{c,a}$}\label{gamma_sim_est:sec}

In Theorem \ref{FSR:thm}, we show that the FASI algorithm can control the FSR at level $\gamma_{c,a} \alpha_c$. This section investigates the deviations of $\gamma_{c,a}$ from $1$. For simplicity, we only focus on $\gamma_{1,a}$. The setup of the simulations is identical to that in Section \ref{section:simulation}. 

Figure \ref{gamma_est_plt} shows the estimates of $\gamma_{1,a}$ for both the Female (green solid line) and Male (orange dashed line) groups. We vary $\pi_{2|F}$ from $0.15$ to $0.85$ while fixing $\pi_{2|M}=0.5$. The y-axis plots the estimate of $\gamma_{1,a}$ averaged over $1,000$ independent simulation runs. In both settings, $\gamma_{1,a}$ is nearly $1$ across both the Female and Male groups. In the most extreme setting ($\pi_{1|F}=0.85$), $\gamma_{1,a}$ deviates away from 1 by $0.01$.

\begin{figure}[htp]
    \centering 
    \scalebox{1.3}{%
    \input{Figures/new_gamma_plt}}
    	\caption{\label{rscore-compare.plot}\small Estimates of $\gamma_{1,a}$ from the simulations in Section \ref{section:simulation}. The solid (green) line represents the estimate of $\gamma_{1,\text{F}}$ for the Female protected group and similarly the orange (long-dashed) line for the Male protected group. }
\end{figure}

\subsection{FASI deployed with other machine learning models}\label{sec:multiple_ML_FSR.plot}

\begin{figure}[tp]
    \centering 
    \scalebox{1.1}{%
    \input{Figures/multiple_ml_models}}
    	\caption{\label{multiple_ML_FSR.plot}\small FSR control for the high risk classification. Left column: The resulting FSR from multiple different ML models that are used to estimate the confidence scores used to calculate the R-value. Right column: The corresponding EPI from different confidence scores. The overall FSR (green / solid) as well as both the Female (blue / dashed) and Male (orange / dot-dashed) protected group FSR's are controlled at the desired 10\% level, for all ML algorithms. The x-axis varies the amount of true proportion of high risk observations from the Female protected group, while fixing the true proportion from the male group at 50\%.  }
\end{figure}

One of the attractive guarantees of our proposed selective inference framework is that we can have the guarantees of Theorem \ref{FSR:thm}, regardless of the machine learning algorithm that is used to generate the confidence scores. In this section, Figure \ref{multiple_ML_FSR.plot} replicates the results of Simulation 1 in Section \ref{section:simulation}, for a variety of machine learning models where the data has two protected groups, Female and Male. In this section we use, logistic regression, GAM, Nonparametric Naive Bayes, and XGBoost \citep{James2013, hastie2009elements, npnb_silverman, chen_xgboost} to estimate the confidence scores that will be converted to the R-values for our FASI framework. 

The left column of Figure \ref{multiple_ML_FSR.plot} plots the FSR for classification group $2$ against a varying proportion of signal $\pi_{2|F}$ from the Female protected group i.e. the true proportion of Females that belong to class $2$. The right column shows the corresponding EPI for each ML model. The goal is to control FSR at the $10\%$ level. 

As we go down the rows, we notice that every model is able to effectively control the False Selection Rate (similar to Simulation 1), however each model has a different EPI. Here, it seems that Logistic Regression, GAM and Nonparametric Naive Bayes have a similar EPI that gets close to $20\%$ in the most extreme case. However, XGBoost has a slightly higher EPI that gets closer to $30\%$ in the worst case. This is a consequence of the accuracy that each ML model has when estimating the true {\color{blue}conditional probability} $P(Y=2|X,A)$ for use in our FASI algorithm. However while some models are more or less accurate than others, they are all able to control the FSR at the desired level.  

\end{document}